\documentclass[12pt]{report}
\usepackage[utf8]{inputenc}
\usepackage[a4paper,right=30mm,left=40mm,top=45mm,bottom=30mm]{geometry}

\usepackage[T1]{fontenc}
\usepackage[spanish]{babel}

\usepackage{helvet} 

\usepackage{lscape,graphicx} 
\graphicspath{ {images/} }
\usepackage{ulem}
\usepackage{caption}
\usepackage{subcaption}
\usepackage[a4paper=true,
            pdfstartview=FitV,
            colorlinks=true,
            bookmarks=true,
            bookmarks=open,
            citecolor=black,
            linkcolor=black,
            bookmarksnumbered=true,
            urlcolor=black,]{hyperref} 
\usepackage[usenames]{color}
\usepackage[dvipsnames]{xcolor}

\usepackage{multirow}

\usepackage{listings}
\definecolor{codegreen}{rgb}{0,0.6,0}
\definecolor{codegray}{rgb}{0.5,0.5,0.5}
\definecolor{codepurple}{rgb}{0.58,0,0.82}
\definecolor{backcolour}{rgb}{0.95,0.95,0.92}
\lstdefinestyle{mystyle}{
    backgroundcolor=\color{backcolour},   
    commentstyle=\color{codegreen},
    keywordstyle=\color{magenta},
    numberstyle=\tiny\color{codegray},
    stringstyle=\color{codepurple},
    basicstyle=\ttfamily\footnotesize,
    breakatwhitespace=false,         
    breaklines=true,                 
    captionpos=b,                    
    keepspaces=true,                 
    numbers=left,                    
    numbersep=5pt,                  
    showspaces=false,                
    showstringspaces=false,
    showtabs=false,                  
    tabsize=2
}
\usepackage{natbib}
\usepackage{apalike}
\usepackage{multirow}
\usepackage{array} 

\usepackage{changepage}   

\usepackage{pdfpages}
\usepackage{pstricks}
\usepackage{a4}
\usepackage{amsmath,amssymb,amsthm,amsfonts} 
\usepackage{changepage} 
\usepackage{bm} 
\usepackage{float} 
\usepackage{fancyhdr} 
\usepackage{color} 
\usepackage[version=3]{mhchem} 
\usepackage{multirow} 
\usepackage{multicol} 
\usepackage{enumitem} 
\usepackage{verbatim} 
\usepackage{etoolbox} 
\usepackage{booktabs} 
\usepackage{notoccite} 
\usepackage[super]{nth}
\usepackage{afterpage}  
\usepackage{mathtools}
\usepackage{listings} 
\usepackage[referable]{threeparttablex} 

\newcommand{\HRule}{\rule{\linewidth}{0.5mm}}
\usepackage{titlesec}
\titleformat{\chapter}[display]
  {\normalfont\bfseries}{}{0pt}{\huge}
\usepackage[detect-weight=true,detect-family=true]{siunitx} 

\usepackage{url}
\begin{document}

\begin{titlepage}

\begin{flushleft}\includegraphics[width=0.2\textwidth]{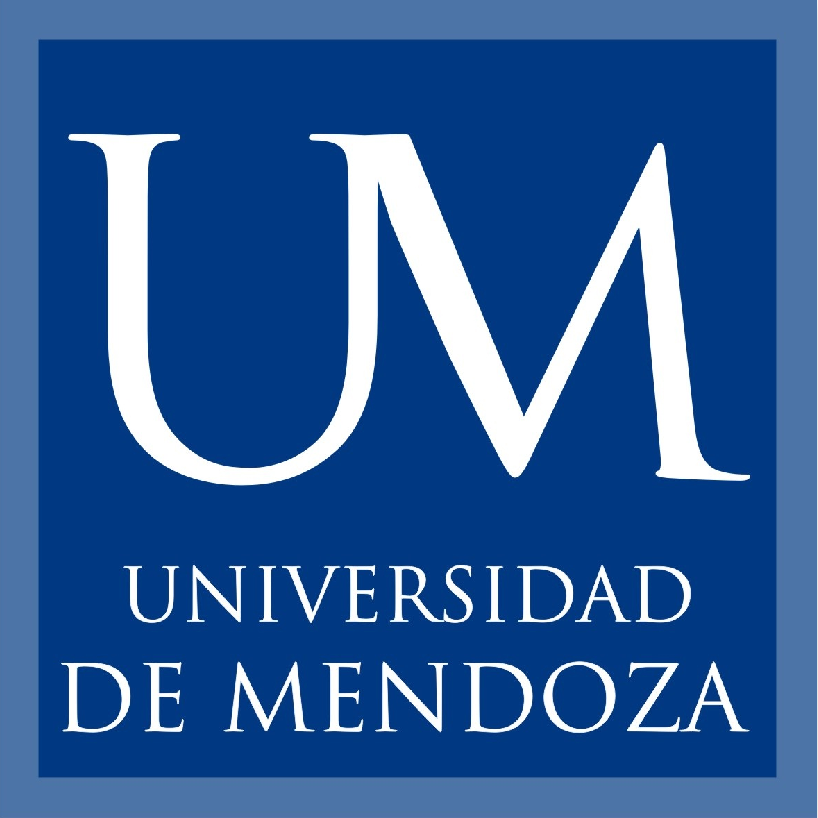} \hspace{8cm}
{\includegraphics[width=0.2\textwidth]{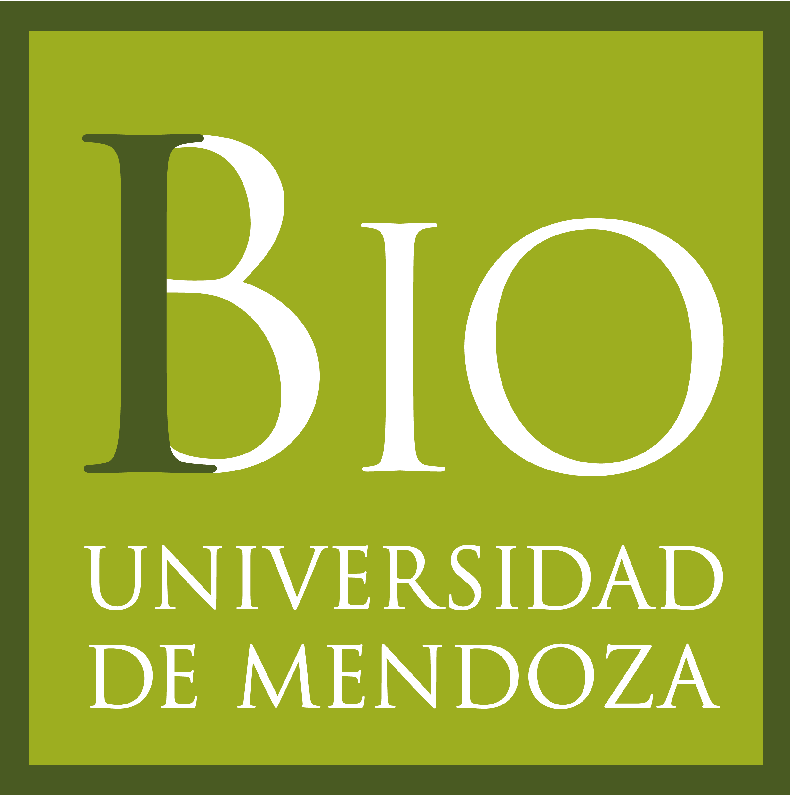}~\\[0.5cm]}\end{flushleft}

\begin{center}
~\\[0.4cm]
\textsc{\LARGE UNIVERSIDAD DE MENDOZA}\\[0.5cm]
\textsc{\LARGE FACULTAD DE INGENIERÍA}\\[0.5cm]
\textsc{\LARGE INSTITUTO DE BIOINGENIERÍA}\\[1cm]

\HRule \\ [0.2cm]
{ \huge{ \bfseries Desarrollo de entrenamientos para el análisis multisensorial de datos}\\[0.5cm]
\HRule \\
\textsc{\Large Trabajo Final}\\[1.25cm]
}\end{center}

\begin{flushleft} \large{
\emph{\textbf{Estudiante:}}
Natasha Maria Monserrat \textsc{Bertaina Lucero}
\\[0.15cm]
\emph{\textbf{Directora: }} 
Bioing. Johanna \textsc{Casado}}
\\[0.15cm]
\emph{\textbf{Asesora: }} 
Dra. Beatriz \textsc{García}

\end{flushleft}
\begin{center}

\vfill

\large{Año 2023}
\end{center}
\end{titlepage}

\clearpage
\thispagestyle{empty}
\null\vfill

\chapter*{Dedicatoria y agradecimientos}
A mis padres Nora y Gabriel, los cuales me brindaron su apoyo infinito durante todos los años de carrera, en especial cuando se tornó difícil. A mis hermanas y hermanos, quienes fueron mi sostén emocional y mi empuje. A Martín Ramírez por su incansable labor a la hora de corregir el presente trabajo. A mi amiga y compañera de estudios Paula Díaz, por el mutuo apoyo y compañía. A mi directora Johanna Casado por el apoyo y guía brindado durante mi última etapa de carrera; por estar disponible para calmar mis inquietudes y ayudarme cuando me encontraba con una dificultad. A Silvina Moyano por apoyarme y alentarme a seguir, confiando en mis habilidades. A Beatriz García, por el apoyo y asesoramiento brindado, además de su guía permanente al grupo de trabajo. Finalmente,
agradecer a Aldana Palma por guiarme en el camino del desarrollo web.

\clearpage
\thispagestyle{empty}
\null\vfill

\chapter*{Resumen}
La percepción es un proceso que requiere una gran cantidad de procesamiento mental, que proporciona los medios por los cuales se crea el concepto que uno tiene del entorno y que ayuda a aprender e interactuar con él. La recopilación de estudios previos a lo largo de la historia ha llevado a la conclusión de que el rendimiento auditivo mejora cuando se combina con estímulos visuales y viceversa. Teniendo en cuenta la consideración anterior, en el presente trabajo se utilizaron las dos vías sensoriales (visión y audición) con la intención de realizar una serie de entrenamientos multisensoriales, los cuales se presentaron en distintas instancias y con el fin de introducir a la sonorización como herramienta en la detección de señales. Se incluyó, además, un desarrollo web para crear un sitio que permitiera ejecutar los entrenamientos diseñados, el cual se encuentra aún en desarrollo debido a dificultades que se presentaron y exceden los límites de este trabajo final. El trabajo que se describe en este informe dio pie a una futura tesis doctoral, la cual cuenta con beca CONICET, donde se plantea el desarrollo de nuevos entrenamientos y el desarrollo continuo del sitio web que permitirá su ejecución.

\textbf{Palabras claves:} Mecanismos de Percepción, Percepción: Entrenamientos, Sonorización de Datos

\clearpage
\thispagestyle{empty}
\null\vfill

\setcounter{page}{6}
\tableofcontents
\listoffigures
\listoftables

\titleformat{\chapter}{\normalfont\huge}{}{20pt}{\huge\bfseries \thechapter. }

\clearpage
\thispagestyle{empty}
\null\vfill


\chapter{Introducción}
\setcounter{page}{1}
\pagestyle{plain}
\pagenumbering{arabic}
La definición de percepción varía de acuerdo con la rama de la psicología que se estudie. A pesar de eso, si se lee la definición que todas ellas dan sobre este concepto, se puede ver que comparten una idea general donde la percepción es considerada como un proceso que se construye con todos los estímulos externos captados por nuestros sentidos, llevando detrás un procesamiento mental complejo para así crear un concepto de nuestro entorno, con el fin de permitir al ser humano desenvolverse en él.

Como la percepción depende de nuestros sentidos, es que a lo largo de los años se ha estudiado cómo nuestro cerebro procesa los estímulos que llegan a él desde las distintas vías sensitivas, siendo la más estudiada la vía visual, en combinación con alguno de los sentidos restantes, corroborando que el grueso de las investigaciones se centran en cómo la visión trabaja mejor cuando se estimula en conjunto con la audición. A pesar de ello, se puede hallar un pequeño porcentaje de artículos en donde se invierte el orden, estudiando cómo el desempeño auditivo mejora al ser acompañado de un estímulo visual. Más aún, dentro de este pequeño grupo se pueden encontrar contados estudios en los que se expone la mejora en la discriminación de frecuencias cuando se somete a una persona a un entrenamiento de 4 a 8 horas, logrando que esta logre una habilidad similar al de un músico profesional.

A pesar de ello, los estudios centrados en entrenamientos auditivos son escasos, llegando a ser casi inexistentes cuando a estos se los especifica más, como en el caso del presente trabajo en el cual se desea elaborar entrenamientos destinados a mejorar la detección de rasgos en señales astrofísicas. Partiendo de lo expuesto anteriormente, es que se elaboraron los siguientes objetivos, para el presente trabajo:

\subsubsection*{Objetivo General}

\begin{enumerate}
 \item Diseñar un programa de entrenamientos que busque mejorar la capacidad de detección de señales, accesible desde una interfaz en la web.
\end{enumerate}

\subsubsection*{Objetivos específicos}

\begin{enumerate}
 \item Desarrollar entrenamientos con el objetivo de capacitar personas en el análisis de datos astrofísicos.
 \item Diseñar dichos entrenamientos teniendo en cuenta personas con diferentes capacidades desde el comienzo del proyecto.
 \item Programar una página web para correr dicho entrenamiento que pueda utilizarse en distintos dispositivos.
\end{enumerate}

Para lograr los objetivos planteados, vinculados con percepción y detección multi sensorial de diferentes señales, es que se eligió utilizar el software multiplataforma Psychopy, el cual permite al usuario elaborar y ejecutar una amplia gama de experimentos en las ciencias del comportamiento, tales como neurociencia, neuropsicología, psicología y lingüística, entre otras. 

A lo largo de la ejecución de los entrenamientos diseñados se encontró la dificultad de la instalación de este software en algunas computadoras. La alternativa era contratar el servicio web creado por Psychopy para ejecutar los entrenamientos en línea, logrando independizarnos de la instalación mencionada. Si bien se utilizó esta plataforma para uno de los entrenamientos, la desventaja de esta alternativa fue su costo. Dado la pronta necesidad de poder ejecutar el entrenamiento en el marco de un proyecto internacional (proyecto REINFORCE, GA 872859, con el soporte de la EC Research Innovation Action bajo el programa H2020 SwafS-2019-1, \url{https://www.reinforceeu.eu}), es que, a pesar de tratarse de un recurso pago, se decidió su uso. Sin embargo, para un proyecto donde se busca desarrollar entrenamientos a largo plazo, esta solución no es la recomendada. Es debido a esto que se comenzó a elaborar un desarrollo web propio para sortear las dificultades que generan los recursos no abiertos al uso público.

Todos los pasos seguidos para llegar a este punto serán relatados en el presente trabajo, el cual fue elaborado a lo largo del año 2022, como una rama independiente dentro del desarrollo del proyecto sonoUno\footnote{ https://www.sonouno.org.ar/}. Este trabajo final de grado, fue llevado a cabo en el Instituto de Bioingeniería de la Universidad de Mendoza, bajo la dirección de la Bioingeniera Johanna Casado y asesorado por la Doctora en Astronomía, Beatríz García.

\chapter{Marco teórico}
\label{cap:marco_teorico}
Para la Psicología Humanitaria la percepción, explicada según la teoría de Gestalt, se entiende como un proceso de extracción y selección de información relevante encargado de generar un estado de claridad y lucidez consciente, que permite un desempeño con el mayor grado de racionalidad y coherencia posible con el mundo circundante \cite{Oviedo}. Siguiendo este lineamiento, la percepción, contrariamente a lo supuesto, no está sometida a la información proveniente de los órganos sensoriales, sino que es la encargada de regular y modular la sensorialidad. En definitiva, esta teoría define a la percepción como una tendencia al orden mental. Primeramente, la percepción determina la entrada de información, para luego garantizar que la información retornada del ambiente permita la formación de abstracciones \cite{Oviedo}.

Por su parte, W. Ittelson \cite{Ittelson}, psicólogo ambientalista, plantea la existencia de dos tipos de percepción: 
\begin{itemize}
    \item Percepción objetual, que enfatiza la búsqueda de las propiedades de estímulos simples, considerando a la persona como un ser pasivo que capta estos estímulos ambientales.
    \item Percepción ambiental, por la cual se focaliza la búsqueda en escenas a gran escala, considerando a la persona como un ser que se encuentra dentro del entorno, moviéndose en su interior como un elemento más. De esta manera, el foco de atención en la investigación de la percepción ambiental es el estudio de las diversas experiencias ambientales que la persona puede tener en relación a su entorno. Con este enfoque se considera que la persona organiza la experiencia obtenida del entorno a partir de determinados propósitos u objetivos.
\end{itemize}

Por otro lado, en la rama de la Psicología Cognitiva nos encontramos con el concepto de que los procesos mentales de una persona son distintos a la conducta que esta demuestra. Para ella existen estados internos en la mente que influyen tanto en la percepción, como en la atención, la memoria y en el pensamiento. 

A su vez existe dentro de la psicología cognitiva como ramificación la neuropsicología, la cual está centrada en la interrelación entre el cerebro y las funciones cognitivas de la mente, buscando determinar cómo funciona la cognición de los individuos con disfunciones o lesiones cerebrales, basándose en los modelos de procesamiento cognitivo normal \cite{Introduccionaneuropsicologia}. Los límites entre esta rama con la neurociencia cognitiva son difusos y difíciles de ver; esto se debe a que en el caso de la neurociencia cognitiva su objeto de estudio son los procesos mentales superiores  denominados como procesos cognitivos, los cuales abarcan el pensamiento, lenguaje, memoria, atención, percepción y movimientos complejos. Si se tuviera que resaltar una diferencia, esta sería la utilización exclusiva de técnicas no invasivas para proceder con su estudio en la neurociencia cognitiva.

En base a lo planteado por las ciencias mencionadas podemos considerar como adecuado definir a la percepción como un proceso que se construye con estímulos externos captados por nuestros sentidos, llevando detrás un procesamiento mental complejo para así crear un concepto de nuestro entorno, permitiéndonos desenvolvernos en él. Este procesamiento puede verse afectado por la atención del individuo, como también por fallas en los canales fisiológicos encargados de llevar la información desde el exterior al interior. En síntesis, la percepción es un proceso complejo que requiere de múltiples disciplinas para ser estudiada.

\subsubsection*{Percepción visual y auditiva}
    Como bien sabemos, la visión es un sentido a distancia ya que nos permite interactuar con el entorno sin tener contacto directo con él. Nuestros ojos solo perciben un pequeño rango del espectro electromagnético (el de la luz visible), que comprende longitudes de onda desde los 400 a los 700 nm, limitando en alguna medida nuestra percepción. Aun con esta limitación, nuestra interpretación del entorno a partir de la información lumínica captada por nuestra visión es suficiente para nuestra supervivencia.
    
    Este proceso de convertir el estímulo lumínico en algo que pueda verse consta de tres momentos: captar, convertir y procesar. En un primer término los receptores de la visión deben captar la radiación electromagnética que lleguen hacia ellos. En segundo lugar, estos deben convertir la información de la luz recibida en señales eléctricas. Por último, se procesan dichas señales para lograr reconocer las características del estímulo. Entre estas características encontramos el tamaño, la forma, el color, la posición, la iluminación y el contraste, la dimensión y el movimiento de lo captado por la visión.
    
    Si especificamos más este proceso de percepción podemos mencionar el modelo de la doble ruta propuesto por Ungerleider y Mishkin \cite{percepcionvisual}, el cual habla de un sistema visual dividido en dos rutas anatómicas y funcionales diferentes. Por un lado la vía visual ventral del qué, que permite percibir la forma de los objetos y su identificación; y por el otro, la vía visual dorsal del dónde y del cómo, que se centra en el procesamiento de las relaciones espaciales de los objetos, promoviendo las acciones motoras en relación a ellos. 
    
    Un factor a tener en cuenta es que la visión no es un sentido que trabaja solo, sino más bien en conjunto con todos los demás. En particular nos centraremos en este trabajo en el conjunto de la visión y la audición. Individualmente, la audición posibilita la percepción del lenguaje y la música, pero en conjunto con la visión permite modular la información a la hora de localizar estímulos que aparecen en el campo visual, ofreciendo cierta información extra como la posición, distancia y dirección de los estímulos en movimiento. Adicionalmente, con la audición podemos percibir estímulos que aún no entran en nuestro campo de visión.
    
    De una forma similar a la visión, el proceso llevado a cabo por el sistema auditivo consta de las mismas tres tareas de captar, convertir y procesar, diferenciándose en el tipo de estímulo que recibe. Primeramente, el estímulo sonoro debe ser captado por los receptores que se encuentran en la cóclea. El siguiente paso es convertir lo captado a señales eléctricas para finalmente procesarlas y reconocer las características del estímulo, logrando identificarlo y localizarlo. Entre las características encontramos la intensidad, el tono, el timbre y la duración.
    
    Otro aspecto explicado de forma similar a la visión es el procesamiento auditivo, el cual también se basa en un modelo de doble ruta con dos vías anatomofuncionales diferenciadas. Por un lado la vía del qué, permite la discriminación e identificación de los sonidos a partir del análisis de frecuencias, dejando una lateralización izquierda para el lenguaje y derecha para la música. Por su parte, la vía del dónde procesa la información del sonido en el espacio, comparando sonidos mono y binaurales \cite{percepcionauditiva}.

    Es entonces que en el trabajo conjunto de la audición y la visión nos encontraremos con que el grueso de las investigaciones se centra en las influencias auditivas con respecto a la percepción visual \cite{Opoku-baah}. Aun así podemos hallar un pequeño porcentaje de artículos que se enfocan en la influencia que tiene la visión en la percepción auditiva. De ellos podemos concluir que el rendimiento perceptivo de la audición mejora cuando a la información auditiva captada se le suma información visual con la que coincida espacial y temporalmente, la cual puede ser relevante o no a la tarea realizada.
    
A pesar de esta última conclusión, también se han encontrado pruebas de la mejora en la discriminación de frecuencias que una persona puede percibir al ser entrenada para reconocer estas diferencias, las cuales pueden ser observadas en cortos períodos de tiempo. Por ejemplo, se sabe que los músicos profesionales pueden discriminar frecuencias mucho mejor que las personas que no se dedican a la música, pero si se somete a estas últimas a un entrenamiento de 4 a 8 horas, su rendimiento puede llegar a ser similar al de los músicos profesionales \cite{Oxenham}.
    
Por lo observado hasta la fecha, no se ha profundizado en el tópico de los entrenamientos auditivos, y menos aún en ejercicios destinados al entrenamiento en la detección de señales astrofísicas. Es por ello que en el presente trabajo ahondaremos en este tema, aportando más datos sobre estos entrenamientos. Para realizar estos últimos se utilizará el software PsychoPy, un programa multiplataforma que permite al usuario ejecutar una amplia gama de experimentos en las ciencias del comportamiento.

Se evidencia esta necesidad de entrenamientos en la tesis doctoral de la Bioingeniera Johanna Casado, titulada \textit{``Investigación sobre acceso, uso y exploración efectiva de datos observacionales y bibliográficos astronómicos a partir de la sonorización''}. A medida que dicho trabajo avanzó en su construcción y más personas usaron el software desarrollado, es que surgió la necesidad, expresada por los usuarios, de elaborar entrenamientos con el fin de aprender a utilizar la sonorización como herramienta en el análisis de datos. La tesis doctoral que se menciona se desarrolló en el mismo equipo de trabajo donde se realizó el presente trabajo final de grado, y se encuentra en su etapa final.

\chapter{Estado del arte}
\label{cap:estado_del_arte}
Como fue expuesto en el capítulo anterior, no hay antecedentes en la elaboración de entrenamientos destinados a mejorar la detección de señales astrofísicas. Pero si hay un antecedente de un pequeño estudio de entrenamiento para mejorar el uso de herramientas de sonorización de datos. En el año 2005 se publicó un artículo titulado \textit{``Effects of Auditory Context Cues and Training on Performance of a Point Estimation Sonification Task''} \cite{Training}, en el que se expone la necesidad de capacitar a los usuarios de software de sonorización en cómo interpretar los datos visuales una vez pasados a sonido. En este artículo no solo se recalca la necesidad de entrenar a los usuarios, sino también de mejorar las sonorizaciones para que sean más fieles a lo mostrado en su despliegue visual. A pesar de implementar un entrenamiento, el mismo fue elaborado con herramientas diferentes a las que se usaron en el presente trabajo. Adicionalmente, los datos sonorizados usados en el trabajo del 2005 se centraron más en la capacitación para poder interpretar distintas marcas en las curvas.

En la búsqueda de posibles usos de la herramienta Psychopy en el tópico de entrenamientos, se llegó a un vacío de antecedentes. Si se incluyen en la búsqueda de antecedentes posibles usos de este software en el área de accesibilidad y rehabilitación, surge un trabajo donde se elaboró un programa de distintos estímulos, destinado al apoyo en la neurorehabilitación de pacientes con acceso limitado a programas de este estilo debido a que se encuentran en comunidades remotas\cite{Angel}.

Finalmente, ampliando aún más el rango de búsqueda, centrando en qué tipos de aplicaciones se hace uso de Psychopy, se llegó a encontrar todo tipo de estudios en neurociencia, ya sea para generar estímulos visuales con distintos fines como también utilizar estímulos visuales y/o auditivos. Tal es el caso de un estudio realizado para evaluar la atención selectiva \cite{Ulloa}, donde se utilizó el software mencionado para el diseño del paradigma a utilizar, basado en el test d2. En la búsqueda de referencia, nos encontraremos mayormente con experimentos realizados con la herramienta, que están orientados a estudios de neurociencia y ciencias del comportamiento.

Sin embargo, conociendo las características que presenta PsychoPy y dada la experiencia previa en el equipo de trabajo, se consideró que la herramienta mencionada es adecuada para los objetivos del proyecto propuesto. Inclusive, es probable que se estén abriendo nuevas posibilidades en su uso con el desarrollo de este trabajo.

\chapter{Herramientas y métodos}
\label{cap:herramientas_metodos}
\section{Psychopy}

\begin{figure} [H]
    \centering
    \includegraphics[width=0.8\textwidth]{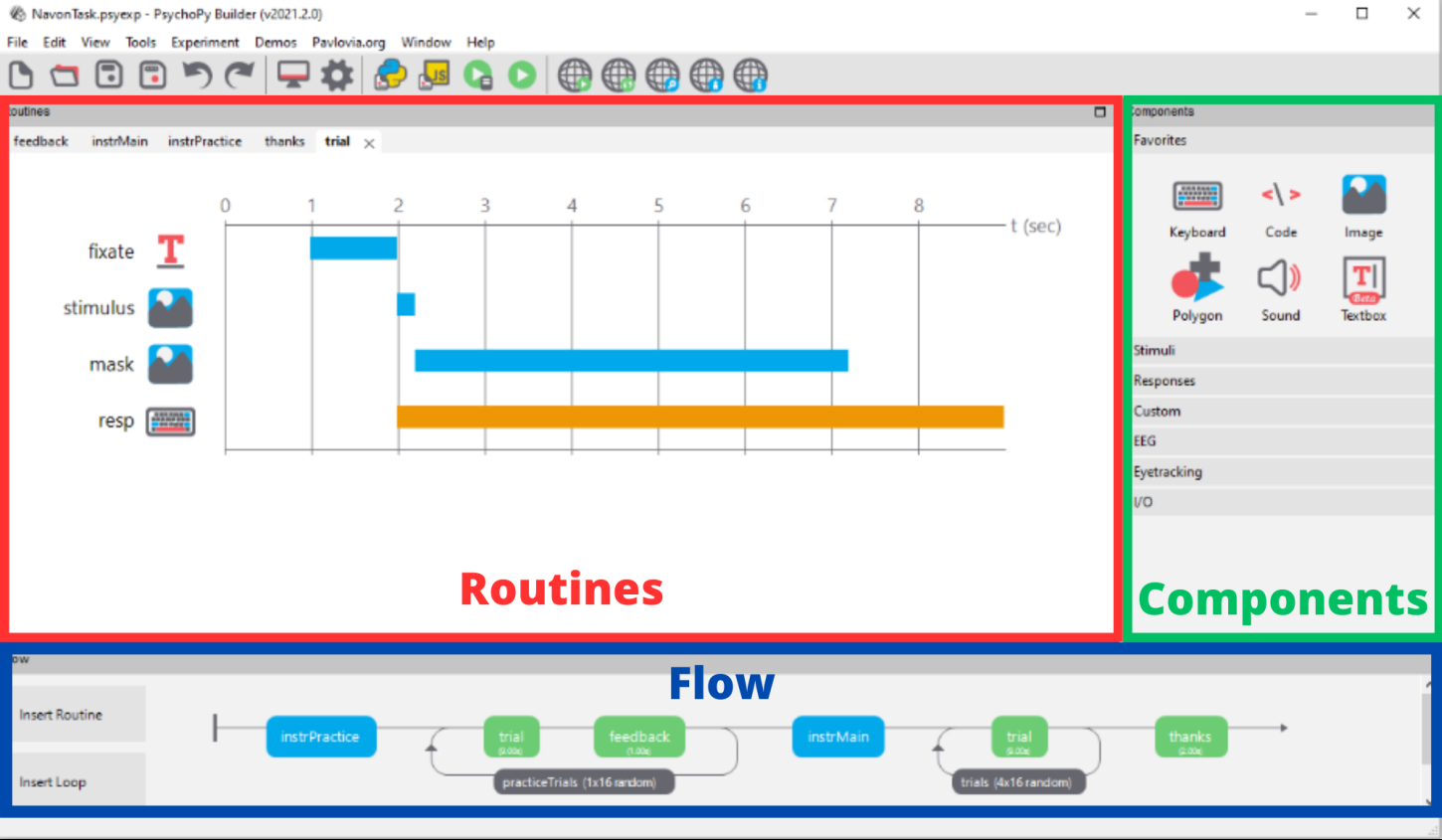}
    \caption{Interfaz `Builder' de Psychopy. Encerrado en verde encontramos la sección de \textit{Components}. En rojo la sección \textit{Routines}. Finalmente, en azul la sección \textit{Flow}}
    \label{fig:1}
\end{figure}

PsychoPy es un proyecto desarrollado desde el año 2002, que nació como una biblioteca del lenguaje de programación Python para realizar experimentos de neurociencia visual, pero creció hasta proporcionar una gran variedad de estímulos y características\cite{Peirce}. Entre sus mejoras, podemos destacar el desarrollo de una interfaz gráfica llamada \textit{``Builder''} (figura \ref{fig:1}), la cual facilita el uso de la biblioteca para usuarios que no son programadores o saben muy poco de programación. La idea central de dicha interfaz es permitir al usuario la creación gráfica de un experimento, generándose un script\footnote{En informática, script es un término usado para referenciar al código de un programa o un fragmento de este.} en lenguaje Python de forma automática por el software gracias a un interpretador de texto propio. Este script se genera con comentarios que guían al usuario en lo que cada parte del código hace, facilitando su comprensión.

\begin{figure} [H]
    \centering
    \includegraphics[width=0.6\textwidth]{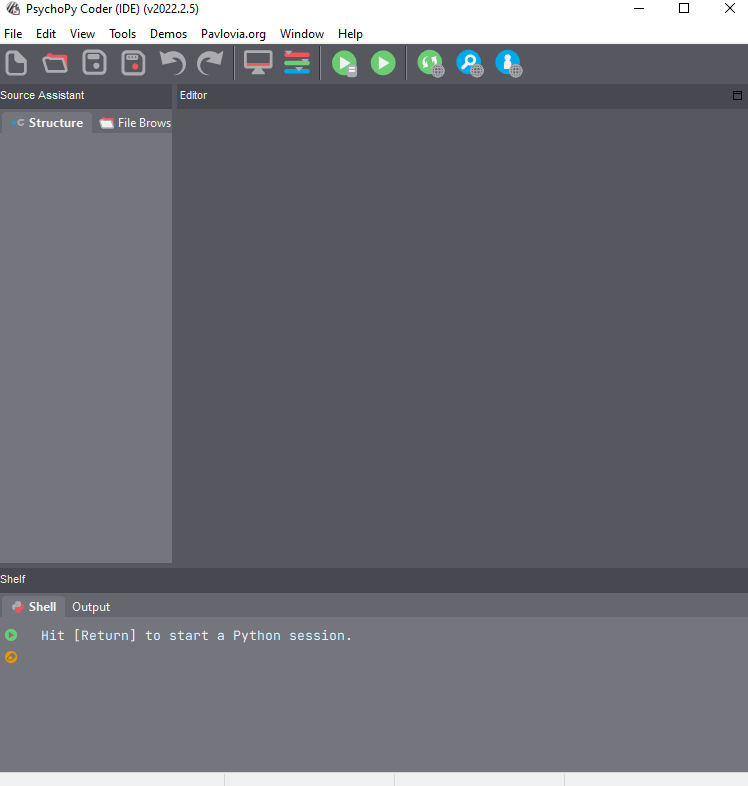}
    \caption{Interfaz `Coder' de Psychopy}
    \label{fig:2}
\end{figure}

Si el usuario es programador o tiene conocimiento de ello, el paquete ofrece un editor de código \textit{``Coder''} (figura \ref{fig:2}) que permite la ejecución y creación de script en Python, generando el experimento desde cero, con un alto grado de personalización del mismo. Otra opción es crear el ``esqueleto'' del experimento desde la interfaz gráfica y personalizarlo desde el editor de código, lo cual es posible ya que ambas opciones se instalan con el paquete. La desventaja de ello es que el camino es unidireccional, es decir, que los script no se vuelven a convertir en una representación gráfica.

Entre las ventajas que proporciona este Software encontramos que es multiplataforma y que está bajo una licencia GPL3\footnote{Licencia Pública General que garantiza que el software es libre y de código abierto. Permite a los usuarios usarlo, modificarlo y compartirlo libremente.}, lo que permite su uso y adaptación libre. Es esta última característica la que permite expandir la capacidad del Builder de generar texto que será interpretado en lenguaje Python para que se generen script en distintos lenguajes. En cuanto a este último punto, en años recientes se agregó la opción de generar un script del experimento en lenguaje JavaScript, permitiendo compartirlo y ejecutarlo en línea. En base a ello es que el proyecto ofrece tanto un repositorio como una plataforma online, donde se pueden almacenar los experimentos creados. Para su ejecución en dicha plataforma se deben comprar créditos, los cuales se comparten con los participantes del experimento, habilitándolos para realizar el experimento online (se consume un crédito por cada participante que completa el entrenamiento).

Retomando la interfaz gráfica, ella se divide en tres secciones bien definidas, tal como se observa en la figura \ref{fig:1}. La primer sección es la de \textit{Components} (recuadro verde de la figura \ref{fig:1}), en donde se encuentran todos los estímulos que podremos utilizar para nuestro proyecto, tales como imágenes y sonidos entre otros. También ofrece componentes de respuesta como entrada por teclado o mouse, y demás. Todos ellos se ordenan por secciones haciendo mas fácil su ubicación y colocación en una rutina. Como segunda sección encontramos las \textit{Routines} (recuadro rojo de la figura \ref{fig:1}), la que nos permite ir diagramando las distintas rutinas que componen el experimento desarrollado. Es en esta sección donde se van dando los parámetros a los distintos componentes seleccionados. Cada componente tendrá parámetros específicos, y algunos en común como es el tiempo que se reproducirá o mostrará por pantalla.

Finalmente, la Tercera sección corresponde al \textit{Flow} (recuadro azul de la figura \ref{fig:1}), en la que se diagrama todo el experimento realizado, definiendo cuantas rutinas se implementarán. Tal como se puede observar su despliegue gráfico es en forma de diagrama de flujo, permitiendo programar loops en rutinas individuales.

Algunos parámetros generales del experimento se pueden programar desde esta ventana, seleccionando el botón \textit{settings}, lo que abre un menú que permite configurar, entre otras cosas: el nombre del experimento; la información pedida al participante antes de ejecutarlo; tamaño de ventana que se desea usar para el despliegue del mismo y en qué tipo de archivos queremos que se almacenen las respuestas obtenidas del participante. 

\section{sonoUno}

\begin{figure} [H]
    \centering
    \includegraphics[width=0.6\textwidth]{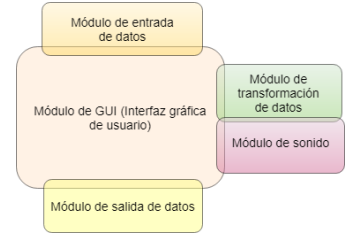}
    \caption{Diseño modular de sonoUno. Extraído de \cite{sonoUno}.}
    \label{fig:3}
\end{figure}

SonoUno es un software multiplataforma que tiene sus inicios en el trabajo doctoral de la bioingeniera Johanna Casado y que fue desarrollado en lenguaje Python. Su objetivo es ser una herramienta accesible que ayude a personas con diversas capacidades sensoriales en el análisis de datos científicos, ofreciendo tanto la sonorización como la visualización de los mismos. Fue elaborado teniendo como eje central el diseño centrado en el usuario, atendiendo a los estándares de accesibilidad dados en la ISO 9241-171:2008 (\textit{Ergonomics of human-system interaction - Part 171: Guidance on software accessibility}). El software presenta un formato modular \cite{sonoUno}, según se observa en la figura \ref{fig:3}, teniendo cada módulo una función específica.

Para iniciar la sonorización y análisis de datos, el usuario debe ingresar una tabla de datos numéricos, que puede ser en formato .csv o .txt, los cuales se despliegan gráficamente, mientras que la sonorización podrá ser iniciada con la función `Play'. Entre las posibilidades que se ofrecen, se encuentra el recorte de una zona específica del gráfico, la elección de distintos instrumentos para sonorizar, cambio del tono del sonido producido o el marcado de puntos específicos en los gráficos, entre otras. Al finalizar con el análisis de los datos, se proporciona la posibilidad de guardar tanto el sonido generado como el gráfico trabajado y las marcas realizadas en el gráfico. 

Una de las características más notable que ofrece el software es su interfaz web, fácil de acceder desde cualquier dispositivo, lo que aumenta las posibilidades de uso. 

\section{Desarrollo Web}  

Un desarrollo web es complejo de realizar, para ello se deben tener claro ciertos conceptos, además de cómo funciona cada parte involucrada en el proceso. Para simplificar la explicación de este proceso, se seguirá la estructura servidor-cliente. Primero se procederá con la explicación front-end (cliente) para concluir con el back-end (servidor). Si bien hay más partes involucradas en el funcionamiento de una página web, para dar estructura a la explicación es que se decidió usar esta división, además de que ella es muy popular en el mundo de la programación web a la hora de distinguir lo que ve el usuario final de lo que se tiene ``por detrás'' en un sitio, encargándose de todos los procesos.

\subsection{Front-end}

Se podría definir al front-end como la parte que se muestra al usuario y con la cual interactúa, por lo que todo su desarrollo está dirigido al diseño de la cara visible de una página web. Las tecnologías más usadas para ello son HTML, CSS y JavaScript.

Para el correcto funcionamiento de un front-end necesariamente se requieren conocimientos en los tres lenguajes mencionados, ya que su programación los necesita. Para el desarrollo del presente trabajo, se decidió usar la biblioteca\footnote{Las bibliotecas agrupan archivos de código que aportan distintas funcionalidades. Tienen como objetivo facilitar la programación.} React para construir dicha interfaz de usuario \cite{REACT}. 

\subsubsection{HTML}

El HTML es un lenguaje de etiquetado de hipertexto (en ingles, \textit{HyperText Markup Lenguage}). Toda página web necesita su archivo \textit{index.html}, ya que es éste el que definirá la estructura contenida en ella. Cabe aclarar que este no es un lenguaje de programación. 

\begin{figure} [H]
    \centering
    \includegraphics[width=0.8\textwidth]{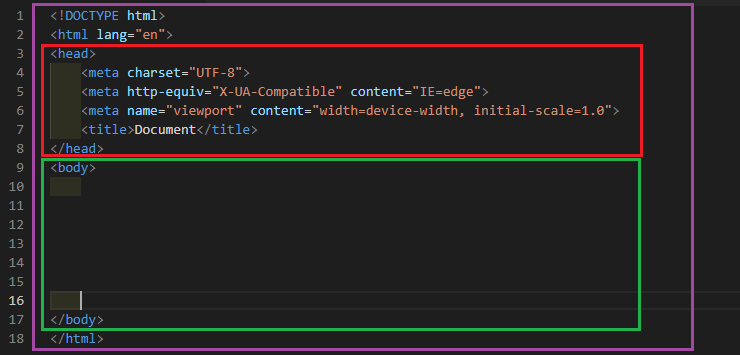}
    \caption{Estructura básica de un archivo html.}
    \label{fig:4}
\end{figure}

La estructura de una página web se define mediante etiquetas, que indican la clase de elemento que tienen contenido para ser mostrado en la web. Un ``esqueleto'' básico de un archivo .html es como se muestra en la figura \ref{fig:4}. En dicha figura se observa la delimitación de dos secciones importantes, de las cuales la primera de ellas (encerrada por el recuadro rojo) empieza y termina con una etiqueta  <<head>> <</head>>. Dentro de esta etiqueta se encierran los metadatos del documento, como por ejemplo el título, o el juego de caracteres que se usarán en la página, entre otros. Esta sección no es visible para el usuario.

Lo que se mostrará al usuario se encierra entre la etiqueta <<body>> <</body>> (recuadro verde de la figura \ref{fig:4}), y puede incluir desde archivos multimedia (audios, videos, imágenes), hasta texto. Es en esta sección donde se estructura todo lo que se quiere mostrar en una página web y cada elemento tiene su etiqueta correspondiente. La forma de visualización es en forma de cascada, es decir, que lo que se escriba primero es lo que aparecerá primero en la página mostrada por el navegador.

\subsubsection{CSS}

El CSS es un lenguaje de estilos (en ingles, \textit{Cascading Style Sheets}), que permite pasar de una página web plana, usando únicamente HTML, a una página web personalizada y más familiar para el usuario. Este código se inserta al archivo .html en la sección del head mediante una  etiqueta <<link>>, ya que debe ser escrito en un archivo distinto con extensión .css. Si bien se podría personalizar los estilos en el archivo .html directamente, esto no se considera una buena práctica, además de que el archivo se vería muy desorganizado. 

Además de permitir cambiar la fuente usada, el color de fondo o del texto y el tamaño de imágenes, es la opción de ordenar el contenido estructurado con HTML en grillas o en bloques lo que hace que sea un lenguaje indispensable a la hora de diseñar una página web. 

\begin{figure} [H]
    \centering
    \includegraphics[width=0.8\textwidth]{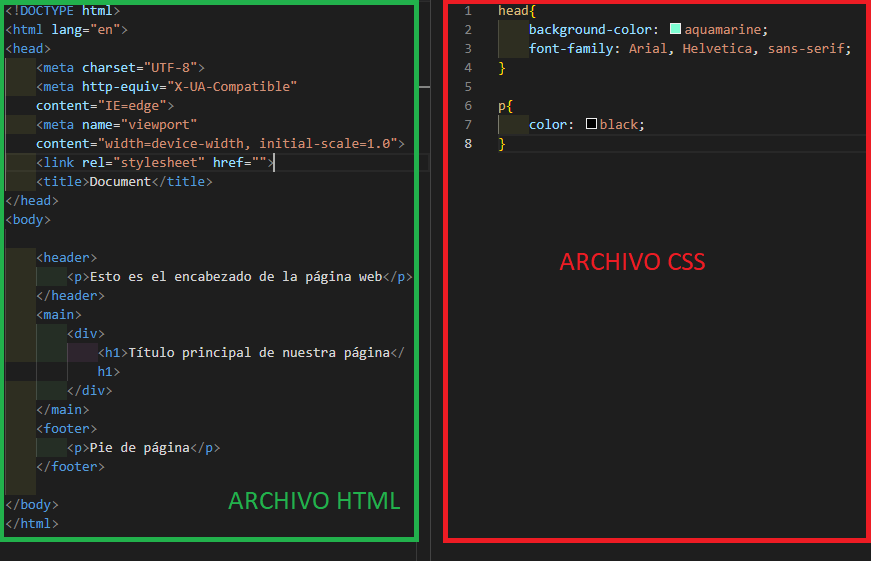}
    \caption{Ejemplo de personalización de un archivo HTML. A la derecha un documento escrito en lenguaje HTML (recuadro verde). A la izquierda un archivo escrito en CSS (recuadro rojo).}
    \label{fig:5}
\end{figure}

Para escribir una hoja de estilos se aprovechan las etiquetas usadas en el archivo .html, tal como se muestra en la figura \ref{fig:5}. El código a la izquierda se estructura por un selector, el cual coincide con la etiqueta de HTML que se desea personalizar, seguida por todas las declaraciones que se deseen. Estas últimas se componen por una propiedad en el archivo CSS (backgraund-color, en el ejemplo) y su valor.

\subsubsection{JavaScript}

JavaScript es un lenguaje de programación que trabaja desde el lado del front-end, o del cliente, con el cual se da funcionalidad a una página web. Este es un paso importante a la hora de crear páginas con las que el usuario pueda interactuar ya que, además de crear o modificar datos, es el que establece la conexión con el servidor. Popularmente se lo conoce como el lenguaje de ejecución de navegadores. Esto está muy ligado a que nació exclusivamente con esa funcionalidad, pero al día de hoy se puede usar en servidores utilizando entornos de ejecución que lo permitan, como es el ejemplo de Node.js que se empleo en el desarrollo back-end del presente trabajo.

\subsection{Back-end}

El back-end es la parte no visible al usuario, la cual trabaja con el servidor y la base de datos. A la hora de programarlo se puede elegir entre varias opciones como Python, PHP, Java o incluso, como ya se mencionó, usar un entorno apropiado para programar con JavaScript. La amplia variedad de opciones se debe a que el back-end será interpretado por un servidor y no por un navegador, este último solo entiende lenguajes de front-end. Para dar un mejor entendimiento de lo que hace un back-end es que debemos aclarar ciertos conceptos, como son: (1) servidor, (2) hosting y dominio.

\subsubsection{Servidor}

Un servidor web es el lugar donde se aloja un sitio web. Este se compone de dos partes que trabajan en conjunto \cite{Servidor}:
\begin{itemize}
    \item Hardware: computadora encargada de almacenar los archivos HTML, CSS, script de JavaScript y multimedia de un sitio web. En él se alojan varios sitios web y tiene conexión a internet, permitiendo el intercambio de datos.
    \item Software: el cual incluye al sistema operativo específico para servidores, sistema de seguridad,  software de servidor web que entienda de direccionamiento y del protocolo HTTP o HTTPS (\textit{Hypertext Transfer Protocol Secure}), el cual está basado en la estructura servidor-cliente para la petición de datos y recursos, entregando los recursos solicitados por el cliente. Según la complejidad de servidor, además del ya mencionado, tendrá mas software encargado de controlar el acceso a los distintos archivos alojados.
\end{itemize}

De forma simplificada, la estructura servidor-cliente funciona de la siguiente manera: al solicitar el usuario una dirección web por medio de un navegador, este lo solicita al servidor mediante el protocolo HTTP/HTTPS. Al llegar la petición al servidor, el software web la acepta, busca el documento requerido y lo envía hacia el navegador. Finalmente el usuario puede acceder a la capa de cliente (front-end) del sitio web deseado \cite{Servidor}.

Se debe hacer una aclaración con respecto a protocolos, reglas para la comunicación entre Cliente-Servidor y se requiere de varios para que esta funcione correctamente \cite{protocols}. En los párrafos anteriores solo se hizo mención de uno: HTTP y de su evolución a HTTPS, el cual puede ser considerado como la base de esta comunicación. Este protocolo es un conjunto de reglas que permite realizar peticiones para el intercambio de archivos multimedia, tales como imágenes y sonidos, además de texto. Al combinarse con un protocolo de seguridad es que pasa a ser HTTPS.

Para cerrar el tema, debemos mencionar el que se ha adoptado como el protocolo de comunicación oficial de internet: TCP/IP  (\textit{Transmission Control Protocol/Internet Protocol})\cite{protocols}. En el se ``fusionan'' las funcionalidades de ambos protocolos, logrando comunicaciones confiables a través de internet, donde con TCP se empaquetan y dividen los archivos que se desean enviar y por medio del IP se envían a la dirección correcta. Cada dispositivo conectado a la red posee una dirección IP, la cual es numérica y única para cada dispositivo, aunque esto último no es tan estricto.

\subsubsection{Hosting y Dominio}

El lugar que ocupa un sitio web en un servidor es conocido como hosting. Se podrían almacenar todos los archivos en computadoras personales, pero esto traería desventajas como por ejemplo, si un usuario desea ingresar a nuestra web y nuestra computadora se queda sin acceso a internet, el proceso de petición queda cortado abruptamente. Este y otros inconvenientes son solucionados con el alojamiento en un servidor web.

Una ventaja que posee el alojamiento de un sitio web en un servidor es el uso de la IP, la cual al quedar fija evita problemas de comunicación entre el cliente y el servidor. Sin embargo, recordar todas las direcciones numéricas de todos los sitios web que una persona quiera acceder no es tarea sencilla, es por ese motivo que se implementó el Sistema de nombres de dominio o DNS (por sus siglas en ingles, \textit{Domain Name System}), el cual asocia la dirección IP numérica con un nombre de dominio textual. Para localizar un sitio web, el usuario hace uso de URLs, las cuales son localizadores de recursos. Una URL normalmente posee las siguientes partes: 
\begin{itemize}
    \item http/https: indica el uso de tal protocolo para realizar peticiones.
    \item www.: nombre del servidor web. No todos los sitios web lo poseen, ya que no todos se alojan en el mismo servidor web.
    \item Nombre de dominio de segundo nivel, como por ejemplo wikipedia, google, youtube.
    \item .com: nombre de dominio de nivel superior TLD (\textit{Top-Level Domain}), el cual puede ser genérico o geográfico.
    \item ubicación jerárquica del archivo, como por ejemplo /home, / , /index.
\end{itemize}

Los dominios pueden comprarse, lo que permite una diferenciación del sitio web, o pueden ser proporcionados por el servidor web que se use para alojarlo. Para el presente trabajo se uso el servicio ofrecido por Heroku para alojar el sitio web durante la etapa de desarrollo, el cual ofrece su propio dominio heroku.com.

\chapter{Desarrollo de Ingeniería}
\label{cap:desarrollo}
 \section{Desarrollo de los entrenamientos} \label{sec:Desarrollo-Ent}
 
En esta sección se describirá el proceso de creación de los distintos entrenamientos hechos a lo largo del año 2022 y cómo fueron evolucionando con tiempo gracias a las sugerencias hechas por los participantes que realizaron los entrenamientos en cada etapa. 

\subsection{Entrenamiento prototipo}
 
Para esta tarea se hizo uso del paquete PsychoPy, más precisamente de su interfaz gráfica. En ella, cada experimento se describe en rutinas, donde dentro de ellas puede haber un conjunto de componentes. En el caso de los entrenamientos desarrollados, en estas rutinas se desplegaron componentes visuales y de audio, como también de texto para dar contexto y brindar información de como se debía proceder.

\begin{figure} [H]
    \centering
    \includegraphics[width=1\textwidth]{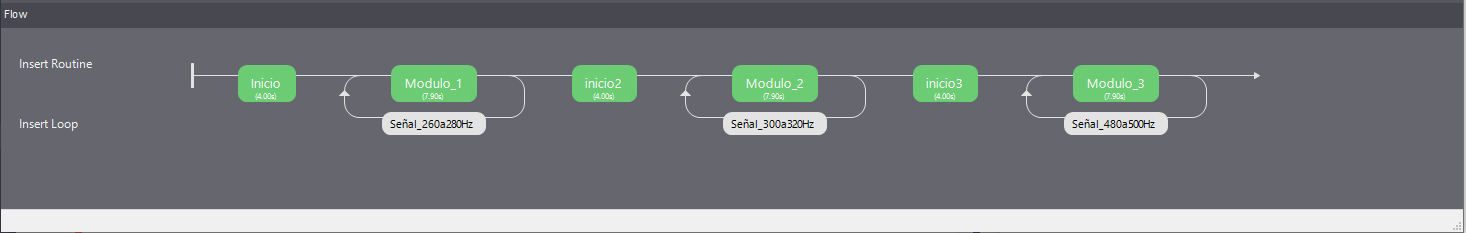}
    \caption{Diagrama de entrenamiento prototipo.}
    \label{fig:6}
\end{figure}

El primer entrenamiento desarrollado con la herramienta fue de prueba y se dividió en seis rutinas con tres loops\footnote{Bucles o lazos cerrados donde las instrucciones contenidas dentro de los mismos se repiten una cantidad determinada de veces. A lo largo del texto se empleará el término en ingles para hacer referencias a los mismos.}, tal como muestra la figura \ref{fig:6}. El fin de este entrenamiento fue probar la herramienta y familiarizarse con la construcción gráfica ofrecida por PsychoPy. Otro objetivo que se buscaba alcanzar era probar la calidad de audios sintetizados a partir de un código propio, mezclado con ruido blanco\footnote{Un ruido blanco es aquel que presenta una densidad espectral constante, es decir, que tiene un espectro con todos los componentes de frecuencia en igual proporción}. Este último objetivo se debió al hecho de que en primer lugar se buscaba probar si el entrenamiento serviría para que los usuarios aprendan a distinguir  distintos tipos de sonidos mezclados con ruido. El script usado para generar estos sonidos se encuentra en el Apéndice \ref{cap:apend}.

Cada rutina cumplía una función específica, tal como se describe a continuación, siguiendo el orden mostrado por el diagrama de la figura \ref{fig:6}:

\begin{itemize}
    \item Inicio: En esta rutina se empleó un único componente de texto el cual señalaba el comienzo del módulo 1, indicando que se mostrarían componentes visuales y de audio al mismo tiempo. Se la programó con 4 segundos de duración.
    \item Módulo 1: Rutina programada en forma de loop para repetir el orden de los componentes por cada audio e imagen listados en un archivo excel. Se usaron audios con un rango de frecuencia comprendida entre 260 Hz a 280Hz. Además de mostrar los archivos multimedia al usuario, se programó la petición de una repuesta por parte de este, la cual debía ser ingresada por teclado según las instrucciones que se mostraban por pantalla. Su duración era de 7,9 segundos.
    \item Inicio2: Similar a la rutina de Inicio del primer ítem, aquí se empleó un único componente de texto que indicaba el comienzo de la próxima rutina con componentes visuales y de audio. Su duración también era de 4 segundos.
    \item Módulo 2: su programación fue igual que el módulo 1, con la diferencia de las imágenes y audios mostrados. Estos últimos correspondían a un rango de frecuencias de entre 300 Hz a 320 Hz. Su duración también era de 7.9 segundos.
    \item Inicio3: Rutina que indicó el comienzo del tercer y último módulo. También se empleó solo un componente de texto para enunciar ésto, además de anunciar que el próximo módulo desplegaría imagen y audio a la vez, como en los módulos anteriores.
    \item Módulo 3: Repetición de lo programado en el módulo 1 y 2, cambiando las imágenes y audios, los cuales correspondían a frecuencias en el rango de 480 Hz a 500 Hz. De la misma manera, su duración era de 7,9 segundos.
\end{itemize}

\begin{figure} [H]
    \centering
    \includegraphics[width=1\textwidth]{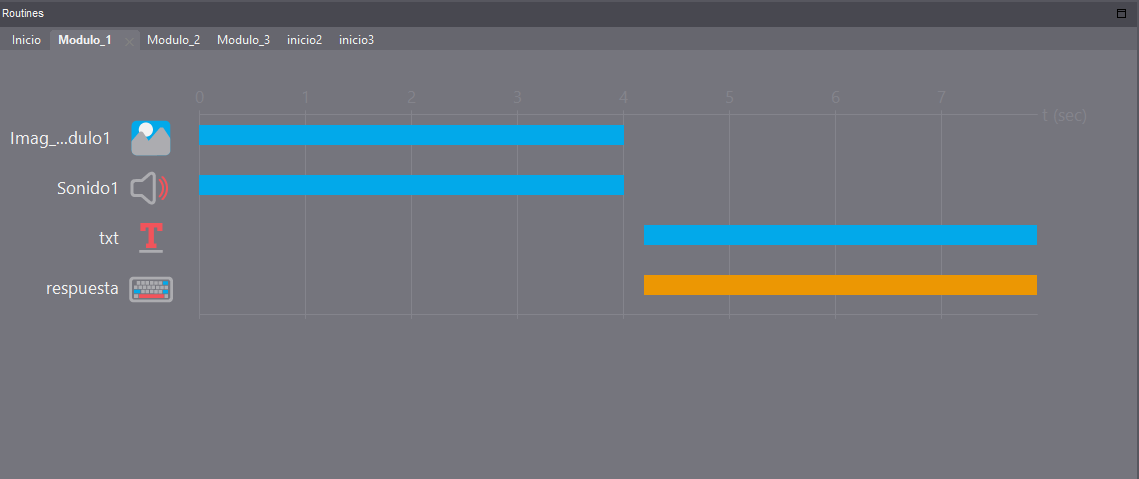}
    \caption{Estructura de los Módulos del entrenamiento prototipo}
    \label{fig:7}
\end{figure}

En la figura \ref{fig:7} se muestra la estructura de los componentes usados en todos los módulos, los cuales varían entre sí únicamente según qué archivos se mostraban. También se debe aclarar, que todos los módulos tenían una duración idéntica debido a que el código usado para sintetizar los audios estaba programado con una duración fija para la sintetización.

\begin{figure} [H]
    \centering
    \includegraphics[width=0.8\textwidth]{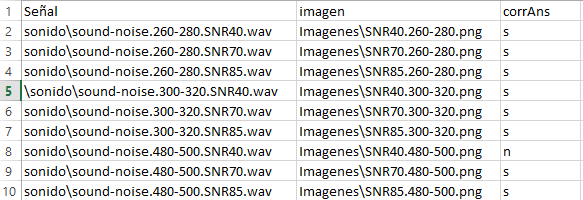}
    \caption{Captura de pantalla de la tabla con los archivos de sonido e imagen usados por la función loop.}
    \label{fig:8}
\end{figure}

Por cada módulo se usaron tres archivos de audio y tres gráficos. Para no insertar la misma estructura tres veces, una debajo de la otra y cambiar los archivos mostrados al usuario, se hizo uso de la función loop. Esta funciona a base de un archivo .xls, en donde se introducen en forma de tabla las rutas de las imágenes y audios que se desean utilizar, tal como se muestra en la figura \ref{fig:8}. Otro dato que se debe colocar en la tabla es la respuesta correcta. Este dato es importante para los archivos de salida del entrenamiento.

\begin{figure} [H]
    \centering
    \includegraphics[width=0.8\textwidth]{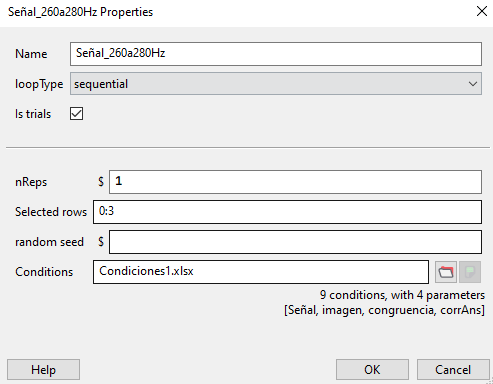}
    \caption{Propiedades del loop.}
    \label{fig:9}
\end{figure}

Una vez armada dicha tabla, lo que sigue es introducirla en la función loop. Esta se debe incluir en la sección \textit{Flow} de la ventana \textit{Builder}, mediante el botón `insert loop'. Una vez activada esta función se elije su inicio y fin, abarcando tantas rutinas como se desee. En el caso desarrollado se eligió insertar tres loops, cada uno abarcando sólo las rutinas de módulos y haciendo modificaciones en sus propiedades según se muestra en la figura \ref{fig:9}. La configuración adoptada fue: colocar un nombre a cada loop; indicar el tipo que se usaría, siendo de tipo secuencial para este caso; elegir una única repetición y seleccionar las filas de la tabla que se usarían.

Una vez programado todo el entrenamiento, se procedió con su ejecución. Al finalizar, se generaron dos archivos con las respuestas ingresadas por teclado. Estos correspondían a un archivo excel y otro csv, con un despliegue de distintos datos, tales como las respuestas ingresadas por teclado y el tiempo que se tardó en responder el participante. Estos archivos se revisaron para poder comprobar su utilidad a la hora de cuantificar los resultados de cada entrenamiento.

Este primer entrenamiento se mostró y discutió en el grupo de trabajo\footnote{Integrado por la directora del presente trabajo, Bioing. Johana Casado, la asesora, Dra. Beatriz García, y la autora y estudiante, Natasha Bertaina.}, el cual concluyó que con el software seleccionado se podría cumplir con el objetivo de programar entrenamientos. En base a que los entrenamientos propuestos se enfocaron en el análisis de datos astronómicos a través de sonorización y como complemento a la visualización de imágenes, junto con el hecho de que el software sonoUno permite guardar archivos de audio e imagen, se decidió utilizar estos archivos como entrada a los módulos de los entrenamientos. Esto además, permite un entrenamiento en sonorización relacionado con la herramienta desarrollada previamente en el grupo de trabajo (sonoUno).

Habiendo definido el mejor modelo para el desarrollo del entrenamiento, se continuó con el diagrama y diseño de un entrenamiento que seria testeado con personas voluntarias en ITeDA (Instituto de Tecnologías en Detección y Astropartículas), sede Mendoza, lugar donde desempeñan sus funciones el grupo de trabajo.

\subsection{Primer Workshop}

El objetivo principal era diseñar un entrenamiento en sonorización que empleara datos astronómicos, los cuales se obtuvieron de la base de datos Sloan Digital Sky Survey (SDSS), que sirviera para que los participantes se familiarizaran con el análisis multisensorial de datos. Para seguir con el lineamiento de despliegue visual y auditivo es que se empleó el programa sonoUno, en su versión de escritorio, para sonorizar los datos, generar sus respectivas imágenes y guardar estos archivos (audio e imagen) para su utilización en el entrenamiento. Cada sonido se generó con una frecuencia máxima de 1700 Hz.

\begin{figure} [H]
    \centering
    \includegraphics[width=1\textwidth]{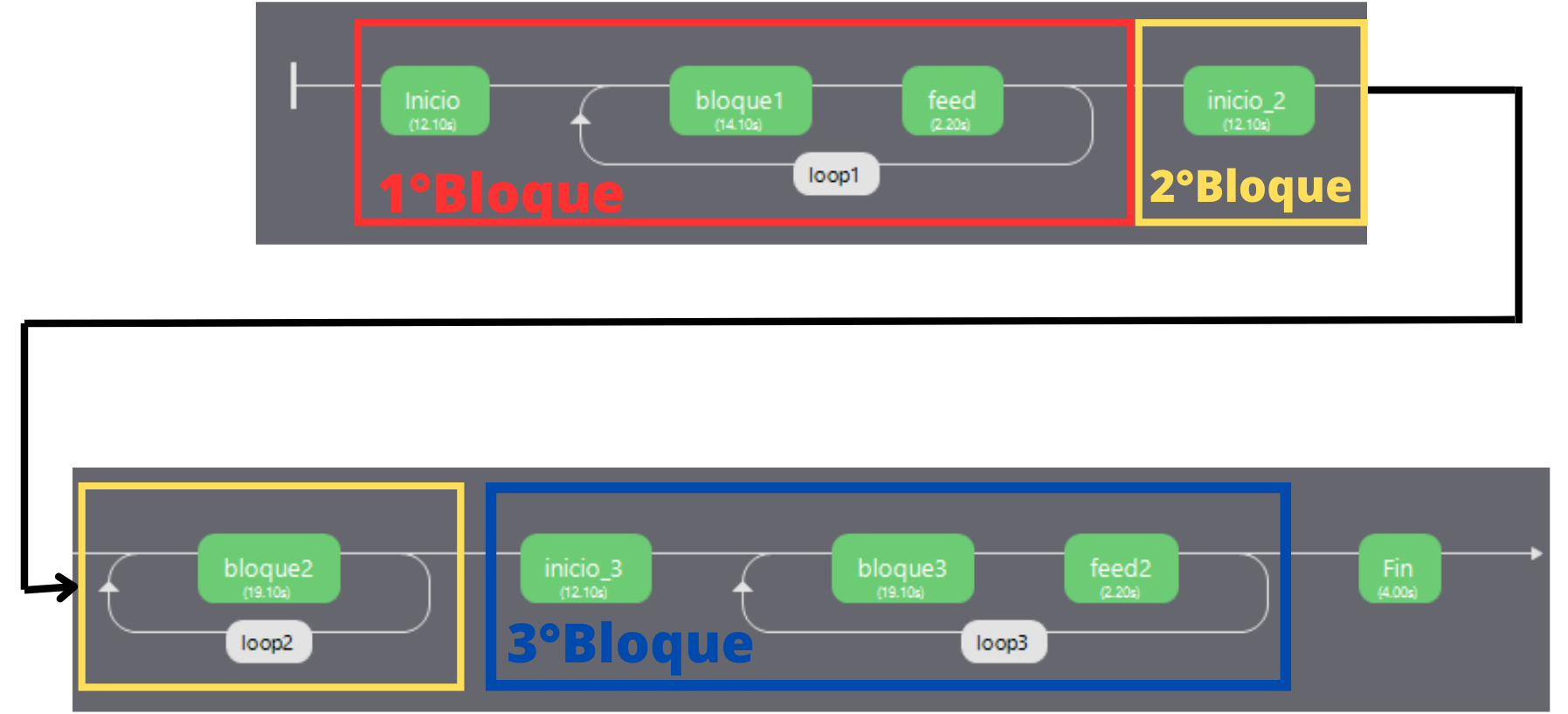}
    \caption{Diagrama del entrenamiento testeado en el Primer Workshop, dividido en tres bloques.}
    \label{fig:10}
\end{figure}

El esquema del entrenamiento diseñado es tal cual se muestra en la figura \ref{fig:10}, en donde se observa que consiste de nueve rutinas. Para facilitar la explicación del diseño se dividirá al diagrama en tres bloques, describiéndose cada uno en subsecciones aparte.

Se hizo uso de la función loop para lograr el cambio de archivos desplegados sin la necesidad de agregar más componentes a las rutinas, tal cual se hizo en el entrenamiento prototipo. La nueva función       
integrada a este entrenamiento es la de feedback, la cual toma la respuesta introducida por el participante, la compara con la respuesta que se indicó ser la correcta en el archivo de datos y le devuelve un correcto o incorrecto al participante.

\subsubsection{Primer Bloque: Funciones simples}

\begin{figure} [H]
    \centering
    \includegraphics[width=0.8\textwidth]{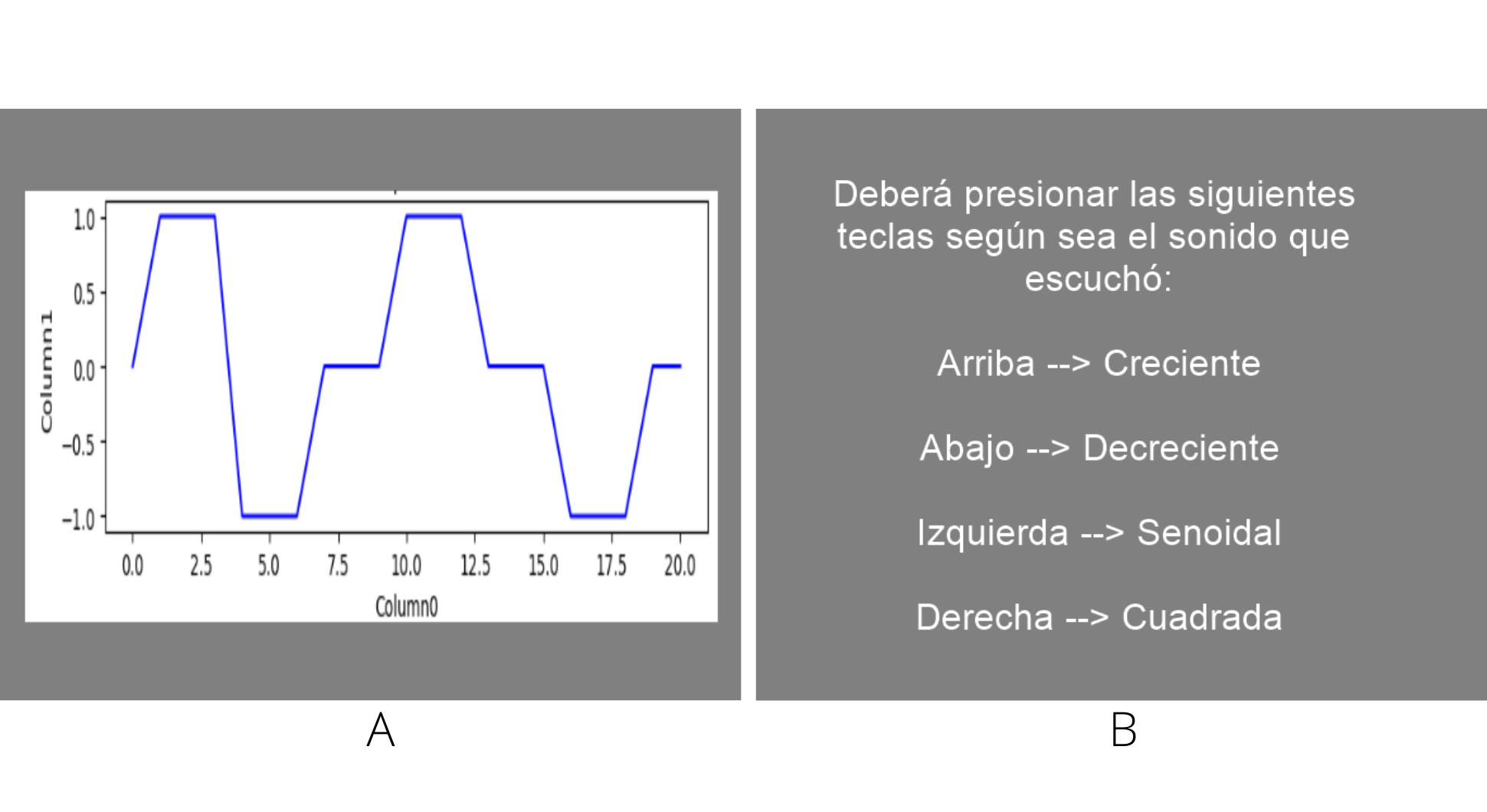}
    \caption{A: Señal mostrada al participante, fue acompañada de su correspondiente señal de audio durante el despliegue. B: Opciones ofrecidas al participante para responder según lo que percibió. Para esta tarea debían usarse las teclas de desplazamiento del teclado (teclas con símbolo de flechas).}
    \label{fig:12}
\end{figure}

El primer bloque consistió en el despliegue tanto auditivo como visual de cuatro funciones matemáticas sencillas como una onda seno, una onda cuadrada, una función linealmente creciente y una decreciente. El objetivo aquí era orientar al participante en el uso de la sonorización, como también ubicarlo en la tarea que debía realizar a medida que el entrenamiento avanzaba. Para lograr este objetivo se implementó la rutina Inicio, que contenía dos componentes de tipo texto que se emplearon para marcar el inicio del bloque y explicar de forma corta y concisa la tarea que debía realizarse. Dicha tarea consistía en observar y oír la función desplegada durante cuatro segundos para luego contestar, mediante el uso del teclado, si lo que escuchó y observó era una función senoidal, cuadrada, creciente o decreciente. La figura \ref{fig:12} muestra como se desplegó la imagen en la pantalla y el mensaje posterior donde se indica la tecla que corresponde a cada función matemática.

La rutina que se encargaba del despliegue es la llamada bloque1, estructurada de la misma forma que los módulos del entrenamiento prototipo, usando los siguientes componentes: audio, imagen, texto y respuesta por teclado. Una característica que se tuvo en cuenta con respecto a los archivos de audio empleados fue su duración: se trató que fuera casi la misma por bloque con el fin de hacer más amigable su despliegue, ya que al usar loops para cambiar los archivos mostrados y programar una única duración para todos estos, si los audios hubieran sido muy dispares en tiempo, se encontrarían largos silencios en algunas repeticiones y no se tendría un sentido de continuidad. Por este motivo, los sonidos empleados en este bloque tuvieron una duración aproximada de cuatro segundos cada uno.

\begin{figure} [H]
    \centering
    \includegraphics[width=1\textwidth]{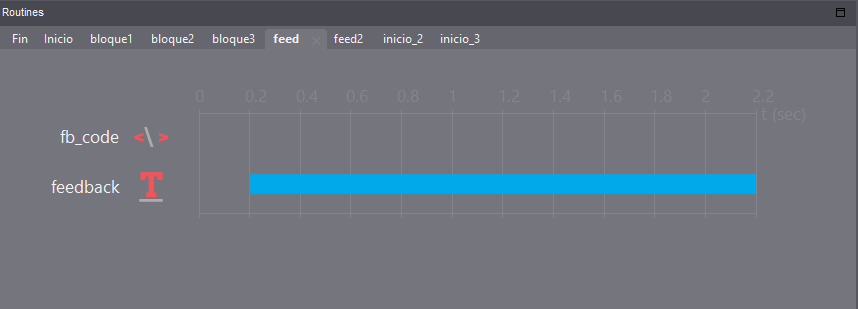}
    \caption{Componentes que se incluyen en una rutina de feedback genérica}
    \label{fig:13}
\end{figure}

Al ingresar una respuesta, se usó la rutina feed para dar una devolución rápida al participante con respecto a si lo que contestó era una respuesta correcta o no. Esta función compara las respuestas marcadas como correctas en la tabla cargada del loop, con la obtenida por teclado. Haciendo uso de un condicional, se programan los mensajes que se muestran por pantalla cada vez que la comparación finaliza. Esto se puede observar en la figura \ref{fig:13}.

\begin{figure} [H]
    \centering
    \includegraphics[width=1\textwidth]{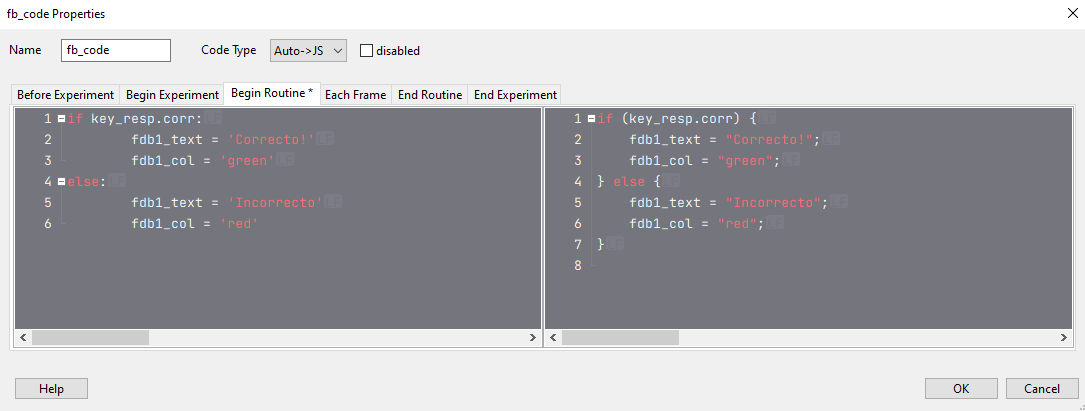}
    \caption{A la izquierda: código escrito en lenguaje Python. A la derecha: código escrito en lenguaje JavaScript}
    \label{fig:14}
\end{figure}

El elemento fb\_code observado en la figura \ref{fig:14} se puede programar tanto en lenguaje Python como en lenguaje JavaScript. Esto es gracias a la posibilidad que ofrece el software de generar un script en lenguaje JavaScript de lo programado en Python, para ser usado en la web. Más adelante, en el presente capítulo, hay una sección que profundizará en esta opción.

\subsubsection{Segundo Bloque: Líneas de Absorción y Emisión}

Para este bloque solo se utilizó un despliegue auditivo de señales generadas en rangos específicos de un archivo de datos de un espectro de galaxia, el cual se descargó de `SDSS J115845.43-002715.7'\footnote{(\url{http://skyserver.sdss.org/dr12/en/tools/quicklook/summary.aspx?id=1237648720693755918})}. El trabajo realizado con el software sonoUno para estos datos fue un poco más complejo ya que consistió en identificar diferentes rangos de longitud de onda que evidenciaban lineas de absorción y líneas de emisión. Finalmente, se seleccionaron los siguientes rangos para:

\begin{itemize}
    \item Líneas de emisión: de 7100 nm a 7300 nm; de 7300 nm a 7500 nm.
    \item Líneas de absorción: de 5500 nm a 5900 nm; de 6000 nm a 6300 nm; de 6300 nm a 6600 nm.
\end{itemize}

En total se obtuvieron cuatro líneas de emisión y seis líneas de absorción, dando un total de diez señales desplegadas para este bloque. Se obtuvieron el doble de archivos porque para cada rango se tomó el flujo respecto de la longitud de onda y una curva de mejor ajuste de dicho flujo con respecto a la longitud de onda (todos datos presentes en el archivo, no se realizó ninguna operación matemática).

Al inicio del bloque se introdujo una rutina llamada inicio2, con la función de indicar el inicio del segundo bloque como también las nuevas instrucciones. Seguido de esto se uso una rutina llamada bloque2, en la cual no se usó un componente de imagen, ya que no se deseaba hacer un despliegue de este tipo de archivo. Se incluye un componente de respuesta ingresada por teclado, como también un componente de texto que repite las opciones de respuestas para que el participante conteste. Como se observa en el diagrama del entrenamiento, este bloque no hace uso de una rutina de feedback, por lo que el participante no sabrá si su respuesta es correcta o incorrecta en la inmediatez en la que ingresaba su respuesta.

\subsubsection{Tercer Bloque: Líneas de Absorción y Emisión}

Los archivos de audio usados en este bloque fueron los mismos que se usaron en el segundo bloque, con la diferencia de que aquí se acompañó el despliegue auditivo con el visual. Para ello al mismo tiempo que se hicieron los cortes y se sonorizaron los datos en sonoUno, se generaron las gráficas correspondientes y se reservaron para ser usadas en este último bloque. El fin de usar las mismas señales pero con distintos despliegues fue comparar los resultados de ambos bloques. Esto se retomará en el apartado Resultados de esta subsección.

La estructura de este bloque es similar al explicado en el primer bloque, usando rutinas llamadas inicio\_3, bloque3 y feed2. La primera tenía como función indicar el inicio del tercer bloque y dar las instrucciones de lo que se debía hacer. La rutina bloque3 es la que se encargó del despliegue visual y auditivo de las diez señales usadas. Finalmente la rutina feed2 era del tipo feedback con la que se indicó si la opción ingresada por el participante a través del teclado era correcta o no.

Al finalizar el tercer bloque se usó una última rutina llamada fin, la cual tenía la función de indicar la finalización del entrenamiento. 

Las opciones de respuesta de cada bloque se mostraban por pantalla al inicio de cada uno, y al finalizar el despliegue combinado de los archivos visuales y de audio o el despliegue único de audio. Esto para que el participante no tuviera que memorizar las teclas que correspondían a cada opción de respuesta disponible.

El entrenamiento tuvo una duración de casi diez minutos para cada persona voluntaria. Esta duración podía ser menor dependiendo de la rapidez con la que el participante ingresaba sus respuestas. El tiempo de espera para las respuestas ingresadas era de 10 segundos, si dentro de este tiempo el participante no ingresaba una respuesta, esta quedaba en blanco y si el bloque estaba acompañado de una función de feedback, se devolvía un mensaje de ''Incorrecto''.

\subsubsection{Testeo}

El testeo se realizó a principios de abril del 2022 en forma presencial, en el mismo participó el equipo de trabajo del instituto ITeDA de la regional Mendoza. El total de participantes fue de seis personas, entre las que se encontraban dos técnicos en electrónica, un ingeniero en física médica, un doctor y una doctora en astronomía y una diseñadora industrial. 

Para proceder con el entrenamiento se hizo uso de dos computadoras donde se instaló previamente el software Psychopy, ya que sin él no era posible ejecutarlo. Al ser limitada la cantidad de ordenadores disponibles, se procedió con la división de los participantes en grupos de dos. Antes de comenzar con el procedimiento, se brindó a cada grupo una breve explicación de lo que se encontrarían y lo que debían hacer a medida que el entrenamiento avanzaba. También se les pidió que usaran auriculares y que comprobaran el nivel de volumen para que no los incomodara, ya que como se mencionó anteriormente, los sonidos usados tenían frecuencias elevadas. Al finalizar con esta pequeña introducción, se dio inicio al entrenamiento, asegurando mantener el lugar lo más silencioso posible para no causar distracciones en los participantes.

\begin{figure} [H]
    \centering
    \includegraphics[width=1\textwidth]{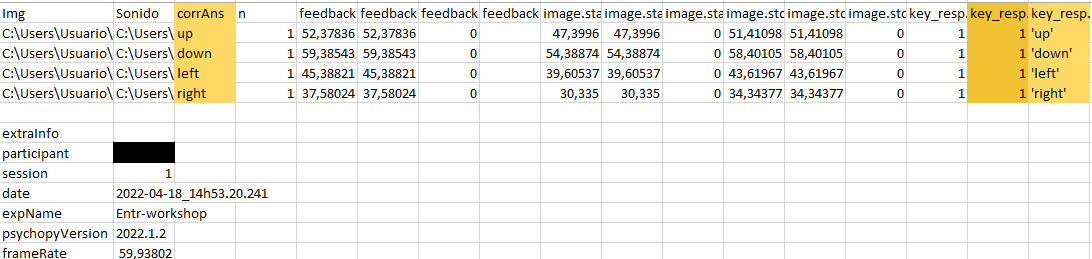}
    \caption{Archivo Excel obtenido al finalizar una sesión de entrenamiento. Las columnas resaltadas con amarillo fueron las respuestas dadas por el participante y las que se usaron para el análisis posterior.}
    \label{fig:15}
\end{figure}

El tiempo total que llevó realizar el testeo fue de una hora y media reloj, y al finalizar este se procedió con la recuperación de los archivos .xls y .csv donde se encontraban todas las respuestas dadas por cada participante. Estos archivos tienen una estructura de tabla en donde los datos se agrupan por loop programado, además de identificar al participante y el número de sesión realizado. La figura \ref{fig:15} muestra las respuestas obtenidas de un participante. Las columnas resaltadas son los datos que se tomaron para el análisis posterior.

En cada archivo .xls los datos se ordenan en hojas separadas de acuerdo a los distintos loops empleados. Dentro de cada hoja se muestra una tabla en la que se indica el archivo mostrado, el orden en el que se exhibió, la respuesta configurada como correcta y la respuesta obtenida por el participante, entre otros datos de menor relevancia. 

Para el análisis se tomaron los datos de cada participante y se trasladaron a un nuevo archivo excel, donde se ordenaron en hojas separadas y en tablas más específicas, las cuales indicaban los aciertos, fallos y no contestaciones de cada bloque. Cuando se obtuvo el estado de detección de cada participante se realizó un análisis más general del entrenamiento en donde solo se tuvieron en cuenta los aciertos de cada voluntario o voluntaria.

\subsubsection{Resultados}

La Tabla \ref{tab:1} muestra la cantidad de aciertos obtenidos en cada bloque. Se recuerda que la cantidad de señales desplegadas era en el primer bloque de cuatro, en el segundo y tercer bloque de diez, además del hecho de que cada participante solo realizó el entrenamiento una única vez. Un resultado interesante que se puede observar con los aciertos del 3er bloque es que los participantes presentaron menor dificultad a la hora de detectar señales auditivas con soporte visual. Por el contrario, cuando solo se debía detectar el sonido sin el apoyo visual, la cantidad de aciertos fue menor, tal como se observa en la columna de aciertos del 2do bloque.

Con estos resultados se obtuvo una primera aproximación a la capacidad de detección de los participantes para distinguir distintas señales utilizando una técnica nueva (como es la sonorización), la cual podría incrementarse si se aumenta el número de sesiones de entrenamiento. 

\begin{table}[H]
    \centering
    \begin{tabular}{| c | c | c | c |}
        \hline
        Participantes & Aciertos 1°Bloque & Aciertos 2°Bloque & Aciertos 3°Bloque \\ \hline
        1°Participante & 3 & 2 & 6 \\
        2°Participante & 1 & 2 & 6 \\
        3°Participante & 4 & 5 & 4 \\
        4°Participante & 4 & 2 & 3 \\
        5°Participante & 4 & 4 & 8 \\
        6°Participante & 2 & 4 & 1 \\ \hline
    \end{tabular}
    \caption{Aciertos de cada participante, por bloque. En el 1° Bloque, el participante debía detectar correctamente cuatro señales. En el 2° Bloque (audio) y en el 3° Bloque (audio e imagen) debía detectar diez señales correctamente.}
    \label{tab:1}
\end{table}

La tabla \ref{tab:2} muestra el porcentaje de aciertos global por bloque, lo que refuerza la necesidad de entrenamientos multimodales, ya que en los bloques donde se recurrió al despliegue visual y auditivo se obtuvieron mejores resultados por parte de los participantes.

\begin{table}[H]
    \centering
    \begin{tabular}{| c | c | c|}
        \hline
        Bloque y Estímulo & Aciertos & Porcentaje de Aciertos  \\ \hline
        Primer bloque: visión y audición & 18 & 75 \\ 
        Segundo bloque: audición & 19 & 31,667 \\
        Tercer bloque: visión y audición & 28 & 46,667 \\ \hline
    \end{tabular}
    \caption{Porcentaje de aciertos por bloque, indicando tipo de estímulos usados.}
    \label{tab:2}
\end{table}

\subsubsection{Conclusión del Primer Workshop}

Los resultados obtenidos reafirmaron la idea de que el rendimiento auditivo mejora cuando este estímulo es apoyado con el correspondiente visual. Gracias a esto es que las experiencias diseñadas posteriormente partieron con un despliegue auditivo y visual como método introductorio. 

Otro punto importante a tener en cuenta para lograr un aprendizaje es el de partir de un plan de sesiones, las cuales pueden extenderse en varios días, ya que es evidente que una única sesión aislada no aporta mayormente al aprendizaje de una persona, sino que es la repetición de la tarea la que promueve la enseñanza. 

Al ser este un testeo, se preguntó a los participantes su experiencia y que es lo que mejorarían del entrenamiento. Entre las sugerencias se destacaron las siguientes:

\begin{itemize}
    \item Incluir la opción 'No estoy seguro' entre las posibles respuestas.
    \item Dar más tiempo al participante para responder.
    \item Usar una mejor combinación de colores en la retroalimentación dada al participante tras su respuesta.
    \item Cambiar las palabras `Correcto' e `Incorrecto' por equivalentes que sean más positivas y ayuden a un mejor intercambio con los participantes.
\end{itemize}

Todas ella fueron tenidas en cuenta para el diseño del siguiente entrenamiento desarrollado. También se tomaron en cuenta las sugerencias de miembros del equipo de trabajo, entre las que destacaba la necesidad de incluir un audio que relatara las instrucciones mostradas en formato de texto, con la finalidad de hacer más accesible el entrenamiento. 

El trabajo hecho en este workshop concluyó en la elaboración de un artículo, que fue presentado en el XXIII Congreso Argentino de Bioingeniería y XII Jornadas de Ingeniería Clínica, realizado en la provincia de San Juan, entre el 13 y 16 de Septiembre del año 2022\cite{paper}. Dicho trabajo, titulado \textit{''The use of sonification in data analysis: a Psychopy training test''}, se encuentra adjunto en el Apéndice \ref{cap:apendB}.

\subsection{Segundo Workshop}

A principios de junio del año 2022, se requirió el desarrollo de un plan de entrenamiento para ser presentado en el marco de un Curso Internacional de capacitación organizado por los integrantes en el proyecto REINFORCE, el cual se llevaría a cabo en julio del mismo año en Grecia. Para esto se sonorizaron datos específicos de demostradores del proyecto, tales como GW-Glitches \footnote{``Ruidos'' en los detectores de ondas gravitacionales} (ver \url{https://www.zooniverse.org/projects/reinforce/gwitchhunters}), partículas del Gran colisionador de Hadrones o LHC (derivado de su nombre en inglés, \textit{Large Hadron Collider}, ver \url{https://www.zooniverse.org/projects/reinforce/new-particle-search-at-cern}) y muones cósmicos\footnote{Partículas producto de la desintegración de rayos cósmicos al entrar en contacto con la atmósfera} (ver \url{https://www.zooniverse.org/projects/reinforce/cosmic-muon-images}).

\begin{figure}[H]
    \centering
    \includegraphics[width=1\textwidth]{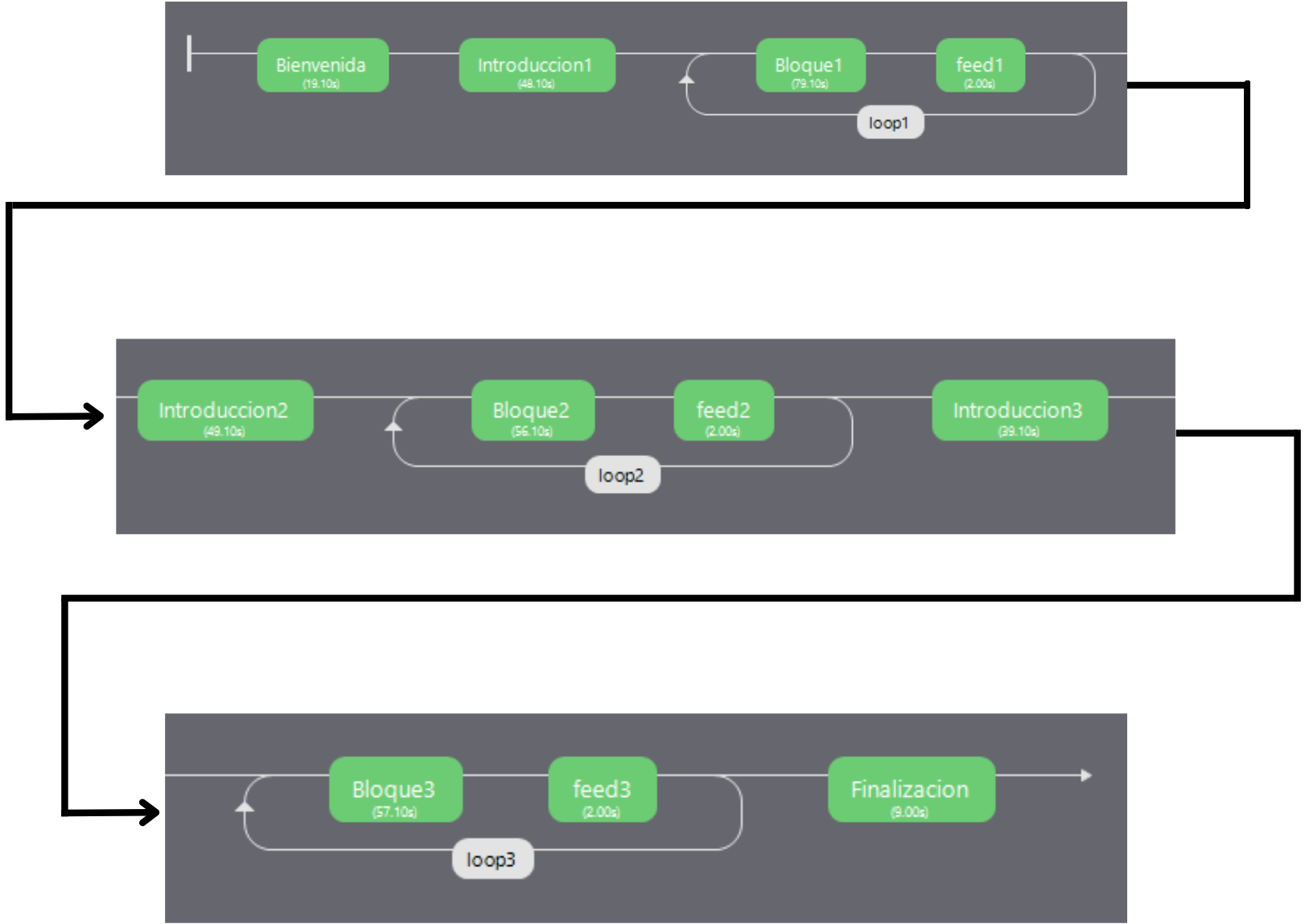}
    \caption{Diagrama para el entrenamiento del día 1. El esquema se repite en el día 2, diferenciándose en la duración de cada loop.}
    \label{fig:16}
\end{figure}

El primer paso que se siguió fue diagramar la cantidad de loops que iban a usarse, como también la cantidad de entrenamientos que debían hacerse. A grandes rasgos, se decidió hacer dos entrenamientos (basados en el tiempo otorgado por los organizadores del workshop), los cuales iban a ejecutarse en días distintos y consecutivos. Cada entrenamiento contaría con tres loops para el despliegue visual y auditivo de los tres tipos de datos mencionados. En esta primera fase de diseño se tuvieron presentes todas las sugerencias mencionadas en la sección anterior. El esquema general de ambos entrenamientos puede observarse en la figura \ref{fig:16}.

La ventaja de implementar un plan de entrenamiento en dos días fue la posibilidad de aumentar la complejidad de los mismos de forma creciente. El entrenamiento de la primera jornada sirvió como introducción para que los participantes conocieran lo que es un entrenamiento, destinándose al reconocimiento de los distintos datos y su clasificación; mientras el de la segunda jornada se diseñó con el doble de señales y de mayor complejidad para poder entrenar al participante en la detección de cada tipo de dato desplegado.

\begin{figure} [H]
    \centering
    \includegraphics[width=1\textwidth]{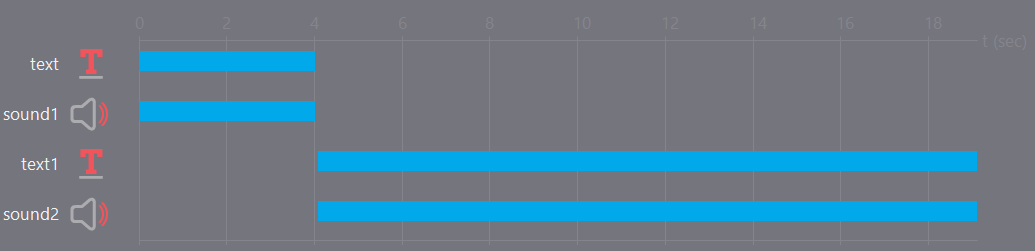}
    \caption{Rutina ``Bienvenida", en la que se implementó texto y audio}
    \label{fig:17}
\end{figure}

Para este diseño se implementaron audios grabados que relataban todo el texto mostrado en pantalla. El fin de esto era lograr un entrenamiento accesible para un mayor número de participantes. Un ejemplo de dicho diseño en PsychoPy se ve en la figura \ref{fig:17}.

Todas las rutinas que incluían un despliegue de texto entre sus componentes fueron acompañadas por su relato grabado. La excepción a esto fue la rutina de feedback, ya que limitaciones del software no permitiieron que esto fuera implementado, a pesar de haber intentado modificar el código en Python directamente.

Otro dato a tener en cuenta fue el idioma desplegado tanto en texto como en forma de audio. Al ser un evento internacional, todo lo mostrado por pantalla fue escrito en inglés, y los audios fueron grabados en dicho idioma.

\subsubsection{Primer bloque: Glitch Classification}

El inicio de este primer bloque contaba con una pequeña introducción en la cual se indicaba el tipo de señal que iba a mostrarse, como también las opciones de posibles respuestas y el botón correspondiente para cada una. 

\begin{figure}[H]
    \centering
    \includegraphics[width=1\textwidth]{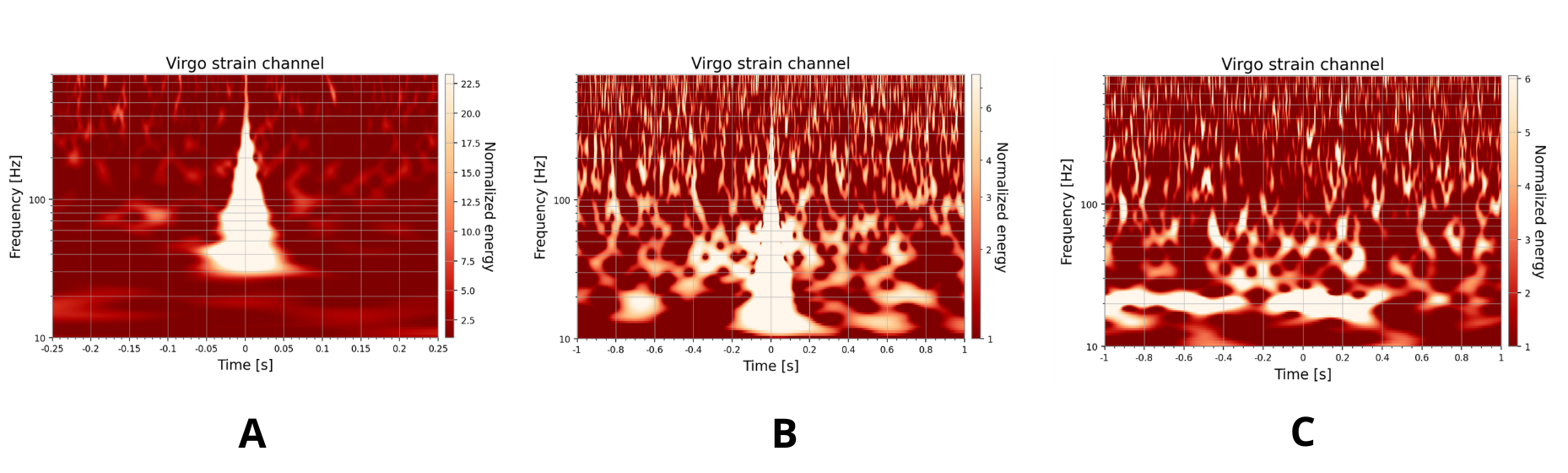}
    \caption{Distintos tipos de Glitches desplegados en el primer bloque de los entrenamientos. A: Tipo blip. B: Tipo koi-fish. C: Tipo scattered light.}
    \label{fig:18}
\end{figure}

En este primer bloque se desplegaron tanto imágenes como sonidos de tres distintos Glitches: del tipo blip, koi-fish y scattered light (Figura \ref{fig:18}), los cuales fueron trabajados con sonoUno. 

El participante tenía la tarea de responder qué tipo de Glitch había escuchado y visto. En total se mostraron tres señales en el entrenamiento del día uno y seis el día dos. El despliegue de cada audio fue de 38 segundos, por lo que al sumar la duración de las tres señales para el día 1, y las seis señales para el día dos, más el tiempo programado para que el participante respondiera, éste se convirtió en el bloque con mayor tiempo de ejecución.

Las respuesta ingresadas por los participantes fueron acompañadas de un feedback en el cual se mostraba por pantalla en forma de texto si lo que habían respondido era correcto (Excellent job!!) o si en realidad de trataba de otro tipo de Glitch (Oops!! This seems to belong to a different glich class). Teniendo en cuenta la sugerencia de los participantes del primer workshop, se cambió el mensaje mostrado por pantalla, para impactar de forma positiva en los participantes.

\subsubsection{Segundo Bloque: Particle Detection}

\begin{figure} [H]
    \centering
    \includegraphics[width=1\textwidth]{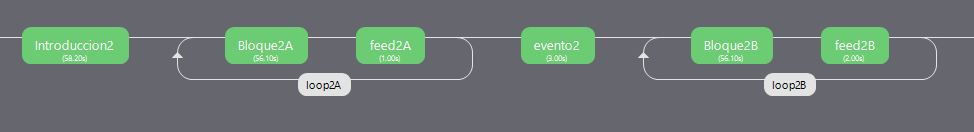}
    \caption{Diagrama entrenamiento día 2, correspondiente al segundo bloque. Para este día se desplegaron dos eventos distintos.}
    \label{fig:19}
\end{figure}

Este bloque también contaba con una introducción, donde se detallaba el tipo de señales que se desplegarían, además de las opciones de respuestas posibles y las teclas a las que correspondían cada una. Es en este bloque donde se encontraba la mayor diferencia entre el entrenamiento del día uno y el entrenamiento del día dos. Para el primer día se diagramó el despliegue de un único evento de señales obtenidas del LHC, mientras que para el día dos se usaron dos eventos separados. Para lograr un mejor despliegue en este segundo día se optó por la implementación de un loop más, tal como se observa en la figura \ref{fig:19}, dando un total de cuatro loops implementados en el entrenamiento del día dos.

\begin{figure} [H]
    \centering
    \includegraphics[width=1\textwidth]{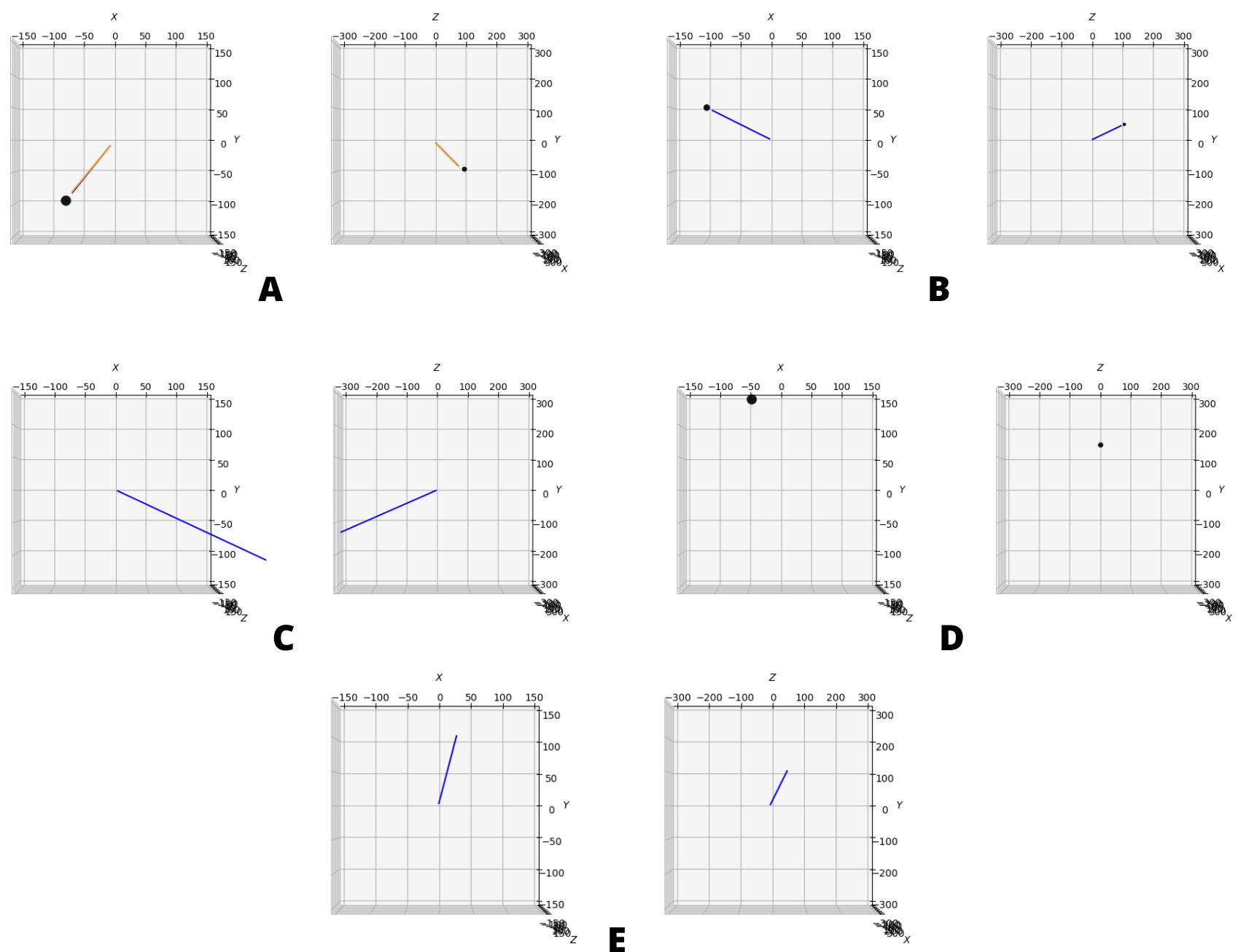}
    \caption{Gráficos de las partículas que debían detectarse. A: fotón convertido (dos trayectorias que apuntan a un mismo depósito de energía). B: electrón (única trayectoria que apunta a un depósito de energía). C: muón (es altamente energético por lo que atraviesa todas las capas del detector). D: fotón (no presenta trayectoria pero si depósito de energía). E: partícula desconocida (solo presenta la trayectoria sin depósito de energía).}
    \label{fig:20}
\end{figure}

En cuanto a las señales desplegadas, estas correspondían a partículas (Fotón convertido, Muón, Electrón, Fotón o Desconocido/Inespecífico) obtenidas en eventos dentro del LHC. En el día uno fueron desplegados los sonidos e imágenes de las cinco posibles partículas, mientras que en el día dos se desplegaron diez en total, cinco por cada evento (en este segundo día se proyectaron dos eventos reales). Vamos a aclarar que por evento no se obtienen todas las partículas mencionadas, ello depende de múltiples factores y es por eso objeto de estudio. Dicho eso, se debe tener en cuenta que a pesar de anunciar todas las opciones en las respuestas, podían no haberse desplegado las señales de todas ellas. La figura \ref{fig:20} muestra las imágenes de partículas desplegadas en este bloque.

Este bloque, al igual que el primero, también implementó un feedback para que el participante supiera si lo que había respondido era correcto o si se trataba de otra partícula.

\subsubsection{Tercer Bloque: Muon Detection}

\begin{figure} [H]
    \centering
    \includegraphics[width=0.9\textwidth]{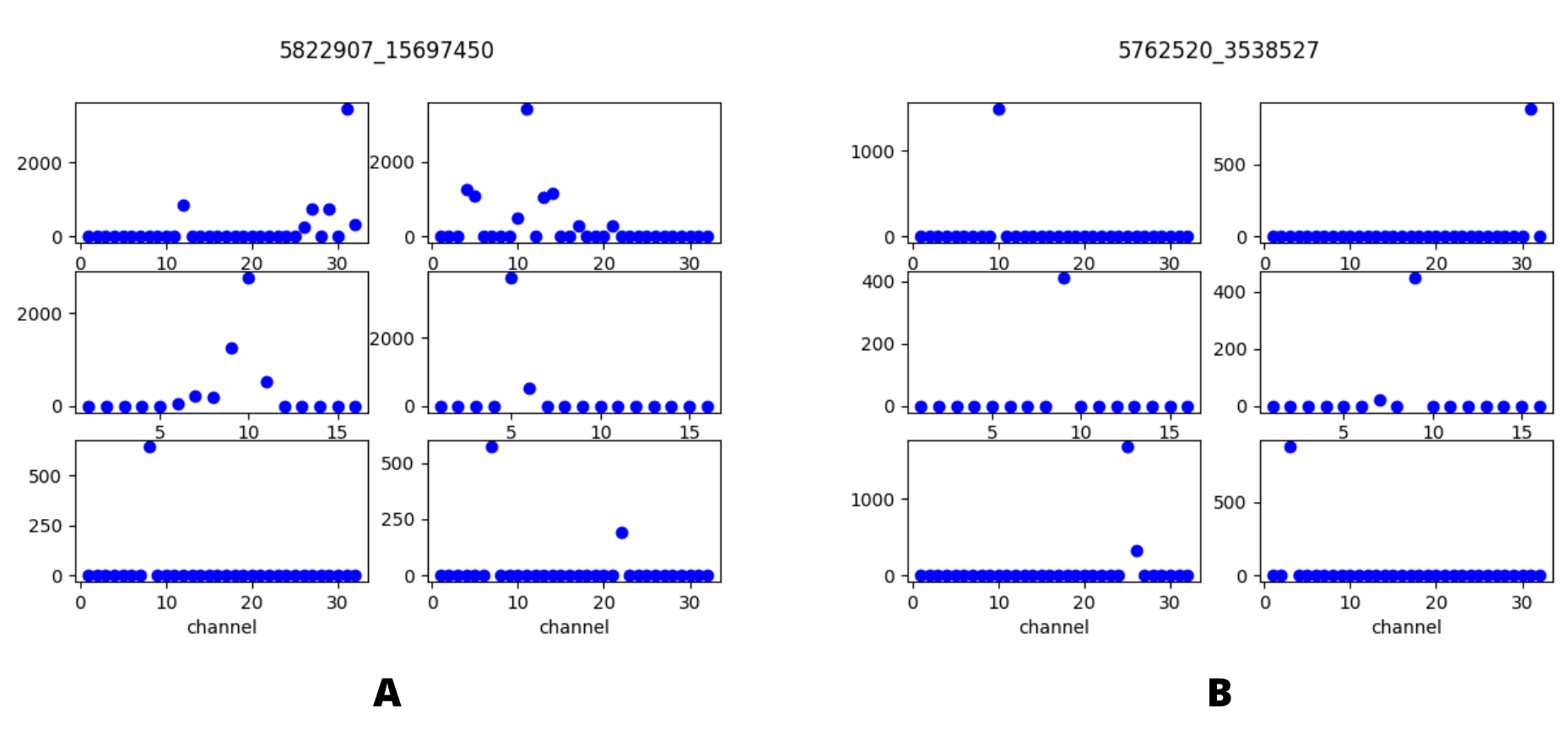}
    \caption{Gráficos desplegados en el tercer bloque. A: corresponde a una inexistencia de muón. B: existencia de un muón que atravesó los detectores (véanse los espacios vacíos en cada capa del detector, si se pueden unir con una recta, esta marca la trayectoria del muón al atravesar los centelladores).}
    \label{fig:21}
\end{figure}

Siguiendo lo diseñado para los bloques anteriores, se agregó una rutina de introducción en la que se especificaba el tipo de señal desplegada como también las opciones de respuestas con sus teclas correspondientes. En este último bloque se usaron muones, los cuales fueron sonorizados y graficados con sonoUno, al igual que las partículas del segundo bloque y los Glitch. En el caso de muongrafía, lo que se debe detectar y por ende entrenar es la capacidad de encontrar la existencia o no de un muón, dejando sólo dos opciones entre las posibles respuestas. Las imágenes desplegadas pueden observarse en la figura \ref{fig:21}.

Como se puede observar en el diagrama de la figura \ref{fig:16}, este bloque también incluyó una rutina de feedback para dar una respuesta inmediata al participante con respecto a lo que contestó a través del teclado. Si había interpretado bien la señal desplegada, se mostraba un mensaje de ``Buen trabajo!'' y si había cometido un error, se mostraba ``Osps!!'' por pantalla para indicar su equivocación.

La cantidad de señales desplegadas fueron de dos en el entrenamiento del día uno y de seis en el entrenamiento del día dos.

\subsubsection{Actividad realizada}

Como se mencionó al inicio, la actividad fue llevada a cabo en el mes de julio del año 2022, en un formato presencial de dos días de duración. Cabe aclarar que en el encuentro no estaba centrado en los entrenamientos, sino que fue un encuentro en el que se presentó el trabajo desarrollado por el grupo de trabajo de sonoUno, donde los entrenamientos en sonorización era una de las actividades a desarrollar. Las personas que asistieron a este evento se desempeñaban en el ámbito de la investigación y la docencia.

En este workshop se trabajó con la exploración multisensorial de datos, agregando a la sonorización impresiones 3D de las imágenes presentadas en los entrenamientos que fueron puestas a disposición de los participantes. Esta herramienta puede aprovecharse en el futuro para ampliar la accesibilidad de los programas de entrenamientos.

Al ser un evento internacional, en una sala de conferencias, apareció una complicación que no se hubiera tenido en cuenta si este se hubiera organizado en el lugar de desarrollo de la propuesta: el proceso de instalación del software Psychopy presentaba ciertos errores en algunos dispositivos. Siendo esto un problema que lleva a desmotivar a las personas que realizan la actividad, y dado que en muchos casos lleva mucho tiempo encontrar una solución para cada problema en particular, se optó en usar el servicio web ofrecido por PsychoPy, llamado Pavlovia (ver \url{https://pavlovia.org/}). Este repositorio ofrece la posibilidad de ejecutar los experimentos diseñados en una interfaz web, como también compartirlos y almacenarlos. La desventaja es que debe comprarse una licencia anual para poder compartir y ejecutar dichos experimentos libremente, la cual tiene un valor de \pounds 1800. Otra opción ofrecida es la de comprar cierta cantidad de créditos, los cuales pueden ser canjeados por un participante para ejecutar una única vez el experimento. Cada crédito tiene el valor de \pounds 0.24, en el caso de la actividad que se describe aquí, teniendo en cuenta la cantidad de posibles asistentes se decidió comprar un total de 50 créditos para ejecutar los entrenamientos.
Si bien el recurso es pago, debe destacarse la practicidad y simpleza en su uejecución.

\subsubsection{Resultados}

Para evaluar la utilidad, usabilidad e impacto de toda la actividad que se llevó a cabo, se realizó una encuesta sobre las distintas temáticas desarrolladas, que fue respondida el último día y se dividía en cinco secciones: antecedentes de los participantes, acceso a sonoUno web, usabilidad de sonoUno web, demostradores de REINFORCE, documentación y conclusiones. Centrándonos en lo que se respondió con respecto a los entrenamientos, se obtuvieron las siguientes sugerencias de mejora: opción para saltar algunas ventanas de introducción, poder volver a escuchar la sonorización, tener una retroalimentación sonorizada junto con la desplegada visualmente y, por último, agregar una sesión donde sólo se use un despliegue auditivo.

Entre algunos de los comentarios que se hicieron con respecto a la sonorización, se mencionó el querer conocer más sobre el sonido, lo que resalta la necesidad de un estudio centrado en la percepción y en más cursos de entrenamiento. Con respecto a este último punto, varios de los participantes comentaron que consideraban necesario un curso de entrenamiento para poder usar la sonorización como herramienta.

\begin{figure}[H]
    \centering
    \includegraphics[width=0.85\textwidth]{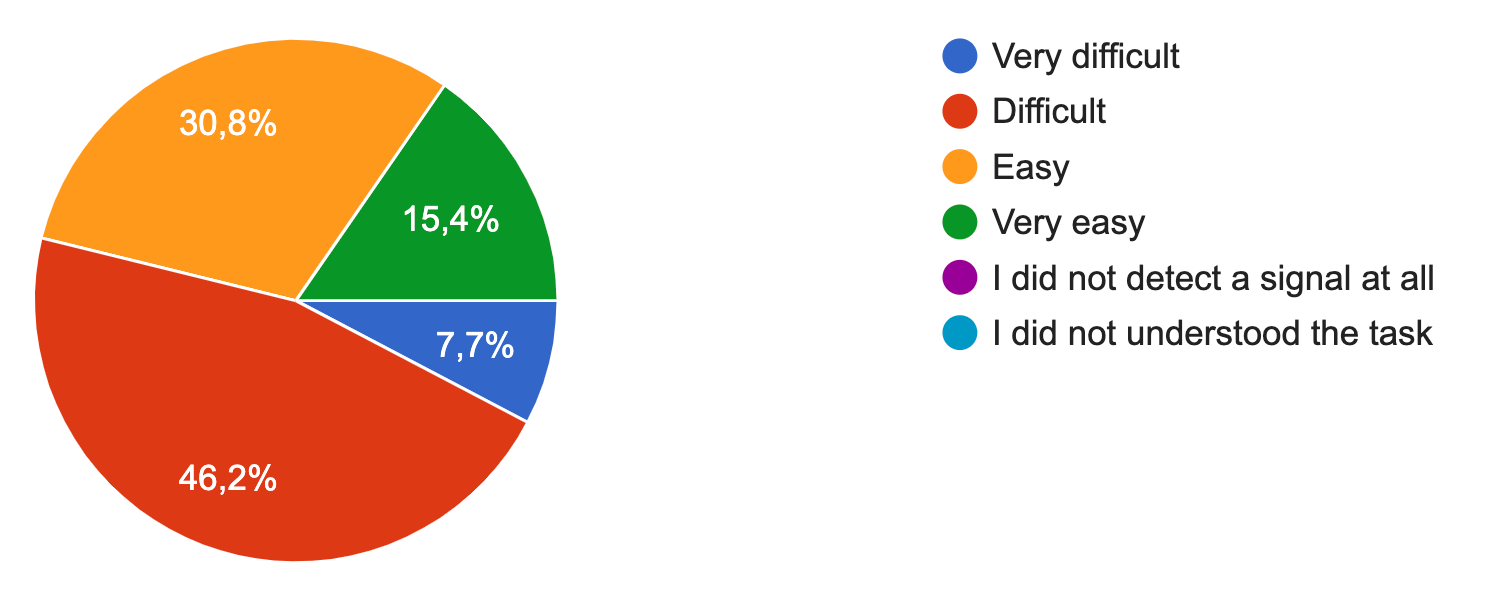}
    \caption{Detección de Glitches}
    \label{fig:22}
\end{figure}

\begin{figure}[H]
    \centering
    \includegraphics[width=0.9\textwidth]{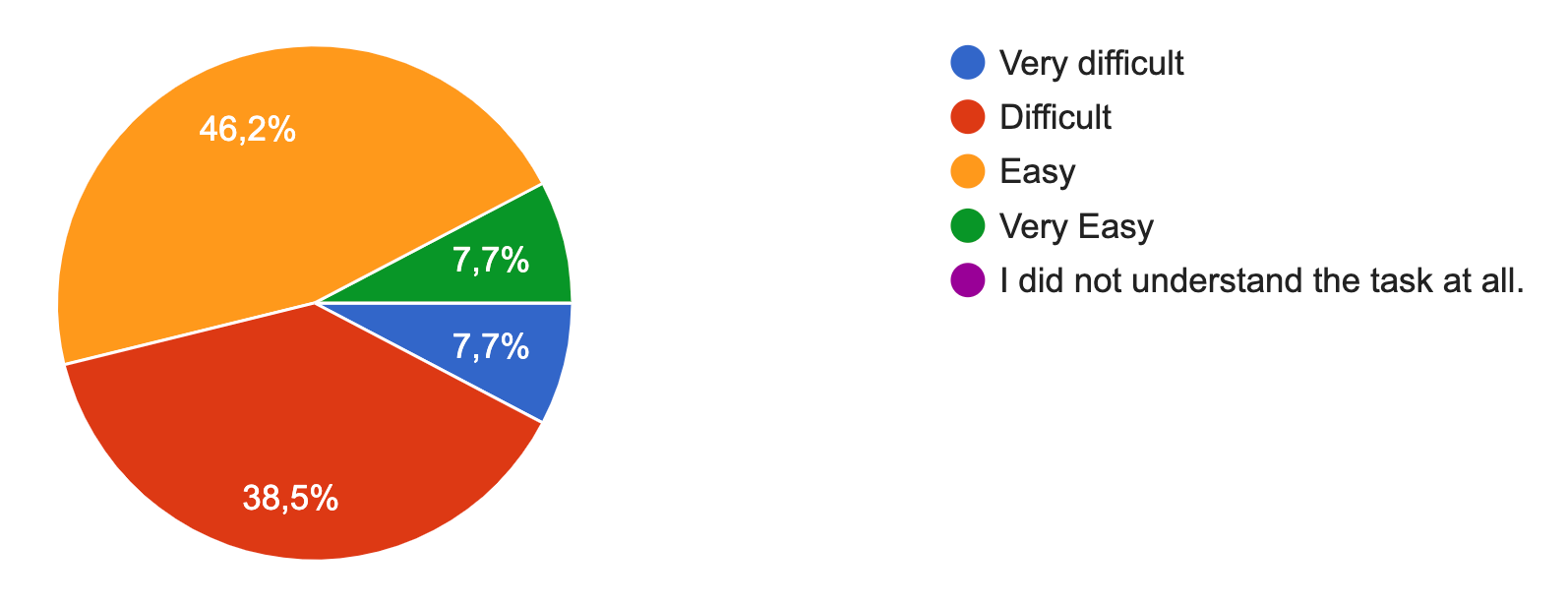}
    \caption{Detección de Partículas}
    \label{fig:23}
\end{figure}

\begin{figure}[H]
    \centering
    \includegraphics[width=0.89\textwidth]{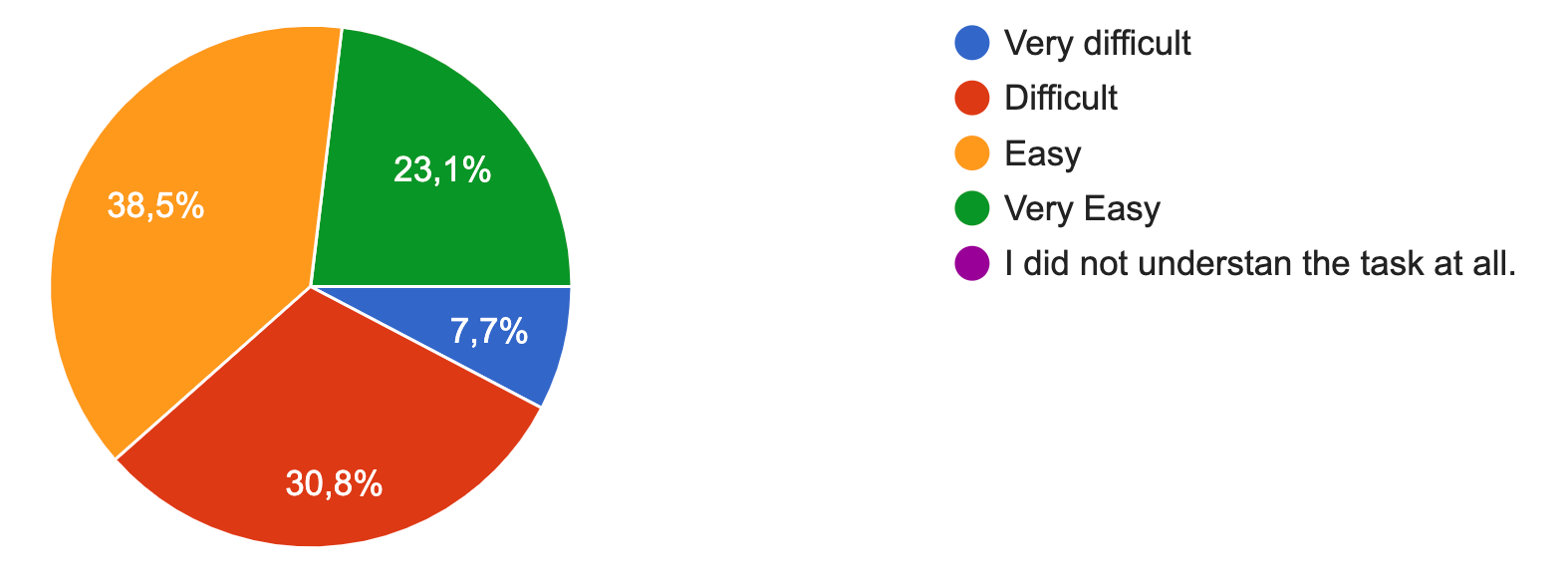}
    \caption{Detección de Muones}
    \label{fig:24}
\end{figure}

La encuesta también incluía una subsección en la que se consultó, para cada tipo de dato sonorizado y usado en los distintos bloques de los entrenamientos, cuanta dificultad se tuvo a la hora de detectar el tipo de dato desplegado. Las respuestas a esto se pueden observar en las figuras \ref{fig:22}, \ref{fig:23} y \ref{fig:24}.


Como se observa, a la hora de detectar patrones para aprender a diferenciar distintos tipos de datos, la mayor dificultad la presentaron los Glitches. Mientras que para detectar muones y partículas del LHC, a la mayoría les resultó más sencilla la detección. Esto podría deberse a que para los muones se usa un arreglo de notas de piano que presenta una relación ascendente o descendente para identificar la existencia de estos. Para el caso de las partículas del LHC también se usan notas de piano con arreglos definidos para cada partícula. En contra parte, la sonorización de glitches se hace a partir de ondas senos puras que responden a la frecuencia de la señal, configurando una frecuencia máxima de 1600 Hz. 

\subsubsection{Conclusiones}

Tal como se muestra en los resultados presentados, es indudable la influencia que la percepción tiene en el aprendizaje, lo que puede verse en la preferencia por sonidos armónicos y musicales. A pesar de ello, se requiere de más estudios para validar los datos obtenidos como también corroborar si un programa de entrenamiento más extenso cambia esta preferencia o la refuerza.

Tanto en este workshop como en el primero, los participantes consideraron el despliegue multisensorial de datos como algo novedoso. También, manifestaron el considerar usarlo en sus investigaciones, ya que permite la exploración de un mismo dato por distintas vías sensoriales (visual, auditiva y táctil). Si esto se bajara a un nivel secundario y primario, se podría lograr una mejor integración de personas con discapacidad. Es por esto que puede considerarse a los entrenamientos como una parte fundamental de este novedoso paradigma.

\section{Desarrollo Web} \label{sec:Desarrollo-web}

Del diseño de los mencionados workshops se desprendió un trabajo inmediato el cual fue el diseño de una página web con una función similar a la ofrecida en Pavlovia, la cual nos permitiera ejecutar los entrenamientos en la web de forma gratuita. A pesar que el precio por crédito no es tan elevado, el pagar por realizar los entrenamientos en línea no sigue el lineamiento de lograr programas de entrenamiento accesibles para un gran número de personas. Nuestro grupo de trabajo tiene objetivo primario el crear herramientas que sean de código abierto y de libre acceso, por lo que tener una barrera económica no es una opción.

Para este desarrollo se trató de aprovechar el script de los entrenamientos generado en JavaScript por el mismo Psychopy, por lo que el desarrollo del back-end girará en torno a eso.

\subsection{Desarrollo front-end}

Como se mencionó en la descripción de las herramientas, para esta parte del diseño web se optó por usar la biblioteca React, con la cual se programó la interfaz de usuario. La ventaja que trae usar esta biblioteca es la interacción que agrega a la interfaz. Esto se vio como una necesidad para poder desplegar el entrenamiento. Pero antes de aplicar esta biblioteca se debe comenzar con una maquetación del sitio web, lo cual se explica a continuación.

En la industria de la programación web, el proceso de maquetado es un paso fundamental a la hora de iniciar el desarrollo front-end. Esto es debido a que en este proceso se empieza a diseñar, usando los lenguajes HTML y CSS, lo que se plasmó en un boceto por el diseñador del proyecto. Es aquí donde se empieza a dar forma a la interfaz con la que el usuario interactuará. Cabe mencionar, que para este trabajo, la interacción que se puede dar es muy baja, y esto se debe a que a la hora de maquetar casi no es requerido usar código escrito en JavaScript. Muchas veces, los equipos de trabajo que se encargan de hacer un front-end se dividen en tres sub-equipos, en donde encontramos:

\begin{itemize}
    \item Diseñador, el cual se encarga de hacer todos los bocetos de cada página que integra el sitio web con el orden de los elementos que se desean mostrar en cada una de ellas. También elige la paleta de colores, tipografía, etc. 
    \item Maquetador, el cual se encarga de diseñar en lenguajes apropiados para ser leídos por un navegador web lo que el diseñador bocetó, creando el esqueleto de la página. También es el que se encarga de cumplir con el estándar de buenas prácticas y emplea el denominado \textit{responsive design} para que el sitio se adapte a distintas vistas, lo que permite obtener un sitio web rápido, amigable con el usuario y usable por este desde cualquier dispositivo. 
    \item Desarrollador, es quien le da la interactividad al sitio usando lenguajes de programación como JavaScript y quien hace de conexión con los encargados del desarrollo back-end.
\end{itemize}

Estas tres tareas pueden ser hechas por una misma persona (es el caso de este trabajo) o combinarse dos de ellas en un solo sub-equipo de trabajo. En la actualidad es cada vez más frecuente que un maquetador sepa programar en los tres lenguajes (HTML, CSS y JavaScript).

\begin{figure} [H]
    \centering
    \includegraphics[width=0.9\textwidth]{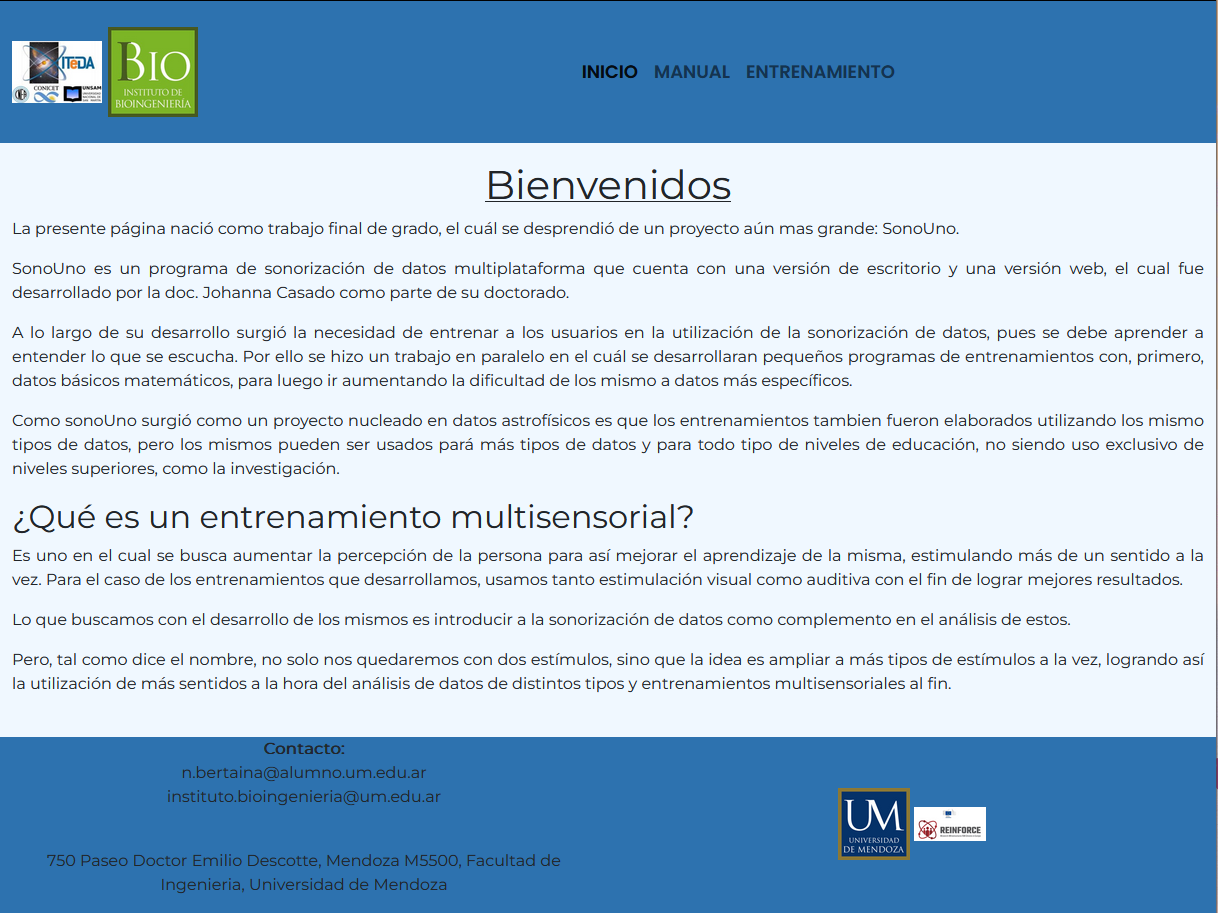}
    \caption{Maquetado de la página index.html}
    \label{fig:25}
\end{figure}

Para el presente trabajo se diseñaron tres páginas web: (1) Inicio, (2) Manual, (3) Entrenamientos. En cada una de ellas el despliegue de elementos y de información es distinto. En la página de inicio se explicó el origen del proyecto, como también lo que es un entrenamiento multisensorial (figura \ref{fig:25}).

Por su parte, en la página titulada Manual, se planeó colocar una breve explicación de lo que el usuario podría encontrarse a la hora de realizar un entrenamiento. Esto acompañado de un breve video que mostrara cómo ingresar a uno y ejecutarlo. Finalmente, la tercer página se maquetó para que los entrenamientos se desplegaran en forma de lista, vinculando cada ítem con un entrenamiento diferente utilizando hipervínculos como conexión. Esta funcionalidad iba a programarse junto con el back-end.

\begin{figure}[H]
    \centering
    \includegraphics[width=0.3\textwidth]{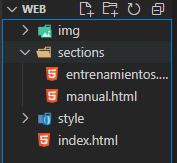}
    \caption{Archivo index.html y carpetas necesarias para el maquetado de un front-end.}
    \label{fig:26}
\end{figure}

El maquetado del sitio se inició creando todas las carpetas que serían necesarias para completar este proceso. En la figura \ref{fig:26} se observa un archivo llamado index.html, el cual corresponde a nuestra página de inicio y que está al mismo nivel que las demás carpetas. Este es el archivo más importante del sitio, ya que sin él no se tendría página web. Cuando se accede a un sitio web desde el navegador, o se busca una página y se obtienen los resultados de la misma, ingresando a la página, es justamente al archivo index al que se tiene acceso. 

Además del archivo index, se debe tener una carpeta que contenga los archivos html de las demás páginas (llamada sections para el presente desarrollo), otra para las imágenes que se deseen usar y por último una carpeta de estilos donde se almacenará el archivo css a la cual se nombró style.

El código de los archivos html puede dividirse en dos grandes secciones para su explicación. Primero se tiene la ``Cabeza'' de dicho archivo, el cual se puede identificar por las etiquetas <<head>> <</head>>. Allí, además de colocar todos los metadatos necesarios para el funcionamiento de la página, se agregaron las etiquetas <<link>> para incluir la hoja de estilos en .css y Bootstrap\footnote{Kit de herramientas para agregar elementos interactivos y responsive al front-end.}. Con la siguiente línea de código se incluyó el kit de herramientas Bootstrap al proyecto, evitando así la instalación del recurso en la computadora.

\begin{lstlisting} [language=HTML]
    <link href=''https://cdn.jsdelivr.net/npm/
    bootstrap@5.1.3/dist/css/bootstrap.min.css'' 
    rel=''stylesheet'' 
    integrity=''sha384-1BmE4kWBq78iYhFl
    dvKuhfTAU6auU8tT94WrHftjDbrCEXSU1oBoqyl2QvZ6jIW3'' 
    crossorigin=''anonymous''>
\end{lstlisting}

La segunda sección es el ``Cuerpo'' de nuestra página, y se puede identificar como todo lo que esté incluido dentro de la etiqueta <<body>> <</body>>. Esta a su vez se puede dividir en tres subsecciones: `Header' o Cabecera de la página; `Main' el cual puede seguir dividiéndose en más secciones y es donde se despliega todo el contenido de la página; `Footer' o pie de página. Para este diseño, se decidió que el `header' incluyera la barra de navegación o `navbar'.

Tanto el `Header', la barra de navegación y el `Footer' de las tres páginas se programaron idénticamente. Con esto se logró mantener un diseño general en todo el sitio. La única diferencia entre las páginas es el cuerpo de estas. Aquí se explayaba el contenido de cada una, se debe tener en cuenta que al variar el cuerpo entre las páginas, su programación también difiere. Esto es algo que puede observarse en los archivos html de cada página que se encuentran en el Apéndice \ref{cap:apend}. 

Para lograr una correcta navegación interna del sitio, los enlaces de cada página, programados en la barra de navegación, se modificaron de acuerdo con su posicionamiento en las carpetas de archivos que contenían todo el maquetado del sitio.

Una vez finalizada la programación de la estructura general de cada página, se procedió con el agregado de los estilos para personalizar el maquetado. Para esto se uso un archivo css. Para lograr una mejor aplicación de estos, se procedió a usar el método de contenedores entre las etiquetas del código html. Este método consiste en ir agregando una propiedad \textit{class} a cada etiqueta, diferenciando las de mayor jerarquía o etiquetas padre de las de menor jerarquía o hijas. El fin de esto es poder aplicar diferentes estilos en distintos elementos, sin usar más de una hoja de estilos css. Esto se considera una buena práctica en el proceso de desarrollo web, ya que optimiza la carga de un sitio web. Un ejemplo de esto puede verse en el siguiente extracto de código:

\begin{lstlisting} [language=HTML]
    <main class="container-fluid">
        <section class="titulo"> 
            <div>
                <h1>Bienvenidos</h1>
            </div>
        </section>
        <section class="about__us">
        ...
        </section>
        <section class="multi">
        ...
        </section>
    </main>
\end{lstlisting}

Puede observarse que la etiqueta ``Padre'' es <<main>>, la cual tiene tres etiquetas hijas con igual jerarquía pero con \textit{class} diferente. Al comenzar con la programación de los estilos, se hizo uso de estas distinciones para poder aplicar propiedades que no afecten a todo el contenido. Sobre todo, se trabajó con el título, el cual se centró y se subrayó. Una muestra de ello puede observarse a continuación:

\begin{lstlisting}
    main{
        grid-area: main;
        text-align: justify;
    }
    .titulo{
        text-align: center;
        text-decoration-line: underline;
        text-decoration-style: solid;
    }
\end{lstlisting}

Aquí puede verse la estructura que debe seguir un archivo css. Primero debe nombrarse la etiqueta que se desea personalizar y luego, entre \{\} se le asignan todas las propiedades. Para el caso que se quiera aplicar un estilo específico a algún elemento, se utiliza un punto seguido del nombre asignado por la función \textit{class}, para luego asignar las distintas propiedades entre llaves. Este método es el que se usó en todo el archivo css que puede verse en el apéndice \ref{cap:apend}. 

\begin{figure} [H]
    \centering
    \includegraphics[width=0.5\textwidth]{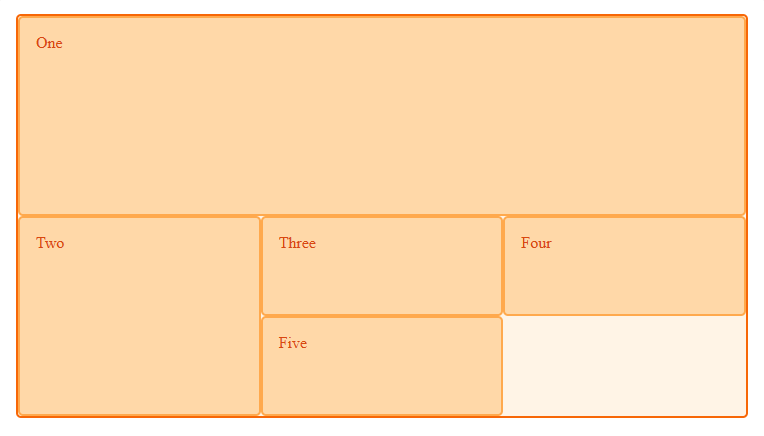}
    \caption{Ejemplo de grilla. Imagen obtenida de \url{https://developer.mozilla.org/en-US/docs/Web/CSS/CSS_Grid_Layout/Basic_Concepts_of_Grid_Layout}, mdn web docs, sección `Grid-Basics concepts of grid layouts-Positioning items against lines'}
    \label{fig:27}
\end{figure}

Una ventaja que trae el uso de contenedores es que permite integrar un esquema de grillas (figura \ref{fig:27}) para posicionar todos los elementos de acuerdo a ella. Para lo cual, en el archivo css debe usarse la propiedad \textit{grid}, especificando las distintas áreas que ocuparán los distintos elementos en la pantalla. Esto se ejemplifica a continuación:

Código en HTML:
\begin{lstlisting} [language=HTML]
    <div class="wrapper">
      <div class="box1">One</div>
      <div class="box2">Two</div>
      <div class="box3">Three</div>
      <div class="box4">Four</div>
      <div class="box5">Five</div>
    </div>
\end{lstlisting}

Código en CSS:
\begin{lstlisting} 
    .wrapper {
      display: grid;
      grid-template-columns: repeat(3, 1fr);
      grid-auto-rows: 100px;
    }
    
    .box1 {
      grid-column-start: 1;
      grid-column-end: 4;
      grid-row-start: 1;
      grid-row-end: 3;
    }
    
    .box2 {
      grid-column-start: 1;
      grid-row-start: 3;
      grid-row-end: 5;
    }
\end{lstlisting}

Además de implementar grillas, se puede optar por usar el modelo de cajas flexibles (\textit{flexbox}) el cual trata con el diseño y posicionamiento de los componentes en una dirección por vez. Este modelo apareció primero y era muy usado hasta que el modelo de grillas expandió las posibilidades de posicionamiento y diseño, volviéndose popular y el más usado al día de la fecha. Por último, existe la posibilidad de usar ambos modelos a la vez, volviendo los diseños mas complejos y personalizados. Esta última opción fue la implementada en el maquetado del sitio web.

Una vez que el maquetado estuvo casi terminado, se implementó el escalado del mismo para distintas medidas de pantalla. Al revisar el archivo css, se puede observar que se hacen tres ``quiebres'', mediante el uso de `Medias Queries'. La mencionada propiedad permite lograr un  \textit{responsive desing}, personalizando los estilos implementados y acomodados para su despliegue en pantallas móviles, tablets o computadora. La forma de usarlo es la siguiente:

\begin{lstlisting} 
    /*pantallas de 500px*/
    @media (min-width:500px){
        .envoltura{
            margin: 0px;
        }
        header{
            grid-template-columns: 0.5fr 2fr;
            grid-template-areas:
            "logoh barra";
            margin-bottom: 1em;
        }
        .logoh{
            grid-area: logoh;
            display: flex;
            flex-direction: column;
            margin-inline: 1em;
        }
        img{
            margin-top: 0.2em;
            margin-bottom: 0.2em;
            width: 4.5em;
        }
        .barra ul{
            display: flex;
            justify-content: space-around;
            margin: 2.3em 0em 2.3em 0em;
            font-family: 'Poppins',sans-serif;
        }
    }
\end{lstlisting}

Para el maquetado del presente trabajo se hizo uso de tres `queries' de tipo media o ``quiebres''\footnote{Estos términos hacen referencia a distintas medidas de pantalla. El uso de `@media' ayuda a crear distintos diseños que se adapten a distintas medidas de pantalla.}: en 320px, 500px y 1024px.

Con este último paso mencionado, se consideró que el maquetado había finalizado satisfactoriamente, por lo que se procedió con la implementación de las distintas funcionalidades, las cuales se agregaron con el lenguaje JavaScript. Para este paso ya se comenzó a modificar lo diseñado para dejar el front-end listo con vista en la comunicación futura con el bak-end. Como el script de los entrenamientos estaba escrito en PsychoJs se decidió encarar el back-end en ese lenguaje, y para lograr una continuidad se procedió a mudar lo programado hasta ahora para el front-end a un nuevo entorno ofrecido por la librería React.

\subsubsection{React}

\begin{figure}[H]
    \centering
    \includegraphics[width=0.9\textwidth]{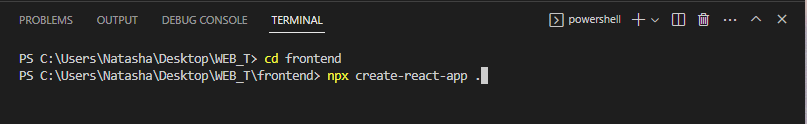}
    \caption{Comando que permite crear el ambiente de desarrollo en React, escrito en consola.}
    \label{fig:28}
\end{figure}

El primer paso a seguir es crear un entorno en React. Para ello, dentro de una misma carpeta se crearon dos sub-carpetas las cuales estaban destinadas a contener el front-end por un lado y el back-end por el otro. Luego de tener la organización general hecha, se procedió a instalar node.js ya que este es uno de los requisitos para poder usar esta biblioteca. Al finalizar con esta instalación, se abrió la carpeta usando Visual Studio Code. Desde la consola de dicho entorno de programación se procedió a crear el ambiente de desarrollo con React en la carpeta ``frontend'', utilizando el comando \textit{create-react-app} (\ref{fig:28}). Este se encarga de instalar todos los paquetes considerados necesarios para iniciar el desarrollo de una aplicación con dicha librería.

\begin{figure}[H]
    \centering
    \includegraphics[width=0.5\textwidth]{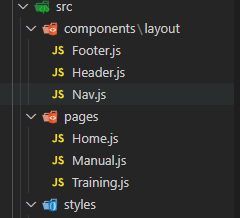}
    \caption{Subcarpetas incluídas en la carpeta src.}
    \label{fig:29}
\end{figure}

Cuando el entorno de React estuvo listo para usarse, se habilitaron las siguientes carpetas las cuales sirven de guía para el desarrollo del front-end: (1) public, conteniendo los archivos que el navegador podrá leer, sobre todo el más importante de ellos, el archivo index.html; y (2) src, donde se almacenan todas las páginas y los estilos aplicados, es decir, el código fuente del sitio (figura \ref{fig:29}).

Una gran diferencia que se tiene al programar en este entorno de desarrollo, es la forma de trabajo. Anteriormente, en el maquetado se crearon archivos únicos por cada página incluida en el sitio web. En React en lugar de páginas se crean componentes, los cuales son módulos que poseen funciones y por ende propósitos únicos. Teniendo esto en cuenta es que se crearon diferentes carpetas dentro de src, las cuales almacenan diferentes componentes. La primera de ellas, llamada \textit{components}, contiene tres módulos que comparten todas las páginas: encabezado (\textit{Header.js}), barra de navegación (\textit{Nav.js}) y pie de página (\textit{Footer.js}). La segunda carpeta, llamada \textit{pages}, reúne los archivos de cada página incluida en el sitio diseñado, los cuales son \textit{Home.js, Manual.js} y \textit{Training.js}. Para finalizar, se creó una tercer carpeta \textit{styles}, encargada de reunir los estilos aplicados a los módulos de la carpeta \textit{components}.

\begin{figure} [H]
    \centering
    \includegraphics[width=0.9\textwidth]{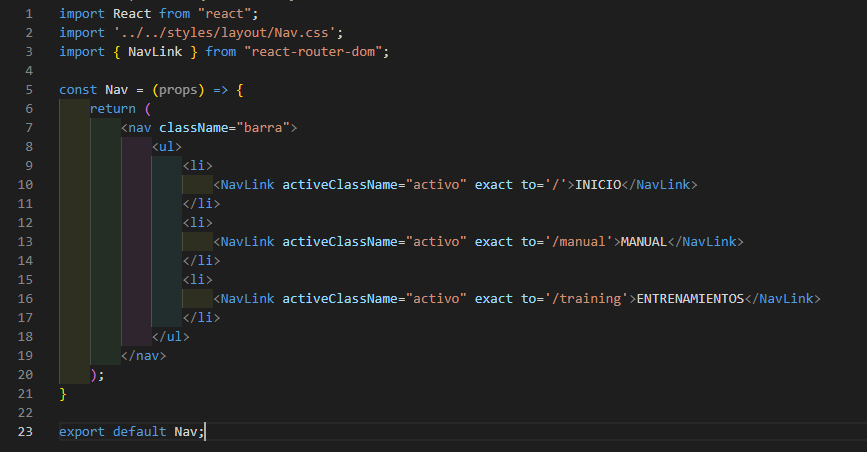}
    \caption{Código del archivo Nav.js.}
    \label{fig:30}
\end{figure}

En cuanto a la estructura de cada módulo o componente, podemos decir que consta de una línea donde se declara el componente con el siguiente formato: \textit{\lstinline !const NombreComponente = (props) => \{...\}!}, seguido del comando \textit{export default} el cual habilita que lo incluido entre las llaves se exporte y muestre por pantalla. Dentro de las llaves anteriormente mencionadas se puede incluir código HTML. Un ejemplo de esto se puede observar en la figura \ref{fig:30}, donde se muestra el código correspondiente al componente \textit{Nav.js}, pudiendo identificarse la estructura mencionada.

Este componente en particular tiene la función de enlazar las distintas páginas que componen el sitio web. Para lograr esto es que se hizo uso de la librería React Router Dom v.5.3.4. 

Para agregarle los estilos que se especificaron en un archivo llamado \textit{Nav.css}, se debieron importar usando la línea: \textit{import '../../styles/layout/Footer.css'}. Esto también se hizo con los otros dos módulos incluidos en la carpeta \textit{components}.

En cuanto a las distintas páginas, estas también fueron programadas como componentes, por lo que en lo que se diferenciaron todos los módulos fue en lo que debían mostrar por pantalla. Un dato importante a mencionar es que los estilos sólo se aplicaron a los módulos de encabezado, barra de navegación y pie de página. El motivo de esta decisión yace en que no hizo falta agregar una personalización aparte para cada página. Si se quisiera establecer un estilo más general, se podría editar el archivo \textit{App.css} que se encarga de esta tarea; el cual no edita lo establecido en cada archivo .css aplicado por módulo, solo se manifiesta en aquellos que no tengan exportado un archivo .css exclusivo.

Una vez que todos los componentes y las páginas estuvieron listas, se procedió a incluirlas en el archivo \textit{App.js}. Dicho archivo es el corazón de la aplicación; es donde se debe incluir todo lo que se desea visualizar en el navegador. El código de este archivo es el que se muestra a continuación (se incluye en el texto porque se harán menciones a ciertas características del mismo).

\includepdf [pages=1]{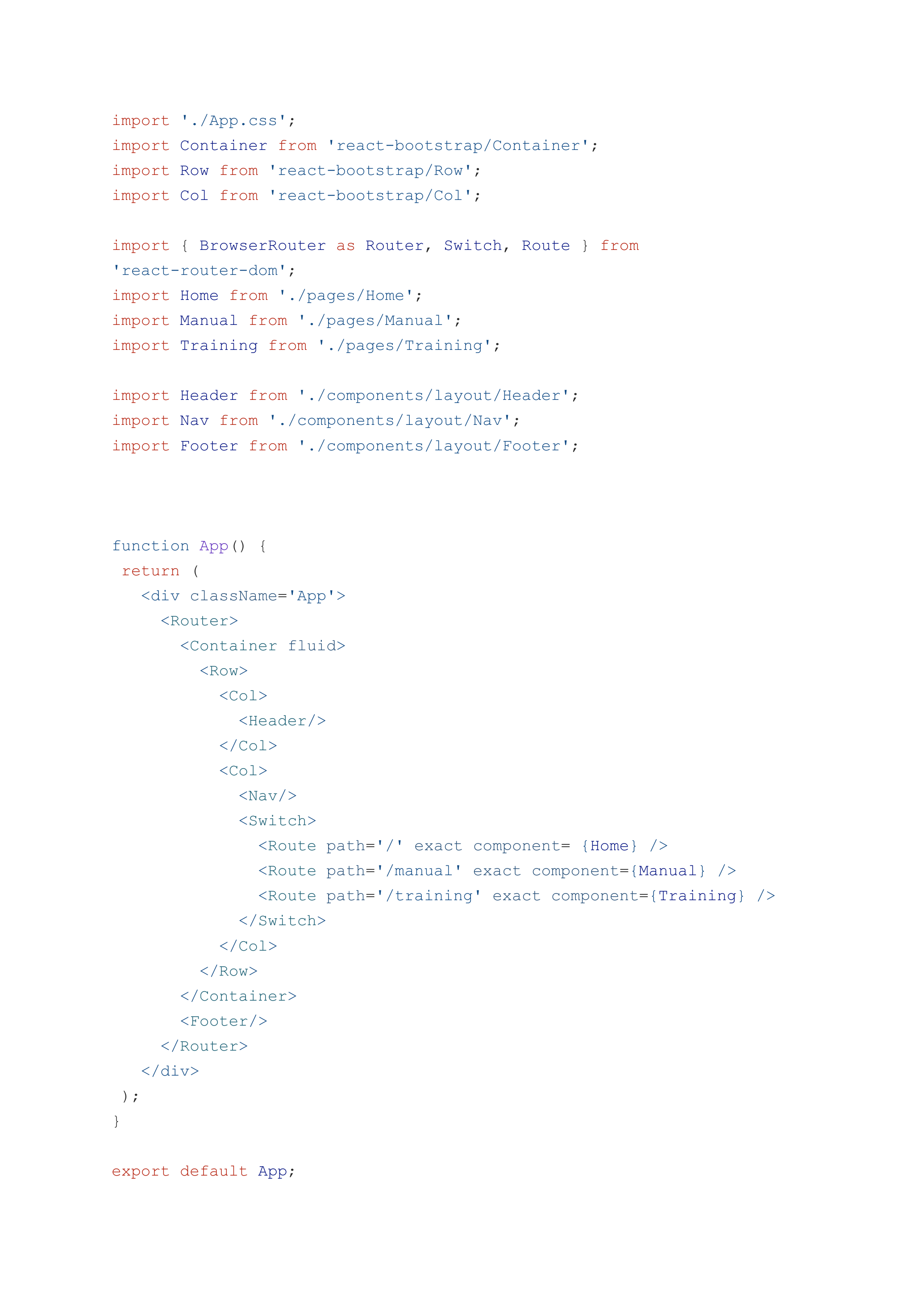}

Como se observa, dicho archivo se compone de declaraciones \textit{import}; de una función llamada \textit{App}, la cual ordena la vista de los componentes que se desean desplegar en la pantalla; y finalmente, de una declaración \textit{export} que permite que el componente \textit{App} esté disponible.

Para que los enrutamientos declarados en el archivo \textit{Nav.js} funcionen correctamente se deben escribir las siguientes funciones:

\begin{itemize}
    \item Router: funciona como una envoltura para \textbf{App.js}. Esto nos da acceso a la API de historia de HTML5, lo que permite una sincronía de la interfaz gráfica con la localización actual. Su hijo es la función Switch.
    \item Switch: permite el cambio y renderización de Route que coincida con la localización actual.
    \item Route: elige que componente renderizar según la localización actual. Se deben escribir tantas funciones Route como enlaces se hayan definido en la barra de navegación.
\end{itemize}

\begin{figure} [H]
    \centering
    \includegraphics[width=0.9\textwidth]{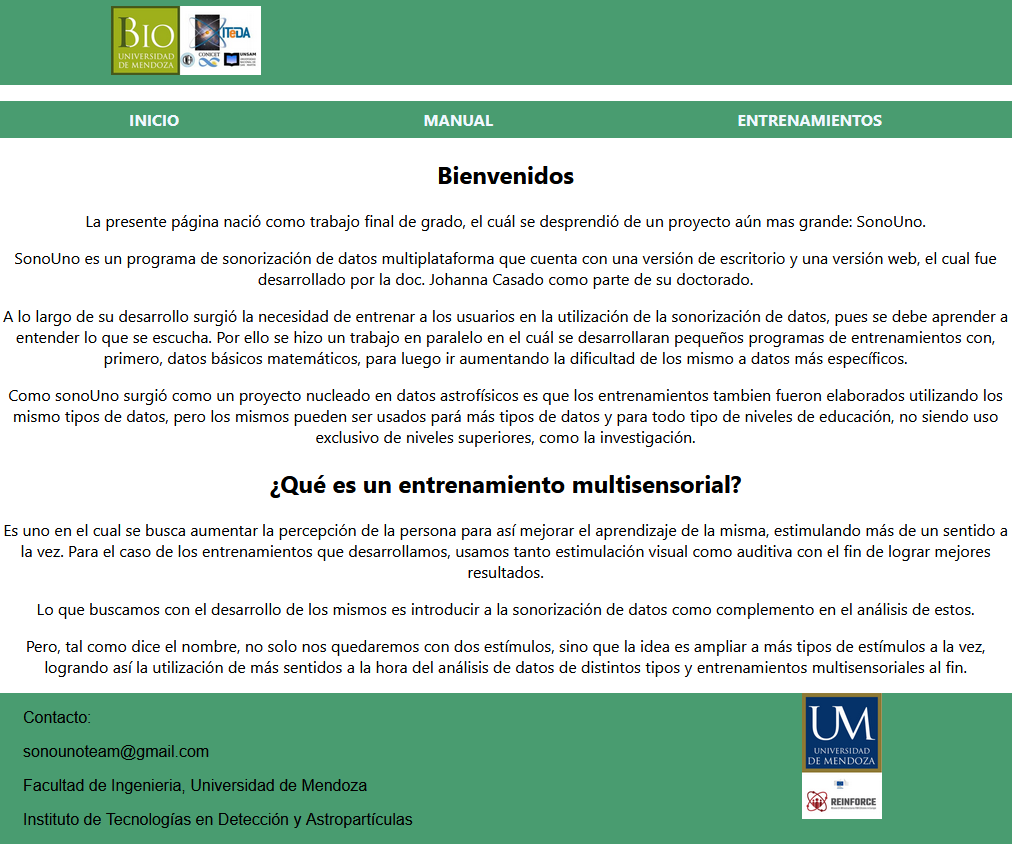}
    \caption{Página inicio del sitio web elaborado}
    \label{fig:31}
\end{figure}

El resultado final de haber mudado todo el sitio a React se muestra en la figura \ref{fig:31}. Como se observa, aún se debería mejorar el aspecto de diseño como también terminar de escribir el manual de uso. Para esto último era necesario trabajar con el back-end para lograr que los entrenamientos corrieran en la web. Luego de tener al menos un entrenamiento funcionando en la página web es que se puede escribir el manual de usuario.

\subsection{Desarrollo back-end}

El proyecto se almacenó en una carpeta, la cual se dividió a su vez en dos subcarpetas. Una de ellas almacenaba todos los archivos del front-end, mientras que la segunda estaba destinada a contener todos los archivos que el back-end necesitaría. Una vez que el front-end estuvo establecido, se procedió a desarrollar el back-end. El primer paso a seguir fue hacer una búsqueda de archivos que fueran de guía en la forma de agregar los archivos escritos en JavaScript de los entrenamientos desarrollados con PsychoPy. 

El primer lugar consultado fue la documentación disponible de psychojs y en ella se informaba que para ejecutar un entrenamiento generado con Psychopy debía usarse el index.html que se obtiene automáticamente al hacer la conversión del archivo escrito en lenguaje Python a lenguaje JavaScript, así como el archivo en Python. Son esos archivos los que se deben alojar en el servidor web, incluyendo en este, además, la biblioteca PsychoJS. 

Al tener la idea general de lo que debía hacerse, se procedió a consultar a personas que trabajan en la elaboración de back-end la forma más eficaz de realizar dicha tarea. Todos estos especialistas informaron lo mismo: ``la inclusión de lo solicitado a un servidor es imposible de hacerse". Esto se debe a que la documentación consultada de Psychojs, fue elaborada cuando el sitio web `pavlovia.com' no estaba desarrollado por completo. Al finalizar con su desarrollo se modificaron esos requisitos para que no pudiera emularse. Esto mismo fue expuesto por uno de los colaboradores del proyecto Psychopy cuando un usuario tuvo una idea similar a la que se quiso lograr en el presente trabajo. Fue esto lo que impidió continuar con la adición de los entrenamientos al sitio web diseñado, al menos de la forma propuesta en este trabajo.

A pesar de ello, el proyecto se subió a un servidor: Heroku. Este fue elegido principalmente por dos razones, la primera de ellas es que es una plataforma de tipo PaaS\footnote{Los servicios ofrecidos en la nube se pueden dividir en tres: IaaS, PaaS y SaaS. La diferencia entre las tres es la forma de presentar el servicio de nube para las distintas empresas. Particularmente, un servicio de tipo PaaS es aquel que ofrece una plataforma como servicio, además de distintas herramientas que ayudan al desarrollador a crear distintas aplicaciones.} que soporta distintos lenguajes de programación, sobre todo Node.js, lenguaje que sería el elegido para programar todo el back-end. La segunda de las razones es que esta plataforma ofrece un año de servicio gratuito, volviéndolo ideal para el despliegue del sitio web al menos durante la etapa de desarrollo. Existen otros servidores que ofrecen un alojamiento gratuito, pero tienen la desventaja de soportar pocos lenguajes de programación, por lo que para elegirlos había que cambiar el lenguaje de programación usado a uno que fuera soportado por el servicio o simplemente descartar ese servidor y buscar otras opciones.

Una vez elegido el servidor, se procedió con la subida de todo el proyecto desarrollado a la nube. Para controlar todos los cambios que se hicieran en un futuro es que se hizo uso de un repositorio Git local, el cual se creo en la carpeta del proyecto. Finalizado esto, desde la consola de comandos se continuó con el login a Heroku para luego crear la aplicación del proyecto allí mediante el comando \textit{heroku create}. Una vez creada dicha aplicación, se subió usando otro comando, el cual devuelve el link de la aplicación.

Una vez que la aplicación estuvo alojada en el servidor es que se siguió con el paso mencionado al inicio del capítulo, es decir, buscar la forma de subir a este tanto la biblioteca PsychoJs, el archivo index.html como el archivo de los entrenamientos en lenguaje JavaScript. Al no ser esto posible es que el desarrollo del back-end sólo llego hasta esta instancia.

Si bien se encontró la descripta problemática, se considera un gran avance para el equipo de trabajo en el que se desarrolló este trabajo final. Se ha logrado diseñar, desarrollar y testear entrenamientos de sonorización, incluyendo la creación de todo el front-end para poder realizar dichos entrenamientos en diferentes entornos y ámbitos de aplicación. Destacando que ya se cuenta con una plataforma de back-end, se detectó la imposibilidad de integración directa de los entrenamientos de PsychoPy, y se continuará evaluando la forma de integrar entrenamientos para poder ser ejecutados mediante la web.

\chapter{Estudio Económico}
\label{cap:Estudio}
El estudio económico en propuestas que se enmarcan en investigación y desarrollo en muchas ocasiones son complejos, sobre todo cuando no se tienen ejemplos comparables al trabajo realizado. En el caso del presente trabajo, la parte del diseño y desarrollo de entrenamiento (o estudios de comportamiento, aplicación que se le ha dado a PsychoPy hasta el momento), al ser una tarea utilizada puramente como técnica de investigación, no se encuentra un promedio de costo asignado a dicha tarea. Sin embargo, el desarrollo de páginas web sí cuenta con montos de referencia, los que pueden consultarse en la web o a personas que estén realizando este trabajo (en el presente estudio se optó por esta última alternativa).

Dado que el presente trabajo final se enmarca como el inicio de una futura tesis doctoral, a realizar por la misma autora y ya habiendo obtenido beca de CONICET, se decidió realizar aquí una estimación de costos a cinco años (considerando que esta es la duración promedio de una tesis). En el presente capítulo se hará una comparativa de los costos que se deberían afrontar si se decidiera contratar la licencia ofrecida para poder ejecutar los entrenamientos en `pavlovia.com' durante cinco años, contra lo que se gastaría en cinco años de desarrollo de back-end para producir un sitio propio de entrenamientos, dejando de lado el uso de psychopy.

En primer lugar se expondrán los distintos gastos que se deberían considerar a la hora de elegir un desarrollo web independiente. El valor de un trabajo de tal magnitud, es decir, con una página principal, una dedicada a un manual de uso y otra que sea un índice con distintos entrenamientos, los cuales están programados cada uno como páginas distintas, y que incluya una base de datos sencilla de usuarios para recolectar los datos de cada sesión de entrenamiento, dependerá principalmente de dos factores. El primero de ellos es el tiempo en el que se desea tener la página funcionando, como también el del equipo de programadores que se hayan contratado para realizar dicha tarea. Hoy en día si se desea ahorrar en costos y se contrata a un programador junior\footnote{Aquellos programadores que se han incorporado recientemente al mundo de la programación} el precio estaría comprendido entre los \$150000 y \$200000, pudiendo demorar el proyecto de tres a cuatro meses en estar listo, con todo lo mencionado anteriormente. Si por otro lado, se desea tener un proyecto más rápido y que se pueda escalar a futuro, entonces conviene contratar a un equipo compuesto de varios programadores. Esto hará subir considerablemente el precio, el cual rondaría entre los \$600000 a \$700000, pero se podría tener el sitio en funcionamiento en cuestión de tres semanas. 

Cabe aclarar que estos valores se pagan una única vez, y el precio más alto incluiría el precio de subirlo a un servidor, como también el hosting. Este a su vez podría variar de acuerdo con el dominio que se esté dispuesto a pagar. Si además se desea contratar un servicio de mantenimiento, este podría variar de acuerdo a la complejidad final del sitio, pudiendo tener un valor mínimo de aproximadamente \$9000 por mes, dependiendo mucho de la empresa o del programador independiente que se contrate. 

Una vez expuesto estos valores, podemos comparar el costo de contratar una licencia en `pavlovia.com' y el costo de desarrollar una web acorde al proyecto presentado a lo largo de este trabajo, programado por un programador junior.

\begin{table}[H]
    \centering
    \begin{tabular}{| c | c | c | c |}
    \hline
       Año & Costo desarrollo web & Costo mantenimiento anual & Costo total \\ \hline
        1° & \$200000 & \$108000 & US\$1557,52 \\ \hline
        2° & \$0 & \$115200 & US\$582,55 \\ \hline
        3° & \$0 & \$115200 & US\$582,55 \\ \hline
        4° & \$0 & \$115200 & US\$582,55 \\ \hline
        5° & \$0 & \$115200 & US\$582,55 \\ \hline
    \end{tabular}
    \caption{Costo del desarrollo web y mantenimiento proyectado cinco años a futuro}
    \label{tab:3}
\end{table}

En la tabla \ref{tab:3} se hicieron las siguientes consideraciones:
\begin{itemize}
    \item Para el costo total se consideró el valor del dólar oficial al día 12 de Febrero del año 2023: US\$197,75.
    \item Todos los costos, a partir del segundo año, son considerados con el precio actual de las prestaciones y sin tener en cuenta las variaciones que podrían haber en su valor anual debido a la inflación.
    \item El costo de desarrollo sólo se abona una única vez, incluyendo el primer año el precio del hosting y dominio\footnote{Para obtener un valor aproximado de este costo se consultó con un programador con conocimiento en el área, como tambien en páginas de servicios freelance (fiverr).}.
    \item El costo del mantenimiento ofrecido el primer año no incluye el valor del dominio y hosting por lo que ya fue considerado en el ítem anterior. Aclarando que el precio mensual considerado es de \$9000\footnote{Se consultaron una gran variedad de páginas (paginaswebsac, berkanadigital) y se sacó un promedio del costo de mantenimiento de una página e-commerce.}.
    \item El costo del mantenimiento a partir del segundo año ya suma al valor anual del mismo, el costo de renovación anual del dominio y el hosting. Para esto se buscaron distintos proveedores de servidores y se compararon los paquetes que ofrecen. De esto se vio que el rango de precios va desde los \$400 hasta los \$900 mensuales. La variación de precios depende del paquete que se haya elegido al inicio del proyecto. El cálculo final se hizo suponiendo la contratación de un plan intermedio\footnote{Precios consultados de Donweb, el día 12 de febrero del 2023.} el cual ofrece almacenamiento de 100 GB, dominio, soporte técnico, ancho de banda sin medición, entre otras características, y cuyo costo es de \$600 mensuales.
\end{itemize}

Una licencia anual en Pavlovia tiene un costo de \pounds 1800. Una libra esterlina equivale a US\$1,21\footnote{Valor consultado de oanda.com, el día 12 de febrero del 2023}; teniendo en cuenta esto, la tabla \ref{tab:4} expresa el valor mencionado de la licencia anual en precio dólar, facilitando de este modo la comparación.

\begin{table}[H]
    \centering
    \begin{tabular}{| c | c | c |}
    \hline
        Año & Licencia anual Pavlovia & Costo desarrollo Web y mantenimiento \\ \hline
        1° & US\$2178 & US\$1557,52 \\ \hline
        2° & US\$2178 & US\$582,55 \\ \hline
        3° & US\$2178 & US\$582,55 \\ \hline
        4° & US\$2178 & US\$582,55 \\ \hline
        5° & US\$2178 & US\$582,55 \\ \hline
    \end{tabular}
    \caption{Comparación de costos con una proyección a cinco años}
    \label{tab:4}
\end{table}

Esto justifica un trabajo de desarrollo y mantenimiento web para crear un sitio que se adapte a las necesidades del proyecto como también al contexto económico que se vive en nuestro país. Debe destacarse el beneficio que representa un desarrollo propio en el marco de la formación de recursos humanos capacitados, la posibilidad de generar nuevas líneas de investigación, consolidar grupos de investigación y desarrollo en ámbitos académicos, presentar resultados en congresos de especialistas en el tema y, finalmente, posicionar al país en un tema de vacancia de alto impacto social.

\chapter{Resultados}
\label{cap:resultados}
En este capítulo se resaltarán los resultados obtenidos en cada etapa del desarrollo:

\begin{itemize}
    \item Primer Workshop: los resultados obtenidos en esta instancia dan un indicio de la mejoría en la detección de señales cuando se utiliza un despliegue multisensorial de los datos. Además, se obtiene realimentación para realizar mejoras en los despliegue de los entrenamientos, como ser: mostrar el texto de forma auditiva y el tiempo de espera de la respuesta, entre otros.
    \item Segundo Workshop: a pesar de lo novedoso de los entrenamientos, estos requieren unas adaptaciones que no pueden ser implementadas usando el \textit{Builder} de psychopy, por lo que deberían tratar de implementarse sobre el código mismo, es el caso de sonorizar el texto de la retroalimentación. Sumado a estas mejoras, se vio la necesidad de realizar un desarrollo web que permita ejecutar en la web los entrenamientos diseñados. Fue en base a este workshop que surgió la segunda parte del desarrollo expuesto en el presente trabajo.
    \item Desarrollo web: en esta etapa se logró realizar el diseño front-end y la implementación de un sitio web, se encontró el problema de integración respecto al código de PsychoPy, pero aún así se dejan sentadas las bases del back-end para continuar con este desarrollo en el futuro.
\end{itemize}

\chapter*{Conclusiones}
\addcontentsline{toc}{chapter}{Conclusiones}
En base a los resultados expuestos al final de cada sección del capítulo \ref{cap:Desarrollo} es que se concluye que se cumplió parcialmente el objetivo general planteado al inicio de este trabajo. Esto es debido a la imposibilidad de concluir el desarrollo back-end para la web planeada. Esta situación no es inesperada en el desarrollo de un trabajo de investigación en temas con antecedentes escasos y en donde lo que se propone es, evidentemente, novedoso. Debe verse como una oportunidad para trabajos a futuro que puedan abordarse en un plazo de tiempo mayor. A pesar de ello, se cumple con la mayoría de los objetivos específicos y se cumplimenta con lo requerido para un trabajo final de grado de la carrera de Bioingeniería.

En cuanto a los entrenamientos desarrollados con Psychopy, se logró comprender el funcionamiento de esta herramienta y obtener el mayor provecho posible de sus recursos, diseñando entrenamientos que cumplieran con los objetivo planteado. Como se detalló, también se encontraron limitaciones en dicho software, como la imposibilidad de agregar una respuesta auditiva en el feedback. Por ello se plantea hacia el final de este trabajo independizarse de este recurso en futuras actualizaciones de esta investigación y propuesta, elaborando una alternativa que cumpla con todo lo pedido por los usuarios. 

A pesar de las limitaciones encontradas, se puede concluir que los entrenamientos elaborados sirvieron como antecedente a proyectos ya planificados, abriendo una nueva línea de desarrollo e investigación, pudiendo a futuro centrarse más en la investigación de cómo la percepción podría verse afectada por el uso de ciertos tonos o melodías usadas en la sonorización de datos astronómicos. Dicha investigación es la que se propone para realizar en la futura tesis doctoral de la autora.

Finalmente, es importante destacar que se ha tenido en cuenta desde el inicio de este desarrollo a usuarios con diversidad funcional. Siguiendo los lineamientos del equipo de trabajo, los diferentes desarrollos obtenidos han tenido en cuenta realizar el despliegue de información de forma multisensorial. Alcanzar la equidad en el acceso a las herramientas y la información debería ser una prioridad siempre, cada persona debería encontrarse con un ambiente equitativo, tanto en lo laboral, como en lo educativo y recreativo.

\chapter{Discusión y trabajo futuro}
\label{cap:TrabajoF}
Del presente trabajo se desprendió un plan de trabajo para una Tesis Doctoral en el cual se continuará con el desarrollo de entrenamientos que puedan ejecutarse en la web, como también adaptarse a las distintas necesidades que el grupo que desarrolla el soft sonoUno requiera. De los resultados obtenidos en el desarrollo web, se puede decir que más que adaptar lo diseñado en Psychopy, se desarrollarán los entrenamientos desde cero para su incorporación en la web, para así sortear las posibles complicaciones y barreras que puedan surgir al usar desarrollos ya establecidos, con las limitaciones que representan los diseños que no atienden al usuario final. 

Los nuevos entrenamientos diseñados se orientarán a personas con diversidad funcional, utilizando en su despliegue datos específicos obtenidos de distintas fuentes, como por ejemplo las ciencias espaciales o ámbitos de la física de partículas. El fin perseguido será aportar datos científicos confiables que aumenten y promuevan la inclusión, tanto en el ámbito científico como de educación y difusión.

En cuanto al diseño web, este partirá del concepto de accesibilidad web, teniendo en cuenta todas las normas establecidas para lograr este punto, como también testear continuamente los avances alcanzados. Se busca obtener retroalimentación que ayude a mejorar tanto el diseño web como el de los entrenamientos. 

Se plantea, dentro de los límites de dicho trabajo a futuro, un análisis de datos en búsqueda de nuevos resultados con respecto a la efectividad de los entrenamientos en sonorización de datos, como también su impacto en la percepción de los usuarios, ampliando la investigación a más ramas de la ciencia, por ejemplo la neuropsicología.

\addcontentsline{toc}{chapter}{Bibliografía}
\bibliography{biblio}

\begin{thebibliography}{}

\bibitem[Casado et~al., 2019]{sonoUno}
Casado, J., Carricondo~Robino, J., Palma, A., Diaz-Merced, W., and García, B.
  (2019).
\newblock Sonouno, un software para el análisis multimodal de datos.
\newblock In {\em 3er Workshop de Difusión y enseñanza de la Astronomía}.

\bibitem[Del Angel~Arrienta et~al., 2021]{Angel}
Del Angel~Arrienta, F., Rojas~Cisneros, M., Rivas, J.~J., Castrejon, L.~R.,
  Sucar, L.~E., Andreu-Perez, J., and Orihuela-Espina, F. (2021).
\newblock Characterization of a raspberry pi as the core for a low-cost
  multimodal eeg-fnirs platform.
\newblock In {\em Annual International Conference of the IEEE Engineering in
  Medicine \& Biology Society}, volume 43rd, pages 1288--1291.

\bibitem[Docs, 2023]{Servidor}
Docs, M.~W. (2023).
\newblock Recuperado de:
  \url{https://developer.mozilla.org/es/docs/Learn/Common_questions/What_is_a_web_server}.
\newblock (accedido: 05-Enero-2023).

\bibitem[Felke-Morris, 2020]{protocols}
Felke-Morris, T. (2020).
\newblock {\em Web Development and Design Foundations with HTML5}.
\newblock Pearson, 10 edition.

\bibitem[Ittelson, 1973]{Ittelson}
Ittelson, W. (1973).
\newblock Environment perception and contemporary perceptual theory.
\newblock {\em Environment and Cognition}, pages 141--154.

\bibitem[Masache and Abad, 2020]{Ulloa}
Masache, E. J.~U. and Abad, R. F.~E. (2020).
\newblock Diseño y validación de un paradigma para evaluar la atención
  selectiva, utilizando el software de código abierto “psychopy”,
  aplicable a la resonancia magnética funcional.
\newblock {\em Revista ecuatoriana de Neurología}, 29(3):55--64.

\bibitem[Natasha María~Monserrat et~al., 2022]{paper}
Natasha María~Monserrat, B.~L., Johanna, C., Beatríz, G., and Gastón, J.
  (2022).
\newblock The use of sonification in data analysis: a psychopy training test.
\newblock {\em Libro de Resúmenes del XXIII Congreso Argentino de
  Bioingeniería y XII Jornadas de Ingeniería Clínica : San Juan, Argentina:
  Septiembre de 2022}, page 155.

\bibitem[Opoku-Baah et~al., 2021]{Opoku-baah}
Opoku-Baah, C., Schoenhaut, A.~M., Vassall, S.~G., Tovar, D.~A., Ramachandran,
  R., and Wallace, M.~T. (2021).
\newblock Visual influences on auditory behavioral, neural, and perceptual
  processes: A review.
\newblock {\em Journal of the Association for Research in Otolaryngology},
  22(4):365--386.

\bibitem[Oviedo, 2004]{Oviedo}
Oviedo, G.~L. (2004).
\newblock La definición del concepto de percepción en psicología con base en
  la teoría de gestalt.
\newblock {\em Revista de estudios Sociales}, 18:89--96.

\bibitem[Oxenham, 2018]{Oxenham}
Oxenham, A.~J. (2018).
\newblock How we hear: The perception and neural coding of sound.
\newblock {\em Annual Review of Psychology}, 69:27--50.

\bibitem[Peirce et~al., 2019]{Peirce}
Peirce, J., Gray, J.~R., Simpson, S., MacAskill, M., Höchenberger, R., Sogo,
  H., Kastman, E., and Lindeløv, J.~K. (2019).
\newblock Psychopy 2: Experiments in behavior made easy.
\newblock {\em Bahavior Research Methods}, 51:195--203.

\bibitem[Portellano, 2005]{Introduccionaneuropsicologia}
Portellano, J.~A. (2005).
\newblock {\em Introducción a la neuropsicología}.
\newblock McGraw-Hill.

\bibitem[Smith and Walker, 2005]{Training}
Smith, D.~R. and Walker, B.~N. (2005).
\newblock Effects of auditory context cues and training on performance of a
  point estimation sonification task.
\newblock {\em Applied Cognitive Psychology}, 19:1065--1087.

\bibitem[Source, 2023]{REACT}
Source, F.~O. (2023).
\newblock Recuperado de: \url{https://es.reactjs.org/}.
\newblock (accedido: 22-Dic-2022).

\bibitem[Triveño~Mosquera et~al., 2019a]{percepcionvisual}
Triveño~Mosquera, M., Bembibre~Serrano, J., and Arnedo~Montoro, M. (2019a).
\newblock {\em Neuropsicología de la percepción}.
\newblock Editorial Síntesis.

\bibitem[Triveño~Mosquera et~al., 2019b]{percepcionauditiva}
Triveño~Mosquera, M., Bembibre~Serrano, J., and Arnedo~Montoro, M. (2019b).
\newblock {\em Neuropsicología de la percepción}.
\newblock Editorial Síntesis.

\end{thebibliography}

\appendix
\chapter{Códigos}
\label{cap:apend}

\section{Python}

\includepdf[pages=1-3]{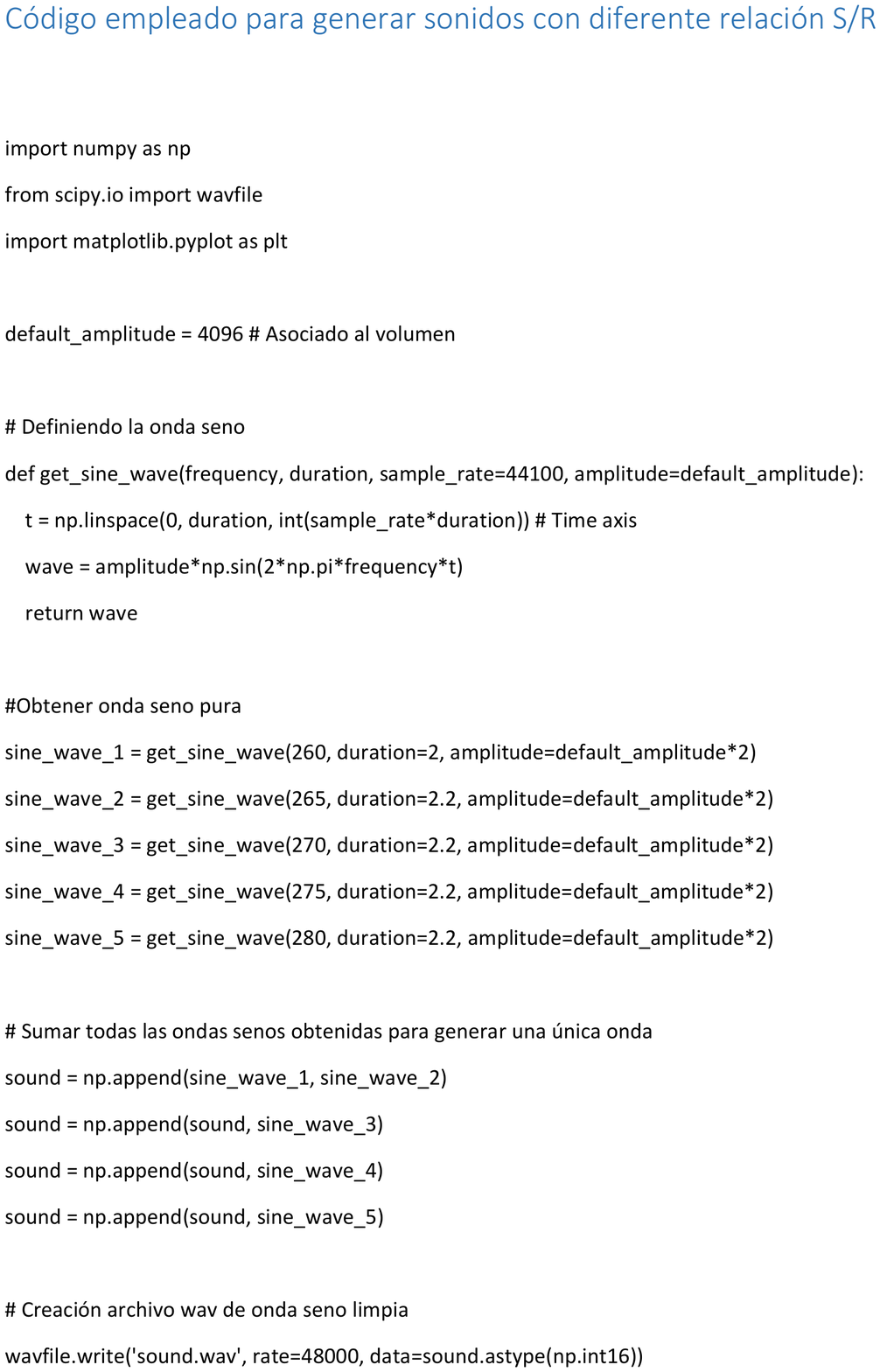}

\section{Maquetado}

\includepdf [pages=1-3]{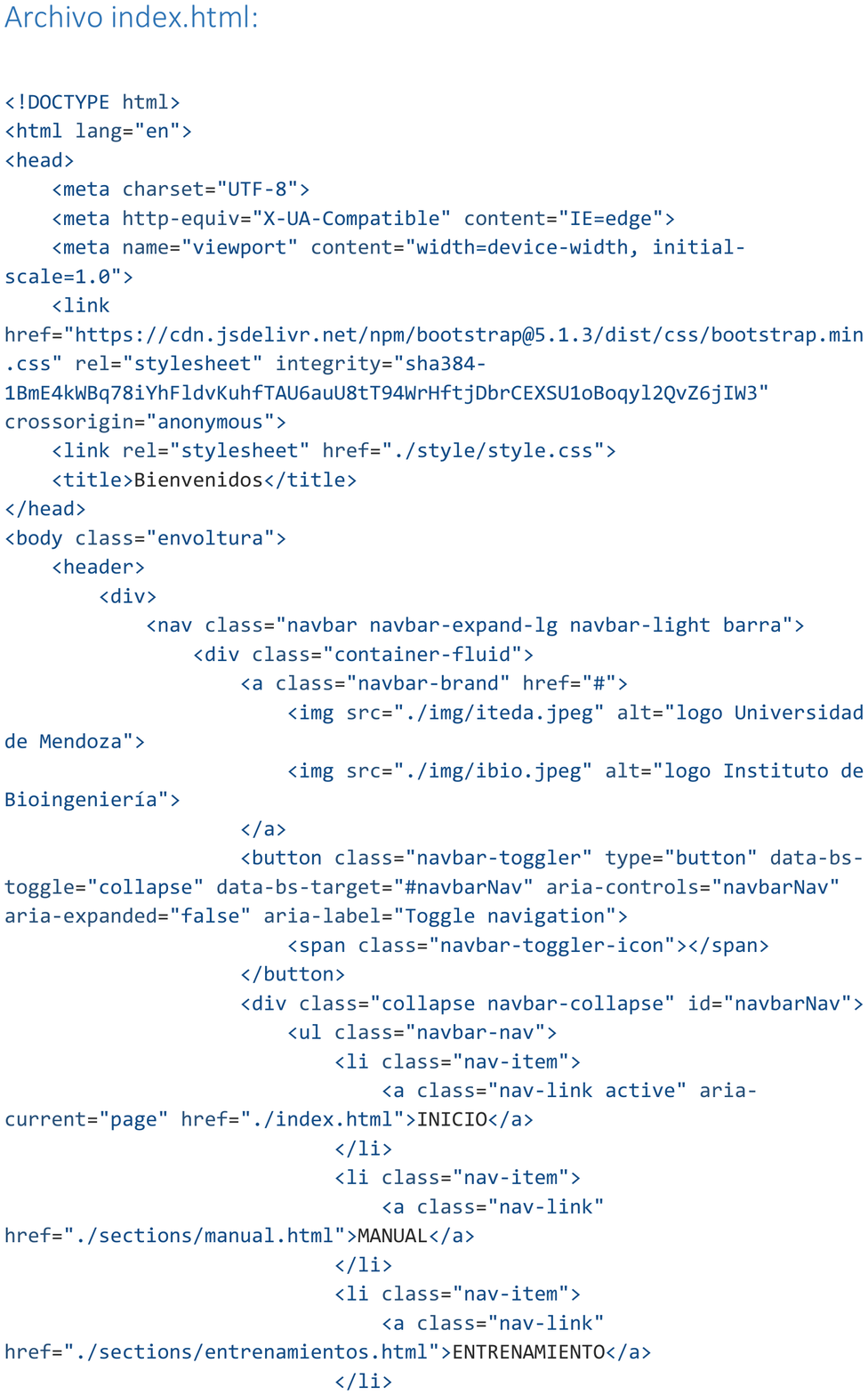}
\includepdf [pages=1-2]{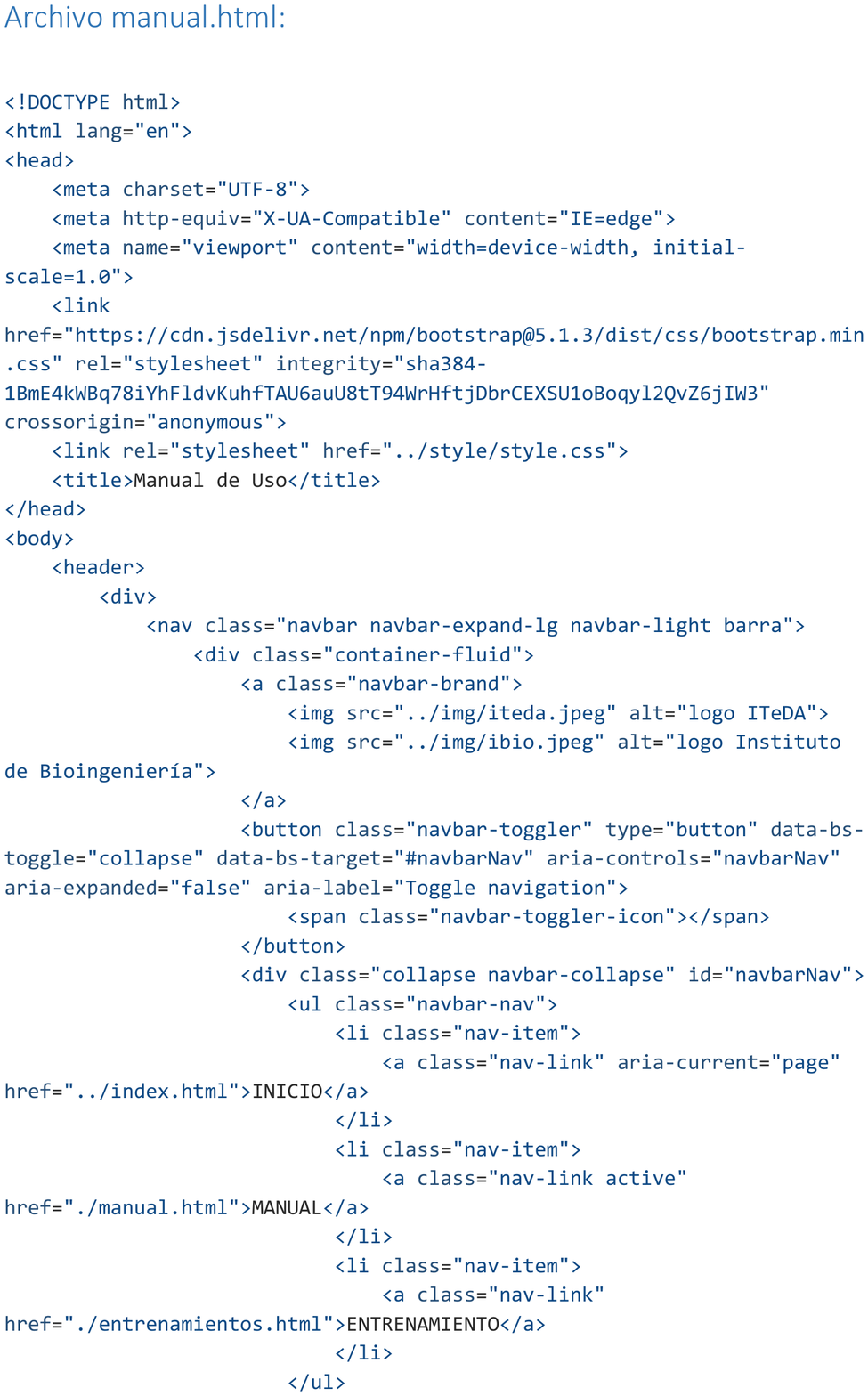}
\includepdf [pages=1-2]{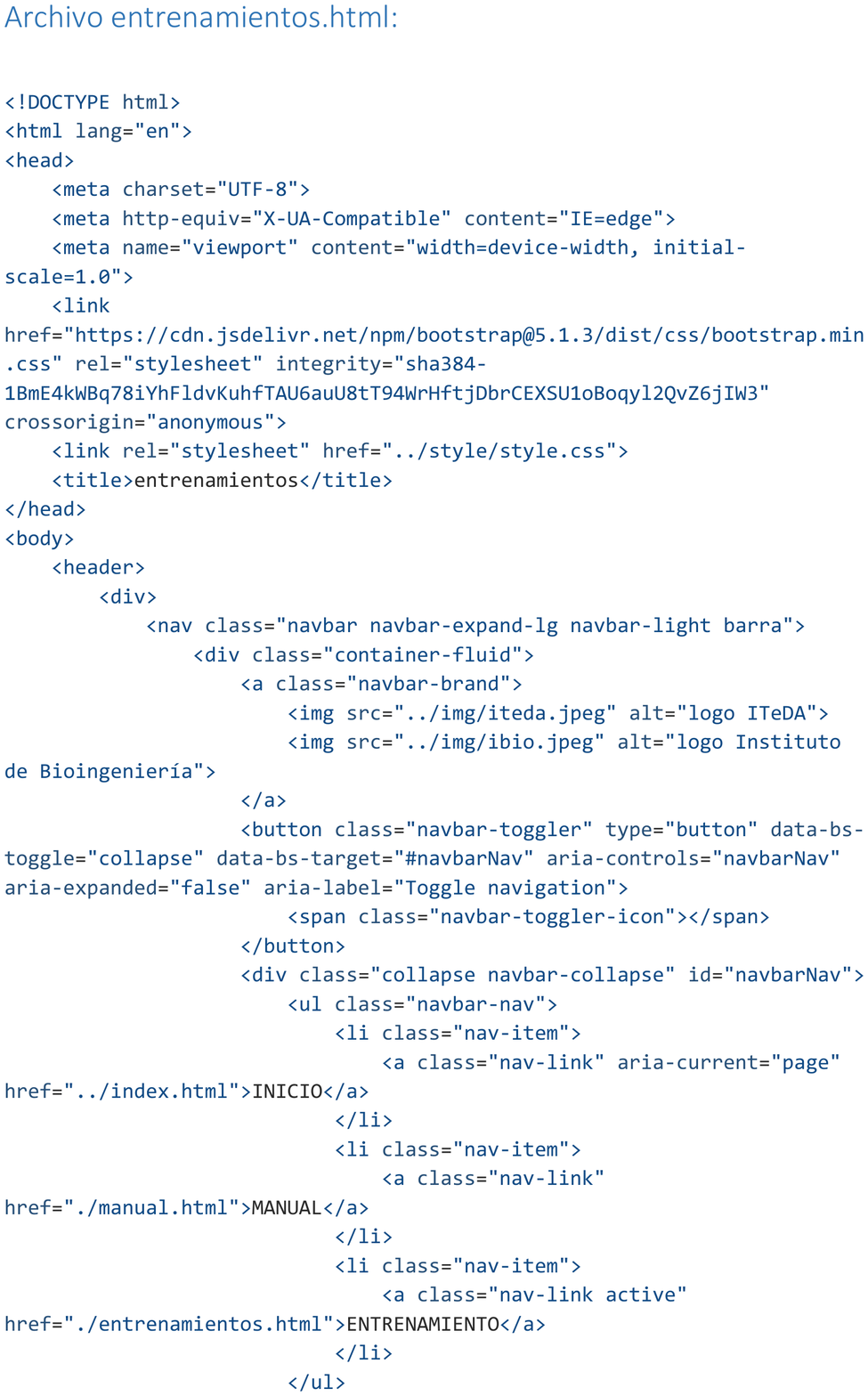}
\includepdf [pages=1-3]{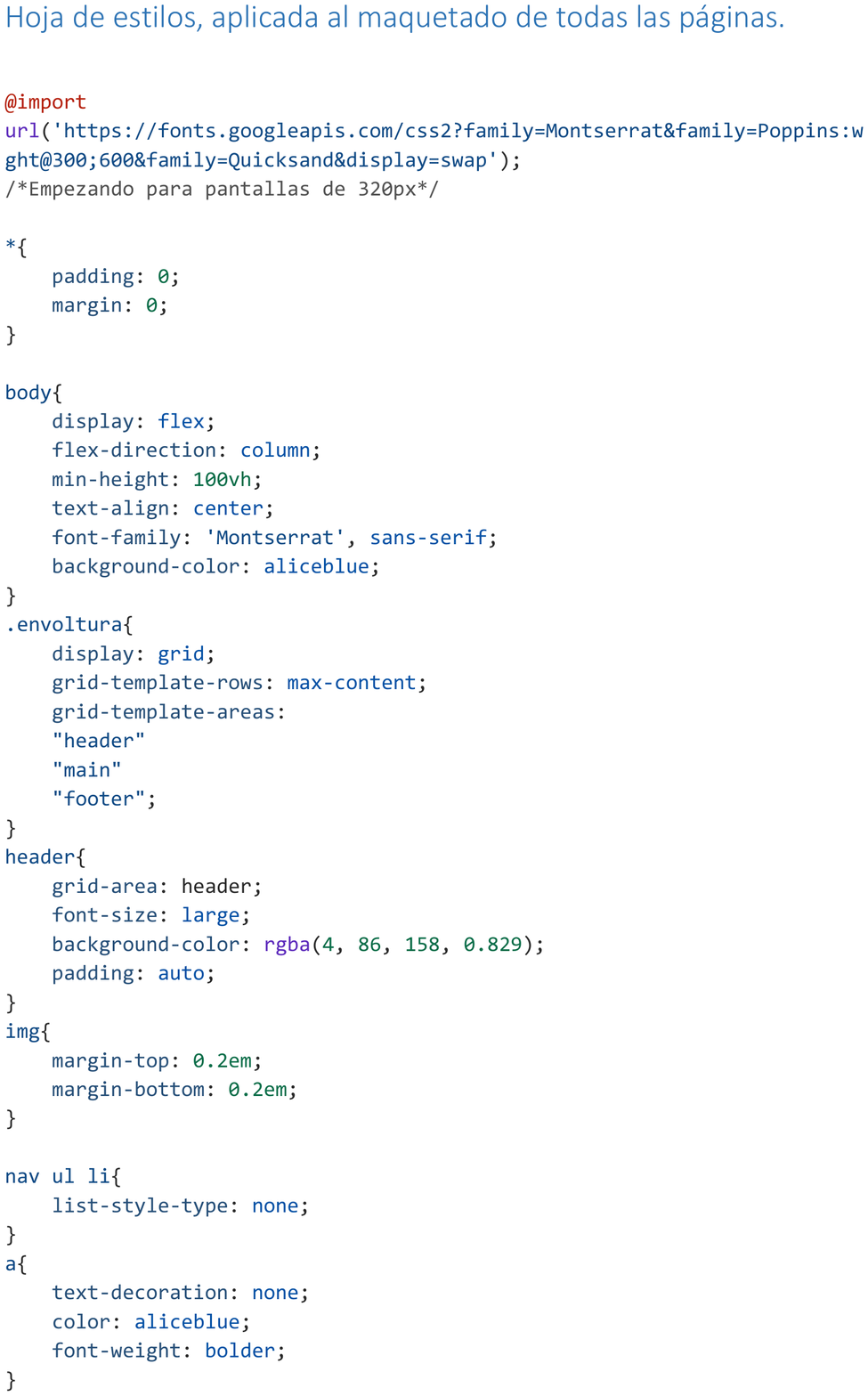}

\section{Códigos en React}

\includepdf [pages=1]{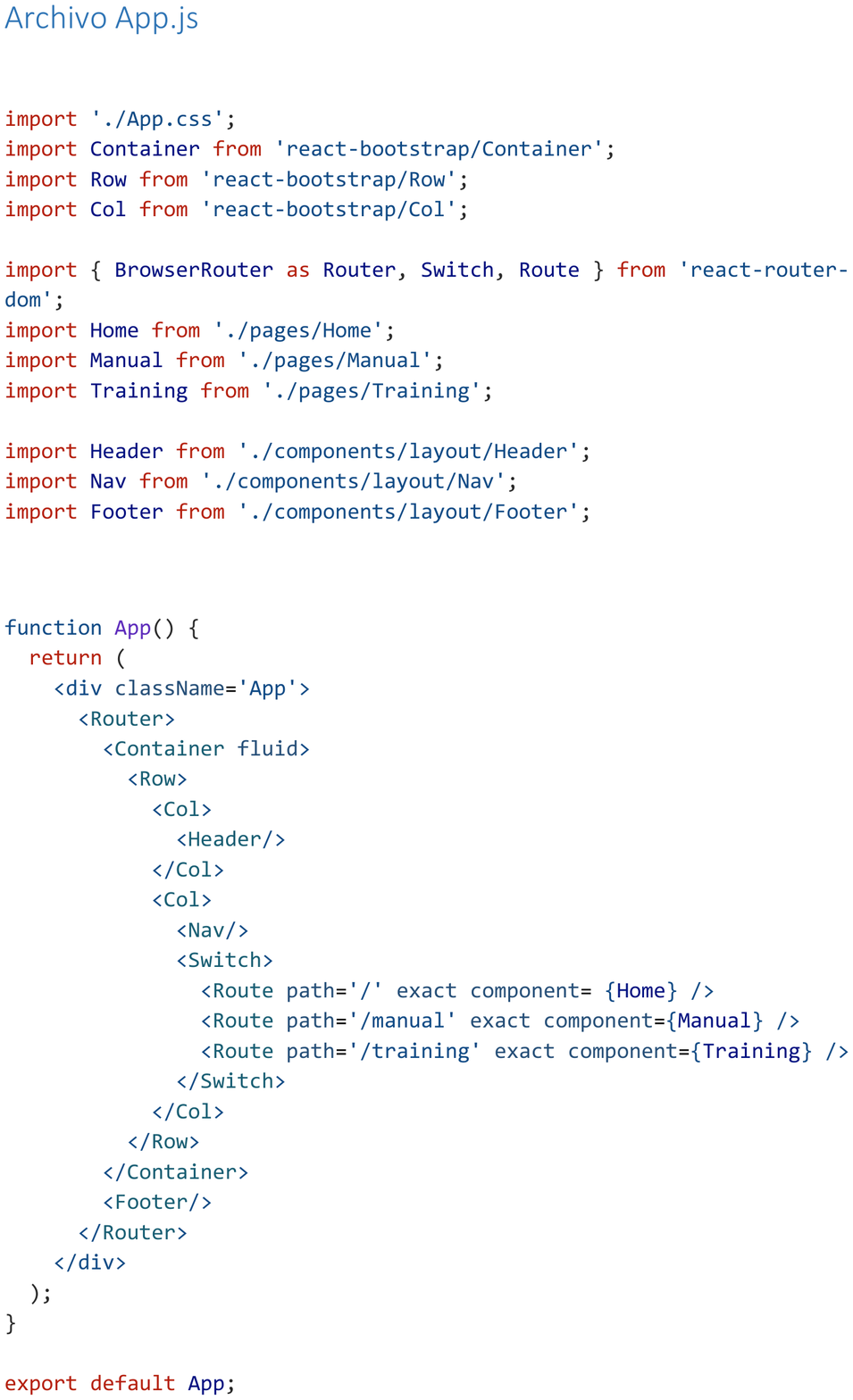}
\includepdf [pages=1]{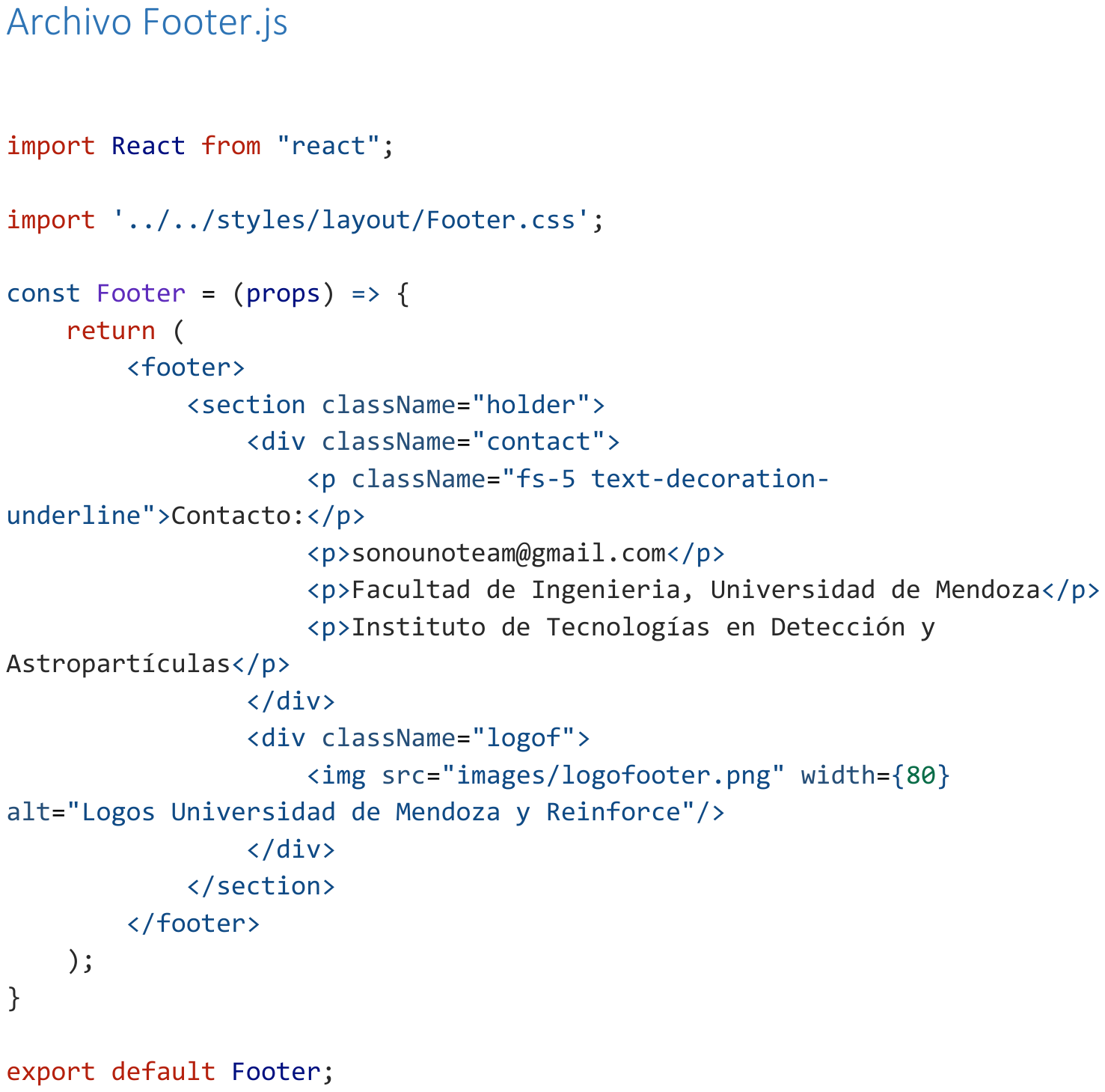}
\includepdf [pages=1]{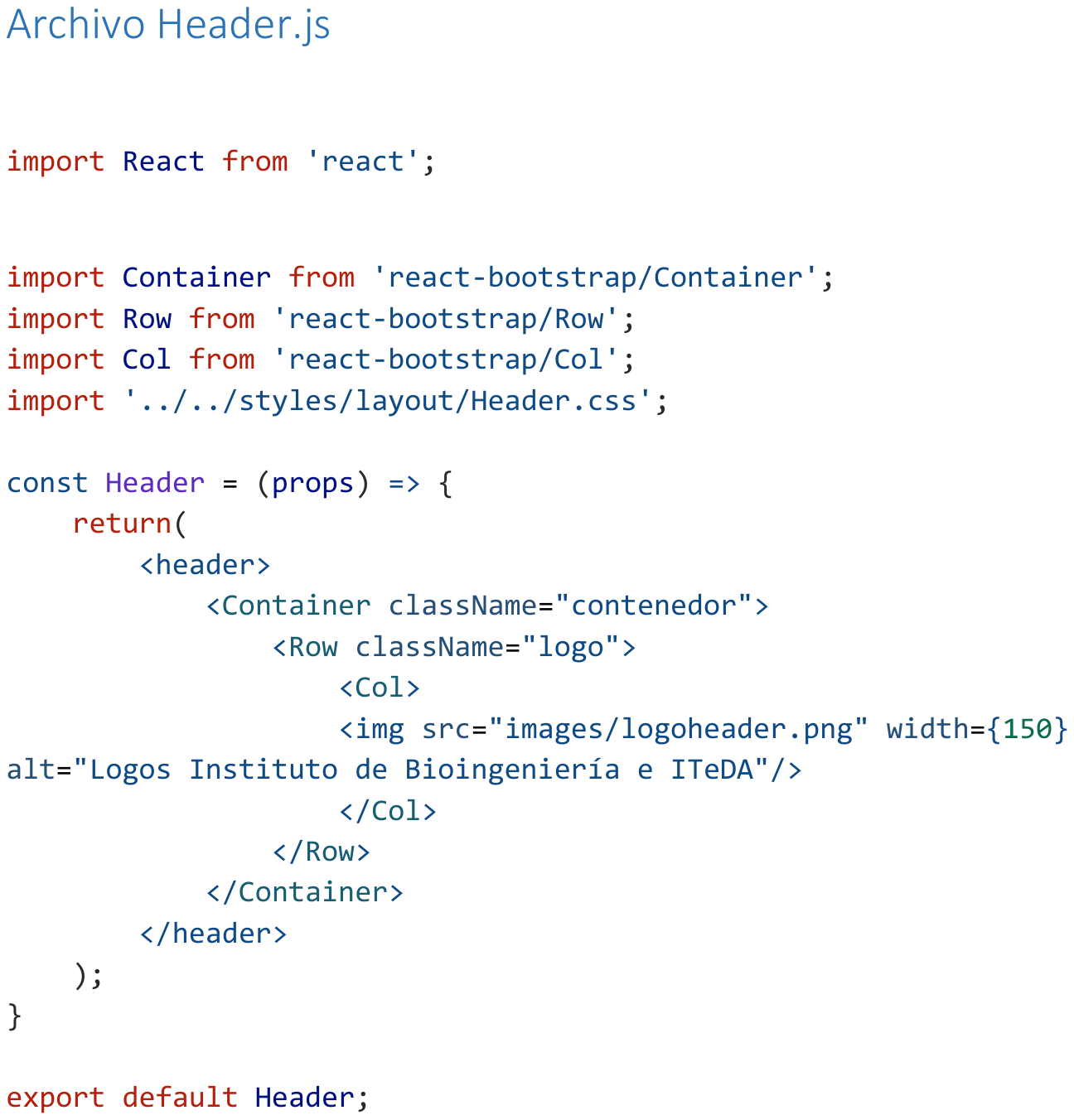}
\includepdf [pages=1]{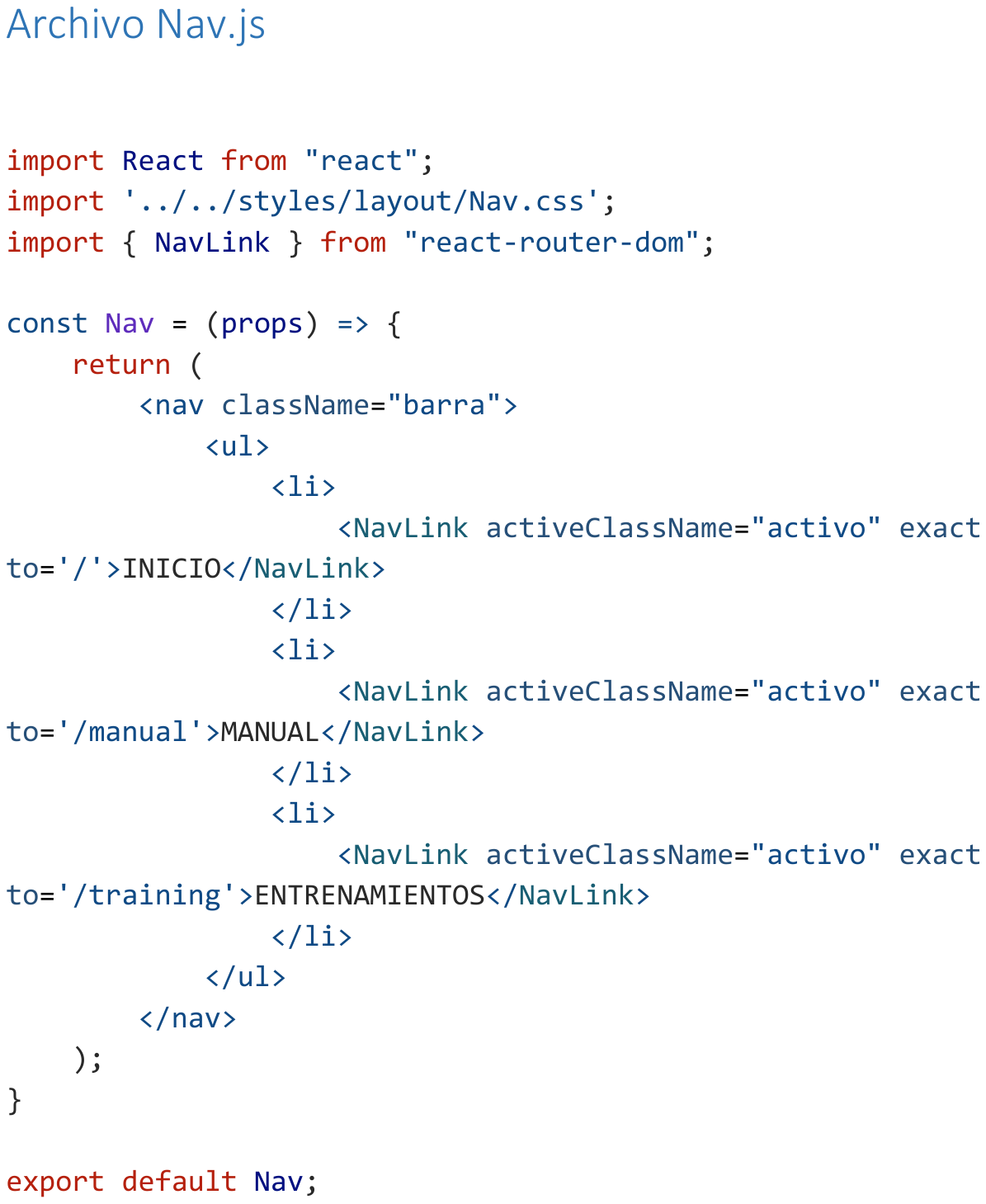}
\includepdf [pages=1-2]{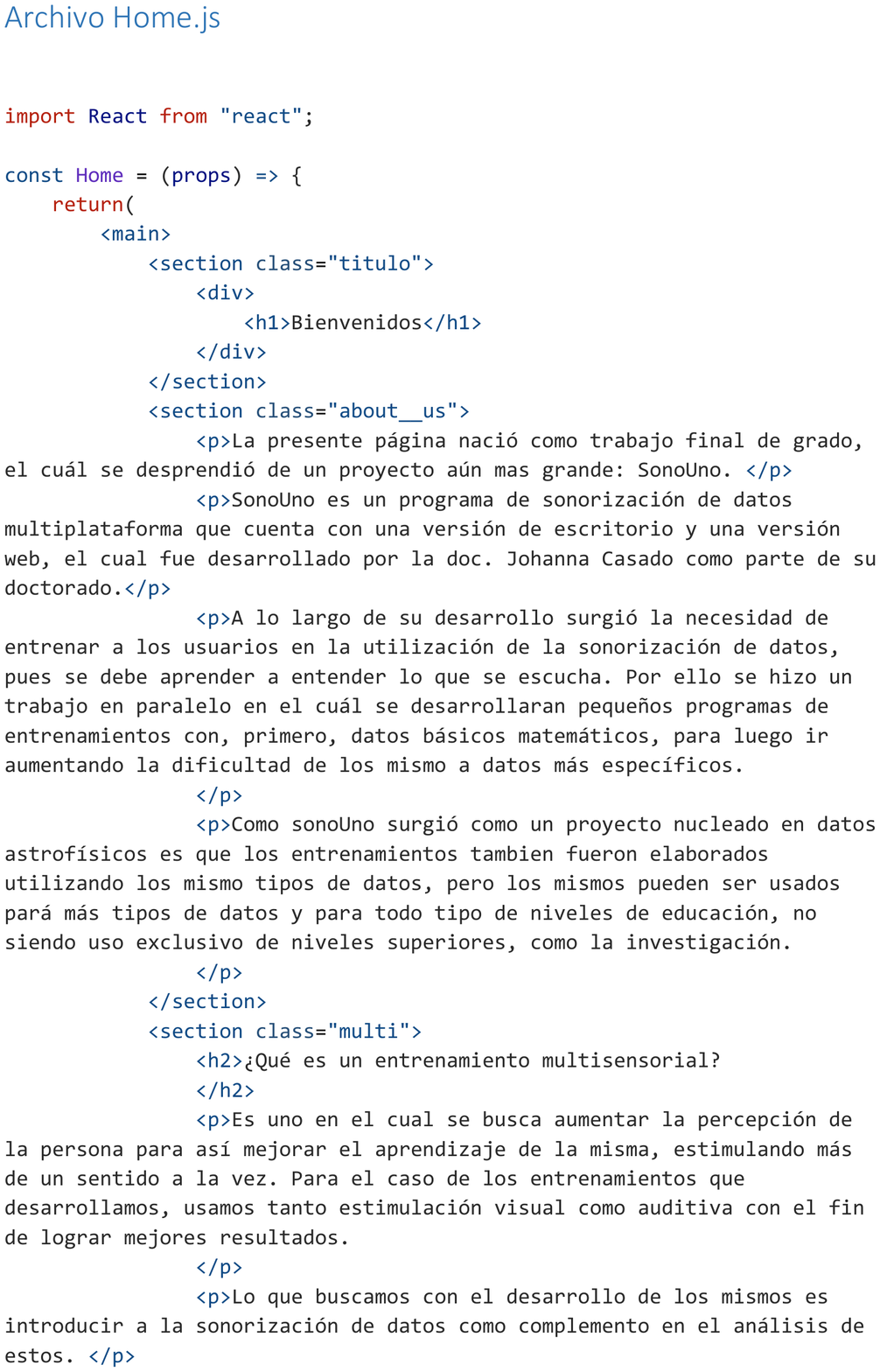}
\includepdf [pages=1]{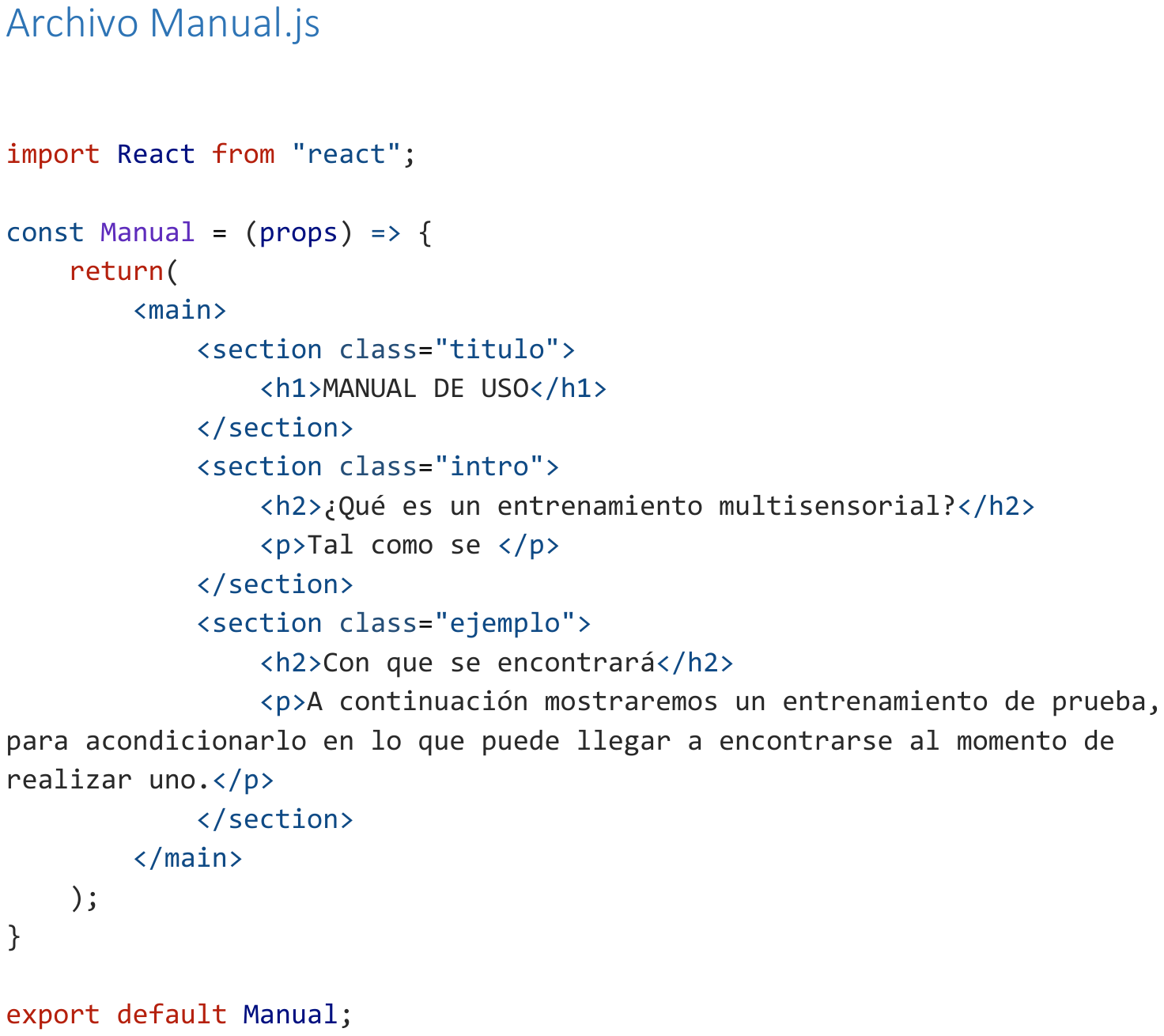}
\includepdf [pages=1-3]{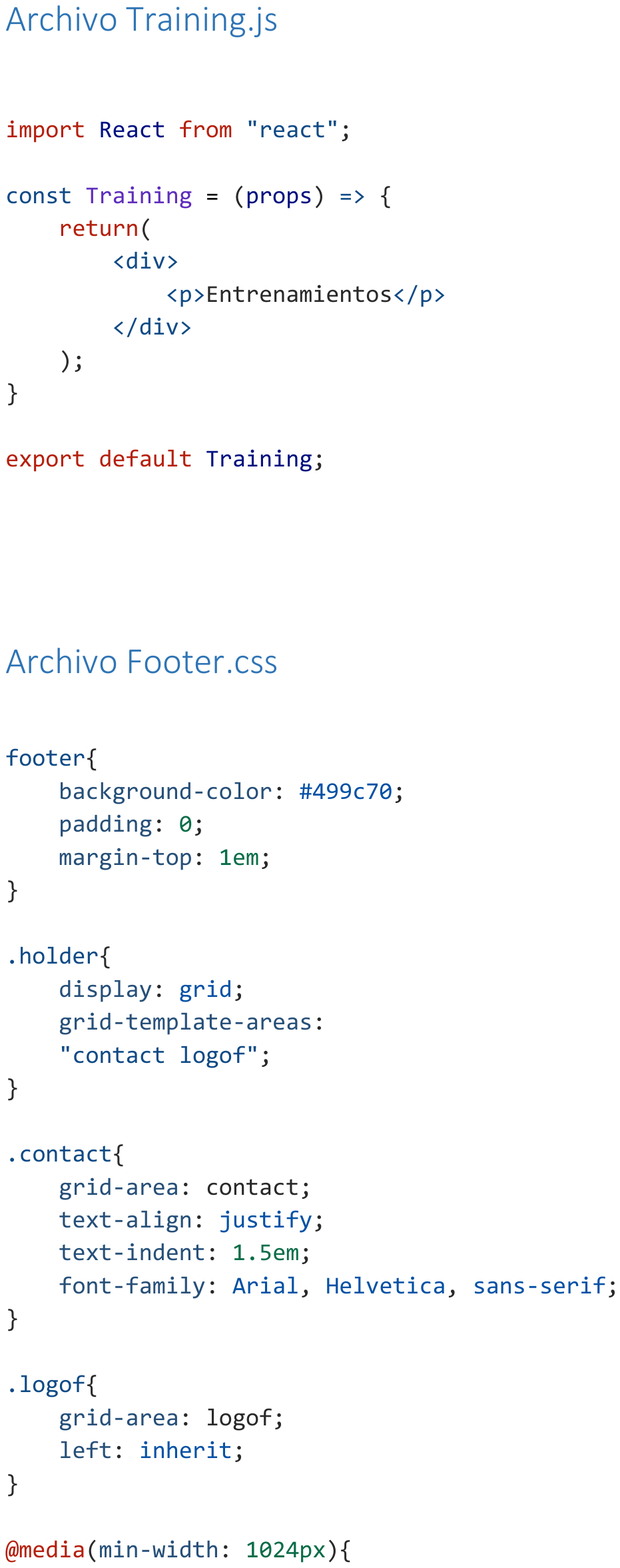}

\chapter{Paper}
\label{cap:apendB}



\section*{Supplementary material}

El software se usó para realizar el diseño de un entrenamiento en sonorización con datos astronómicos obtenidos de la base de datos Sloan Digital Sky Survey (SDSS) y sonorizados con el software SonoUno [9].

SonoUno es un programa de sonorización de datos multiplataforma que cuenta con una versión de escritorio y una versión web. En el caso de este trabajo se utilizó su versión escritorio, donde se abrieron los datos previamente descargados, se realizaron diferentes cortes en eje de abscisas y se guardaron las imágenes y sonidos con los cuales fue diseñado el entrenamiento.

### B.  Diseño del Entrenamiento

El objetivo principal del entrenamiento fue familiarizar a los participantes con el análisis multisensorial de señales, en particular con datos astronómicos. En general, la aproximación a este tipo de datos es netamente visual, en esta propuesta se agrega el uso de sonorización como herramienta para el análisis, la cual es nueva en comparación con la visualización.

El entrenamiento fue dividido en tres bloques con tareas distintas. El primero consistió en el empleo de señales matemáticas: función senoidal, cuadrada, lineal creciente o decreciente, generadas en excel y salvadas como archivos csv. Dichas señales fueron desplegadas en el programa sonoUno, donde se estableció la frecuencia máxima de sonido de 1700 Hz, y se guardaron tanto el gráfico como el audio. Dichos archivos se desplegaron auditiva y visualmente durante el entrenamiento. La función principal de este primer bloque era orientar a la persona en el uso de la sonorización. El participante debía identificar y seleccionar (el software permite desplegar botones para la selección por teclado) la señal que veía y oía.

En el segundo bloque se desplegó solo la señal sonora generada a partir de rangos específicos de un archivo de datos de un espectro de galaxia descargado de Sloan ('SDSS J115845.43-002715.7') (Figura 1). Dicho archivo se trabajó con sonoUno, seleccionando la frecuencia máxima de sonido en 1700 Hz para generar los archivos de audio y luego se fueron identificando diferentes rangos de longitud de onda, donde se evidenciaban líneas de absorción y líneas de emisión.

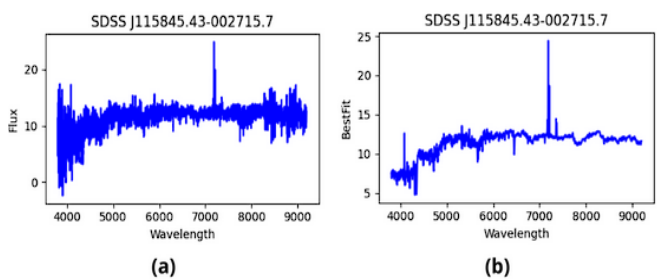

Figura 1. Imágenes obtenidas del sonoUno, utilizando los datos SDSS J115845.43-002715.7. (a) Grafico de Flux con respecto a Wavelength . (b) Grafico de BestFit con respecto a Wavelegth.

Los rangos seleccionados con líneas de emisión fueron: 7100-7300; 7300-7500. Los rangos seleccionados con líneas de absorción fueron: 6300-6600; 5500-5900; 6000-6300. Dichos rangos se seleccionaron tanto en la (Figura 1(a)) y con reducción de ruido (Figura 1(b)). En total, la persona tuvo que clasificar 10 señales, donde se encontraban cuatro líneas de emisión y seis líneas de absorción.

Finalmente, el tercer bloque reutilizaba las señales ya escuchadas en el bloque anterior pero sumando la ayuda visual. En todos los casos los participantes debían identificar si lo que oían y veían se trataba de una línea de emisión o de absorción.

La duración total del entrenamiento fue de aproximadamente 10 minutos. Los diferentes bloques estaban articulados de manera consecutiva, sin tiempo de descanso. Luego de cada señal sonora o combinación de señal sonora y visual, se mostraba un mensaje recordando las opciones entre las cuales estaba la señal presentada y las teclas con las que se podía seleccionar cada opción.

### C.  Realización del entrenamiento

A principios de abril, ya con el entrenamiento diseñado, se ejecutó el primer test en un grupo focal, integrado por miembros del Instituto de Tecnologías en Detección y Astropartículas (ITeDA), sede Mendoza, en formato presencial. El grupo de participantes estuvo formado por un ingeniero en Física Médica, 2 técnicos en electrónica, un doctor en astronomía experimental, una diseñadora industrial y una doctora en astronomía con experiencia en sonorización.

Las y los participantes fueron divididos en grupos de dos para que cada uno de ellos realice el entrenamiento en las dos computadoras disponibles; la duración total del testeo fue de una hora y media. Antes de comenzar, se les dio una breve explicación de las tareas a realizar en cada bloque y se les mostró un entrenamiento de prueba para que se familiarizaran con el recurso. El test consistió en escuchar y ver cada señal (Figura 2), o escuchar solamente en el caso del bloque dos, y responder a qué señal correspondía. Como última precaución, se les pidió a cada grupo que revisaran el nivel de volumen para que no fuera incómodo para ellos, ya que los sonidos que se utilizaron tenían unas frecuencias medianamente elevadas, hasta 1700 Hz.

Luego de la introducción, se procedió a dar inicio al entrenamiento, manteniendo el lugar donde se realizó lo más silencioso posible para no producir distracciones.

### D.  Adquisición y análisis de datos

Las respuestas de cada participante, a medida que el entrenamiento corría, fueron guardadas en archivos excel y archivos csv; una opción ofrecida por el software para luego hacer el análisis de datos.

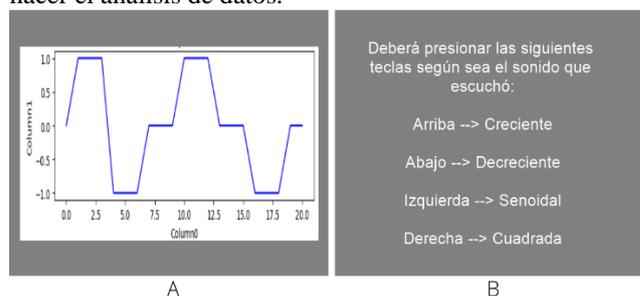

Figura 2. A. Señal mostrada al participante, acompañada de la señal de audio, correspondiente al 1°bloque. B. Opciones ofrecidas al participante, debiendo éste utilizar las teclas para responder según le parezca correcto.

Dichas respuestas se agruparon por cada bloque programado en hojas separadas, con los respectivos archivos mostrados. De esta forma, por cada participante, luego de la finalización del entrenamiento, se obtuvieron dos archivos con los datos ordenados, identificados con el nombre del voluntario y las respuestas obtenidas por el participante para cada sonido escuchado y la correspondiente respuesta correcta.

Todos los datos obtenidos de cada participante se procesaron en planillas excel, donde se volcaron los datos aportados por todos participantes, organizándolos en hojas

separadas y en tablas con la cantidad de aciertos, fallos y no contestaciones de cada bloque. El análisis general del entrenamiento, se realizó teniendo en cuenta solamente los aciertos de cada participante.

### III. RESULTADOS

Los resultados de este estudio se muestran en la Tabla 1. En ella se observa la cantidad de aciertos por cada bloque, recordando que cada participante realizó el entrenamiento una única vez y que el total de señales por bloque era de cuatro para el primero, diez para el segundo y diez para el tercero. Podemos destacar que los participantes presentaron menos dificultad a la hora de detectar señales auditivas con soporte visual, como fue el caso del tercer bloque. Por otra parte, cuando se debía detectar el sonido sin apoyo visual, la cantidad de aciertos fue menor en la mayoría de participantes (ver resultados del segundo bloque).

TABLA 1

| Participantes | Aciertos en 1°Bloque | Aciertos en 2°Bloque | Aciertos en 3°Bloque |
|---|---|---|---|
| 1°Part. | 3 | 2 | 6 |
| 2°Part. | 1 | 2 | 6 |
| 3°Part. | 4 | 5 | 4 |
| 4°Part. | 4 | 2 | 3 |
| 5°Part. | 4 | 4 | 8 |
| 6°Part. | 2 | 4 | 1 |

Aciertos por bloque de cada participante. En el 1°bloque, el participante debía detectar correctamente cuatro señales. En el 2°bloque y el 3°bloque debía detectar diez señales correctamente.

Estos resultados nos dan una primera aproximación a la capacidad que tienen los participantes de este test para detectar distintas señales, la cual podría incrementarse al aumentar el número de sesiones de entrenamiento.

En la Tabla 2 se observa el porcentaje de aciertos global por cada bloque, potenciando los comentarios previos: con multimodalidad, se pasa de 32% a 47% de detección correcta. En la presente investigación no fue posible realizar un análisis estadístico ya que se trabajó con un solo grupo de voluntarios.

TABLA 2

| Bloque y estímulo | Aciertos | Porcentaje de Aciertos |
|---|---|---|
| Primer bloque: visión y audición | 18 | 75 |
| Segundo bloque: audición | 19 | 31,667 |
| Tercer bloque: visión y audición | 28 | 46,667 |

### IV. CONCLUSIONES

En base a los resultados obtenidos en este testeo para el entrenamiento diseñado, podemos reafirmar que el rendimiento auditivo fue mejor cuando se apoyó al estímulo auditivo con uno visual como complemento. Para futuras experiencias se deben planificar entrenamientos que aprovechen esta ventaja inicial como método introductorio, para luego dejarla de lado, logrando un aumento del rendimiento auditivo y la independencia de éste con respecto a estímulos visuales que lo acompañen.

Cabe resaltar que para lograr un aprendizaje en la detección de señales, no basta con una sesión aislada de entrenamiento, sino que debe planificarse un plan de sesiones extendidas en varios días, dejando el tiempo óptimo de descanso entre ellas para así obtener resultados óptimos, asociados con los mecanismos de percepción.

Un trabajo a futuro podría centrarse en la comparación de planes de entrenamientos que inicien con apoyo visual versus planes de entrenamiento que solo se apoyen en estímulos auditivos, logrando llegar a más conclusiones sobre el apoyo visual. Integrar estos entrenamientos a un rango más amplio de personas, no limitándose únicamente a personas que soporten el estímulo visual, son parte de propuestas relacionadas con la astronomía inclusiva.

### REFERENCIAS

[1] Oviedo, G. L. (2004). La definición del concepto de percepción en psicología con base en la Teoría de Gestalt. Revista de estudios Sociales, 18, 89-96.
[2] Ittelson, W. (1973). *Environment and cognition.* (141-154) Nueva York: Seminario.
[3] Portellano, J. (2006). *Introducción a la neuropsicologia.* España-Madrid: McGraw-Hill Interamericana.
[4] Triveño Mosquera, M., Bembibre Serrano, J., & Arnedo Montoro, M. (2019). Neuropsicología de la percepción visual. En Triveño Mosquera, Bembibre Serrano, &. Arnedo Montoro, Neuropsicologia de la percepción (págs. 60-63). Madrid: Editorial Síntesis.M.
[5] Triveño Mosquera, M., Bambibre Serrano, J., & Arnedo Montoro, M. (2019). Neuropsicología de la percepción auditiva. En M. Triveño Mosquera, J. Bambibre Serrano, & M. Arnedo Montoro, Neuropsicología de la percepción (pp. 112-116). Madrid: Editorial Sintesis.
[6] Opoku-Baah, C., Schoenhaut, A., Vassall, S., Tovar, D., Ramachandran, R., & Wallace, M. (2021). Visual Influences on Auditory Behavioral, Neural, and Perceptual Processes: A Review. Journal of the Association for Research in Otolaryngology: JARO, 22(4), 365-386.
[7] Oxenham, A. J. (2018). How We Hear: The Perception and Neural Coding of Sound. Annu Rev Psychol, 69(1), 27-50.
[8] Peirce, J. W. et al (2019). PsychoPy2: Experimentos en el comportamiento facilitados. Behav Res 51, pp195-203.
[9] Casado, J., De La Vega, G., Díaz-Merced, W., Gandhi, P., & García, B. (2019). SonoUno: a user-centred approach to sonification. *Proc. of the International Astronomical Union*, *15*(S367), 120-123.

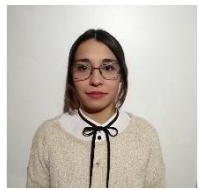

**Natasha Bertaina,** estudiante avanzada de la Carrera de Bioingeniería de la Universidad de Mendoza. Actualmente me encuentro realizando mi trabajo final bajo la dirección de la Bioingeniera Johanna Casado.

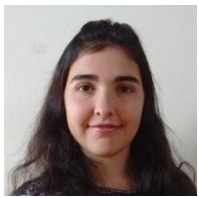

**Johanna Casado,** estudiante avanzada de doctorado, trabajo en el Instituto de Tecnologías en Detección y Astropartículas (ITeDA) y en el instituto de Bioingeniería de la Universidad de Mendoza. Mi tema principal de investigación es el desarrollo de aplicaciones accesibles para personas con discapacidad visual y centradas en el usuario, para mi tesis de doctorado el trabajo está aplicado al campo de la astronomía bajo la dirección de la Dra. Beatriz García.

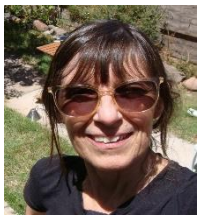

**Beatriz García,** Astrónoma, Investigadora del CONICET. Miembro de las colaboraciones internacionales en el Observatorio Pierre Auger y QUBIC. Ex presidente de la Comisión de Educación y Desarrollo de Astronomía de la IAU (2018-2019). Docente Universitaria y de nivel medio. Vicedirectora de ITeDA.



\end{document}